\documentclass{aa}  

\usepackage{graphicx}
\usepackage{txfonts}
\begin{document}


\title{A revised moving cluster distance to the Pleiades open cluster}
\subtitle{}

\author{
P.A.B. Galli \inst{1,2,3}
\and
E. Moraux\inst{1,2}
\and 
H. Bouy\inst{4}
\and
J. Bouvier\inst{1,2}
\and
J. Olivares\inst{1,2,5}
\and
R. Teixeira\inst{3}
}

\institute{Univ. Grenoble Alpes, IPAG, 38000, Grenoble, France
\and
CNRS, IPAG, F-38000 Grenoble, France 
\and
Instituto de Astronomia, Geof\'isica e Ci\^encias Atmosf\'ericas, Universidade de S\~ao Paulo, Rua do Mat\~ao, 1226, Cidade Universit\'aria, 05508-900, S\~ao Paulo - SP, Brazil
\and
Centro de Astrobiolog\'ia, INTA-CSIC, PO Box 78, ESAC Campus, 28691 Villanueva de la Ca\~nada, Madrid, Spain
\and 
Dpto. de Inteligencia Artificial, UNED, Juan del Rosal, 16, 28040, Madrid, Spain
}

\date{}

 
\abstract
{The distance to the Pleiades open cluster has been extensively debated in the literature over several decades. Although different methods point to a discrepancy in the trigonometric parallaxes produced by the \textit{\textit{Hipparcos}} mission, the number of individual stars with known distances is still small compared to the number of cluster members to help solve this problem. }
{We provide a new distance estimate for the Pleiades based on the moving cluster method, which will be useful to further discuss the so-called Pleiades distance controversy and compare it with the very  precise parallaxes from the \textit{Gaia} space mission. }
{We apply a refurbished implementation of the convergent point search method to an updated census of Pleiades stars to calculate the convergent point position of the cluster from stellar proper motions. Then, we derive individual parallaxes for 64~cluster members using radial velocities compiled from the literature, and approximate parallaxes for another 1146~stars based on the spatial velocity of the cluster. This represents the largest sample of Pleiades stars with individual distances to date. }
{The parallaxes derived in this work are in good agreement with previous results obtained in different studies (excluding \textit{\textit{Hipparcos}}) for individual stars in the cluster. We report a mean parallax of $7.44\pm 0.08$~mas and distance of $134.4^{+2.9}_{-2.8}$~pc that is consistent with the weighted mean of $135.0\pm 0.6$~pc obtained from the non-\textit{\textit{Hipparcos}} results in the literature.}
{Our result for the distance to the Pleiades open cluster is not consistent with the \textit{\textit{Hipparcos}} catalog, but favors the recent and more precise distance determination of $136.2\pm 1.2$~pc obtained from Very Long Baseline Interferometry observations. It is also in good agreement with the mean distance of $133\pm 5$~pc obtained from the first trigonometric parallaxes delivered by the \textit{Gaia} satellite for the brightest cluster members in common with our sample.}

\keywords{stars: distances, open clusters and associations: Pleiades }

\maketitle

\section{Introduction}

The problem of distance determination has always played a central role in astronomy. For example, distances are necessary to determine the dimensions of celestial objects, the physical properties of stars, and their true motion (i.e., spatial velocity). In this context, determining the distance to open clusters is also vitally important when calibrating the astronomical distance scale. The stars associated with open clusters are assumed to have similar properties (age, distance, kinematics, and chemical composition), which makes them ideal targets to refine stellar evolution models. 

As the closest open cluster to the Sun in terms of age and richness, the Pleiades is an important cornerstone to many studies related to star formation from physical models to observational properties of young stars.  Given the crucial role of the Pleiades for calibration purposes, one would expect its distance to be well established. However, there is still a current debate in the literature regarding the distance to the cluster.  

The Pleiades distance controversy began when the first studies in the pre-\textit{\textit{Hipparcos}} era using the isochrone fitting method \citep{Nicolet(1981),Giannuzzi(1995)} delivered a distance estimate for the cluster that roughly exceeds by 10-15\% the distance determination of $118.3\pm 3.5$~pc \citep{vanLeeuwen(1999)} that was obtained from the trigonometric parallaxes of the \textit{Hipparcos} catalog \citep{ESA(1997)}. Later studies using the same method but different samples of stars \citep{Pinsonneault(1998),Stello(2001),Percival(2005),An(2007)} confirmed the previous results for the distance of the cluster, and the new distance estimate of $120.2\pm 1.9$~pc \citep{vanLeeuwen(2009)} that was based on the re-calibrated parallaxes of 53 stars from the new version of the \textit{Hipparcos} catalog \citep{vanLeeuwen(2007)} did not change the situation. 

In the meantime, two independent studies returned trigonometric parallaxes for a few Pleiades stars. \citet{Gatewood(2000)} measured the parallax of seven stars in the cluster yielding a weighted mean distance of $130.9\pm 7.4$~pc, and \citet{Soderblom(2005)} find a distance of $134.6\pm 3.1$~pc based on trigonometric parallaxes for another three stars. 

Distance determination obtained from orbital modeling of eclipsing binaries is another possibility for estimating the distance to the Pleiades, and the results vary from $132.0\pm 5.0$~pc \citep{David(2016)} to $139.1\pm 3.5$~pc \citep{Southworth(2005)}. Unfortunately, this method is restricted to a few systems that have been discovered to date and that can be used in this regard \citep{Zwahlen(2004),Munari(2004),Groenewegen(2007)}. 

Recently, \citet{Melis(2014)} measured the trigonometric parallaxes of four stars (or stellar systems) in the Pleiades based on multi-epoch Very Long Baseline Interferometry (VLBI) observations. The weighted mean distance of $136.2\pm 1.2$~pc obtained in their work represents the most precise distance determination for the cluster to date, and the parallaxes for individual stars have a precision that is better than 1\%. Indeed, the VLBI trigonometric parallax method is the most precise and accurate technique for measuring distances nowadays, and it has already proved to deliver good results in nearby associations of young stars \citep[see e.g.,][]{Loinard(2007),Loinard(2008)}. 

In a recent paper, \citet{Madler(2016)} use a different approach to derive the distance of 15 stars in the cluster based on stellar twins. This method assumes that the difference in brightness between two stars with different locations in the sky, but identical physical properties (i.e., stellar twins), results directly from the difference in their distances. Thus, by knowing the distance of one star from other methods, they calculate the distance to its stellar twin. Doing so, they found an average distance of $134.8\pm 1.7$~pc for the Pleiades open cluster that clearly supports the non-\textit{Hipparcos} results. 

The moving cluster method is another alternative to compute distances of comoving groups of stars. The method takes the observables proper motions and radial velocities into account to compute the distance to individual cluster members \citep[see e.g.,][]{Galli(2012)}. Although a variant of this methodology has already been used in previous studies to investigate the distance of the Pleiades, the admittedly large error bars in the distances results of $130.6\pm 10.8$~pc \citep{Narayanan(1999)} and $125.9\pm 7.5$~pc \citep{Roeser(2013)} make it difficult to resolve the current dispute between the \textit{Hipparcos} and VLBI distance determinations of the cluster. 

In this context, we decided to revisit the Pleiades based on the moving cluster method by using a refurbished implementation of the method, an updated census of cluster members and more precise data to provide a revised distance determination to the cluster with this technique. The results obtained in this work are independent of previous distance determinations of the cluster in the literature and they will be useful for double-checking the upcoming state-of-the-art parallaxes delivered by the \textit{Gaia} space mission in the near future. 

This paper is organized as follows. In Sect.~2 we describe the moving cluster method and present a new implementation of the method that is used in this work. In Sect.~3 we present the sample of Pleiades stars and the dataset (proper motions and radial velocities) used in this analysis. Section~4 describes our convergent point analysis using different samples of cluster members to investigate our solution. In Sect.~5, we present the individual parallaxes (i.e., distances) for Pleiades stars obtained in this work, and discuss our results in Sect.~6. We summarize our results and conclusions in Sect.~7.

\section{The moving cluster method}

\subsection{Kinematic parallaxes}

Moving groups are kinematic aggregates of stars whose members share the same space motion. The low velocity dispersion, typically of a few km/s \citep{Mathieu(1986)}, allows them to be easily identified as overdensities in the velocity space \citep[see e.g.,][]{Antoja(2008)}. As discussed in the seminal work of \citet{Blaauw(1964)}, the proper motions of comoving stars in such kinematic groups converge to a vertex referred to as the convergent point (CP) of the moving group, because of their parallel space motion or the state of linear expansion.   

In this context, individual parallaxes for members of young moving groups can be derived from their proper motions, radial velocities and the CP position. They are given by, 
\begin{equation}\label{eq_plxind}
\pi=\frac{A\,\mu_{\parallel}}{V_{r}\tan\lambda}\, ,
\end{equation}
where $A=4.74047$ km yr/s is the ratio of one astronomical unit in km to the number of seconds in one Julian year, $\mu_{\parallel}$ is the proper motion component directed parallel to the great circle that joins the star and the CP \citep[see][for more details]{Galli(2012)}, $V_{r}$ is the radial velocity of the star, and $\lambda$ is the angular distance between the star and the CP position.  

Alternatively, for binaries (or multiple systems) and group members with unknown (or poor) radial velocities one can derive the kinematic parallax of the star from the spatial velocity of the moving group under the assumption that all members share the same space motion. In this case, it is given by
\begin{equation}\label{eq_plxapp}
\pi=\frac{A\,\mu_{\parallel}}{V_{space}\sin\lambda}\, ,
\end{equation}
where $V_{space}$ is the average spatial velocity of the moving group. In both cases, the parallax uncertainty is calculated from error propagation of the equations above, and takes the error budgets owing to proper motions, radial velocities, (or the spatial velocity) and the CP into account. The parallaxes derived in this way are obviously not as precise as those obtained from VLBI observations, but errors in the same order of the \textit{Hipparcos} catalog and recent trigonometric parallax results from the ground \citep{Weinberger(2013),Ducourant(2014)} can be achieved \citep[see e.g.,][]{Bertout(2006),Galli(2013)}. 

\subsection{A new implementation of the convergent point search method based on Markov chain Monte Carlo methods}

The computation of kinematic parallaxes using the moving cluster method requires prior knowledge of the CP position. The technique that we use to calculate the CP of a moving group is based on methods developed by \citet{deBruijne(1999)} and \citet{Galli(2012)} (see these papers for a more detailed description about the basic concepts, algorithms, and implementation of the method). As discussed in these papers, the convergent point search method (CPSM) simultaneously determines the most likely CP position and performs a membership analysis to select the moving group members. 

The CPSM takes stellar proper motions, the velocity dispersion, and a distance estimate of the cluster as input parameters. However, in general, the results are not sensitive to the assumed distance of the cluster \citep[see also][]{Mamajek(2005)}. We anticipate that the results and conclusions for the Pleiades cluster presented in the upcoming sections are rather insensitive to the assumed distance of the cluster ranging from the \textit{Hipparcos} distance ($\sim$120~pc) to the VLBI distance ($\sim$136~pc).  On the other hand, the velocity dispersion is one important input parameter in the CPSM to identify the moving group members. While a low velocity dispersion value would not be sufficient to recover all cluster members, a high velocity dispersion would enable the method to include field stars (interlopers) in the solution. 

\begin{figure*}[!htp]
\begin{center}
\includegraphics[width=14cm]{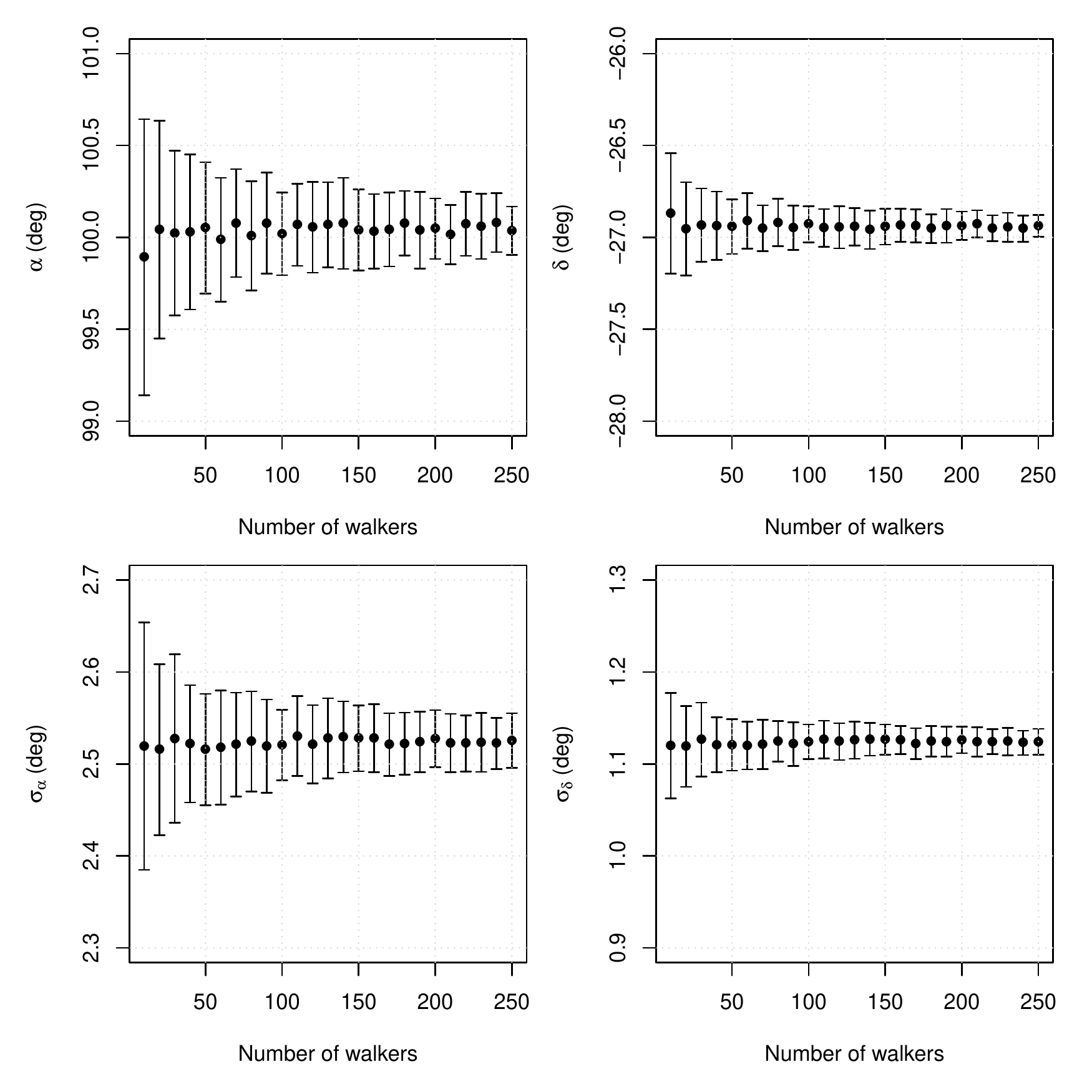}
\caption{
\label{fig1}
CP coordinates (\textit{upper panel}) and associated errors (\textit{lower panels}) for the TW Hydrae association as a function of the number of walkers (with 200 iteration steps) used in the MCMC-based version of the CPSM . Each point is an average of 1000 MCMC realizations of the method and the error bars indicate the scatter of the solution given for each ensemble of walkers.}
\end{center}
\end{figure*}

The original implementation of the CPSM \citep{deBruijne(1999),Galli(2012)} uses an analytical minimization routine to solve the least-square equations and return the CP position with its uncertainty. Although this procedure has already proved to deliver good results for nearby associations with tens of stars \citep{Galli(2013),Ducourant(2014)}, it was clearly less attractive when dealing with rich clusters, such as the Pleiades, which require more computing time to converge to a final solution. Thus, we developed a new version of the CPSM that exploits parallelism and exhibits good performance at low computational cost. This was achieved using the \texttt{emcee} package \citep{Foreman(2012)} written in Python programming language, which implements the Markov chain Monte Carlo (MCMC) method proposed by \citet{Goodman(2010)}. The algorithm was adapted to our purposes and applied to the general problem of computing the CP coordinates $(\alpha_{cp},\delta_{cp})$ of a moving group. Briefly, the new implementation of the CPSM that we use in this work exploits the parameter space using a number of so-called walkers and iteration steps, which are defined by the user to search for the CP position via Bayesian inference. The walkers move around the parameter space $(\alpha,\delta)$ and take tentative steps towards the lowest valley of $X^{2}$ that roughly defines the CP position \citep[see][]{deBruijne(1999),Galli(2012)}. Our final CP solution obtained after one MCMC realization is computed from the distribution of individual solutions given by the ensemble of walkers.   

In the following, we apply the CPSM to the TW Hydrae association whose CP position is well established \citep{Mamajek(2005),Ducourant(2014)} to calibrate our new implementation of the CPSM and illustrate its application.  In this context, we use the sample of 30~stars identified as kinematic members of the association by \citet{Ducourant(2014)}.  We vary the number of walkers from 10 to 200  and run the MCMC version of the CPSM using 50, 100, and 200 iteration steps for each walker to compare the results. Figure~\ref{fig1} shows the CP coordinates with their errors obtained for 200 iteration steps. We conclude that the formal errors on the CP coordinates are significantly larger than the observed scatter in the individual CP solutions obtained with the ensemble of walkers used in each case. This makes our choice of the minimum number of walkers to be used in our analysis rather arbitrary. In practice, we verified that our final solution is not sensitive to the number of walkers in the range of 100 to 200 if we run a significant number of MCMC realizations. We also confirmed that convergence of the Markov-chains of the ensemble of walkers was attained after 50 iterations where the mean of both CP coordinates is clearly bounded by the variance of the sample.  So, we decided to work with 100 walkers and 200 iteration steps as a more conservative approach. In addition, we emphasize that all CP solutions presented in this paper are calculated from a total of 1000~MCMC realizations of the CPSM, and the results listed are averaged values obtained from the distribution of the CP coordinates. 

\begin{figure*}[!htp]
\begin{center}
\includegraphics[width=0.45\textwidth]{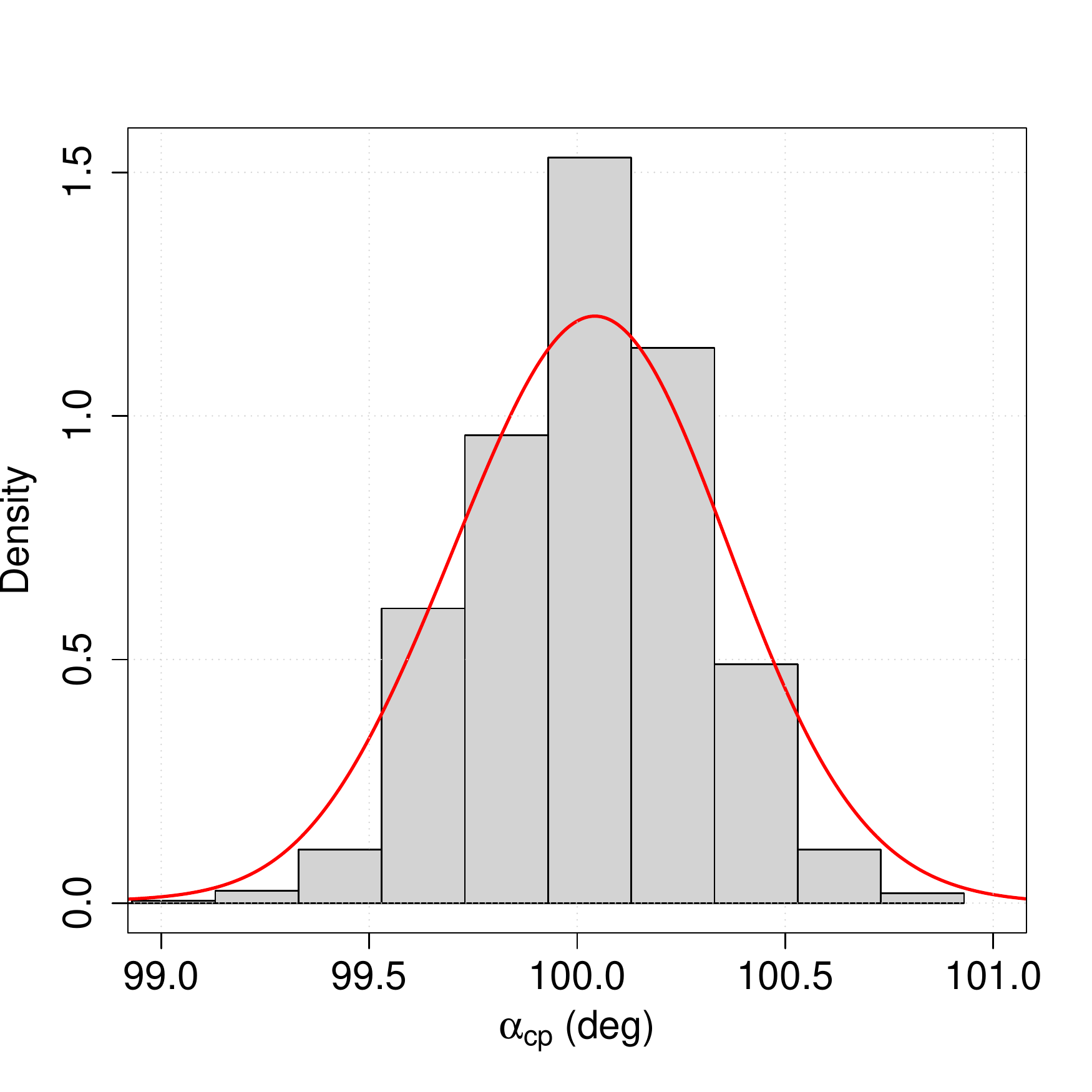}
\includegraphics[width=0.45\textwidth]{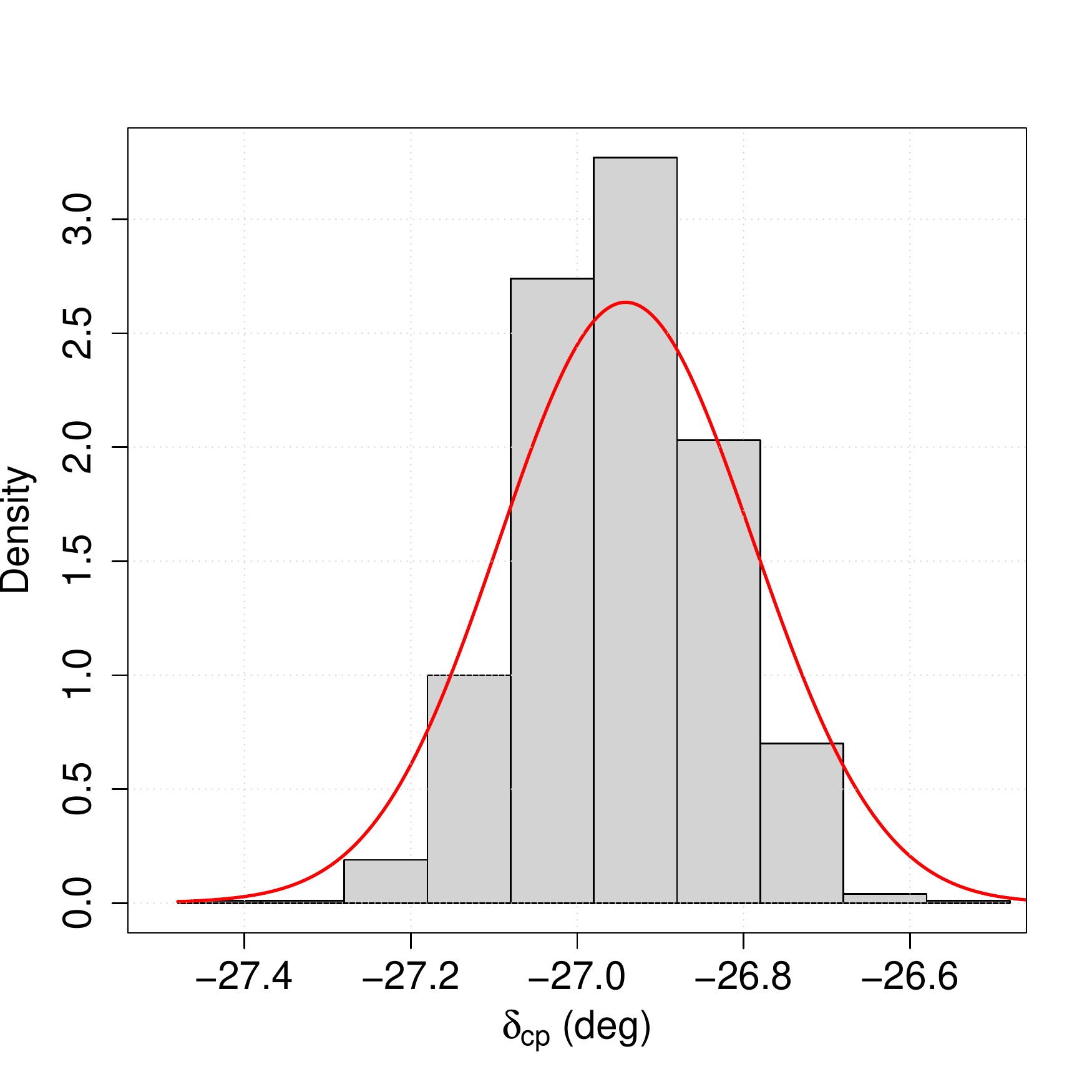}

\caption{
\label{fig2}
Distribution of the CP coordinates after 1000 MCMC realizations of the CPSM for the TW Hydrae association. The red solid line indicates the kernel density estimator.}
\end{center}
\end{figure*}

Doing so, we run the CPSM for the TW Hydrae association using a velocity dispersion of 0.8~km/s and distance estimate of 50~pc \citep[as described in Sect~4.1 of][]{Ducourant(2014)}. The resulting CP solution is located at (see also Fig~\ref{fig2})
\begin{equation*}
(\alpha_{cp},\delta_{cp})=(100.0^{\circ},-26.9^{\circ})\pm(2.5^{\circ},1.1^{\circ})\, .
\end{equation*} 
We conclude that our result is in very good agreement with the more recent CP solution obtained by \citet{Ducourant(2014)}, $(\alpha_{cp},\delta_{cp})=(100.1^{\circ},-27.1^{\circ})\pm(3.0^{\circ},1.3^{\circ})$, but the formal errors of our solution are smaller. Thus, we have tested and calibrated our new version of the CPSM based on the  MCMC method, and it will be used in the upcoming sections to investigate the CP position of the Pleiades cluster.

\section{Sample of Pleiades stars}

\subsection{An updated census of the cluster (Bouy et al. 2015)}

In a recent work, \citet{Bouy(2015)} apply a probabilistic method based on multivariate data analysis \citep[see][]{Sarro(2014)} to select high probability members of the Pleiades open cluster in the DANCe \citep{Bouy(2013)} and Tycho-2 \citep{Hog(2000)} catalogs. They identify a total of 2107 high probability members of the Pleiades open cluster making it the most complete census of the cluster to date. This list includes most cluster members identified in previous surveys \citep[e.g.,][]{Stauffer(2007),Lodieu(2012)} and 812 previously unknown members. As discussed in Sect.~2 of their paper, the absence of some cluster members in this sample is justified by their different properties implying low membership probabilities, while a few of them may have been missed because they fall in the domain of incompleteness between the DANCe survey ($i\geq 14$~mag) and the Tycho-2 catalog ($V\leq12$~mag). 

The catalog of Pleiades stars given by \citet{Bouy(2015)} contains stellar positions, proper motions, multi-band photometry (filters $u,g,r,i,z,Y,J,H,K_{s}$) and membership probabilities. This is the initial sample of stars that we use in our forthcoming analysis to investigate the distance of the Pleiades.

\subsection{Proper motions}

Our initial sample of Pleiades stars consists of 2010 stars (prob. $\geq75\%$) from the DANCe survey and 207 stars (prob. $\geq48\%$) from the Tycho-2 catalog with 110 stars in common \citep[see][]{Bouy(2015)}. Thus, the proper motions that we use in this work come mostly from the DANCe survey (1900 stars), and we use the proper motions from Tycho-2 for the remaining stars (i.e., the brightest stars in our sample). The proper motions from the DANCe catalog provided by the Vizier/CDS tables are not anchored to an absolute reference system.  We applied the procedure described in Sect.~7.10 of \citet{Bouy(2013)} to tie these proper motions to the International Celestial Reference System (ICRS). The corrected measurements are presented in the upcoming tables, together with other results from our analysis. 

\begin{figure*}[!htp]
\begin{center}
\includegraphics[width=0.49\textwidth]{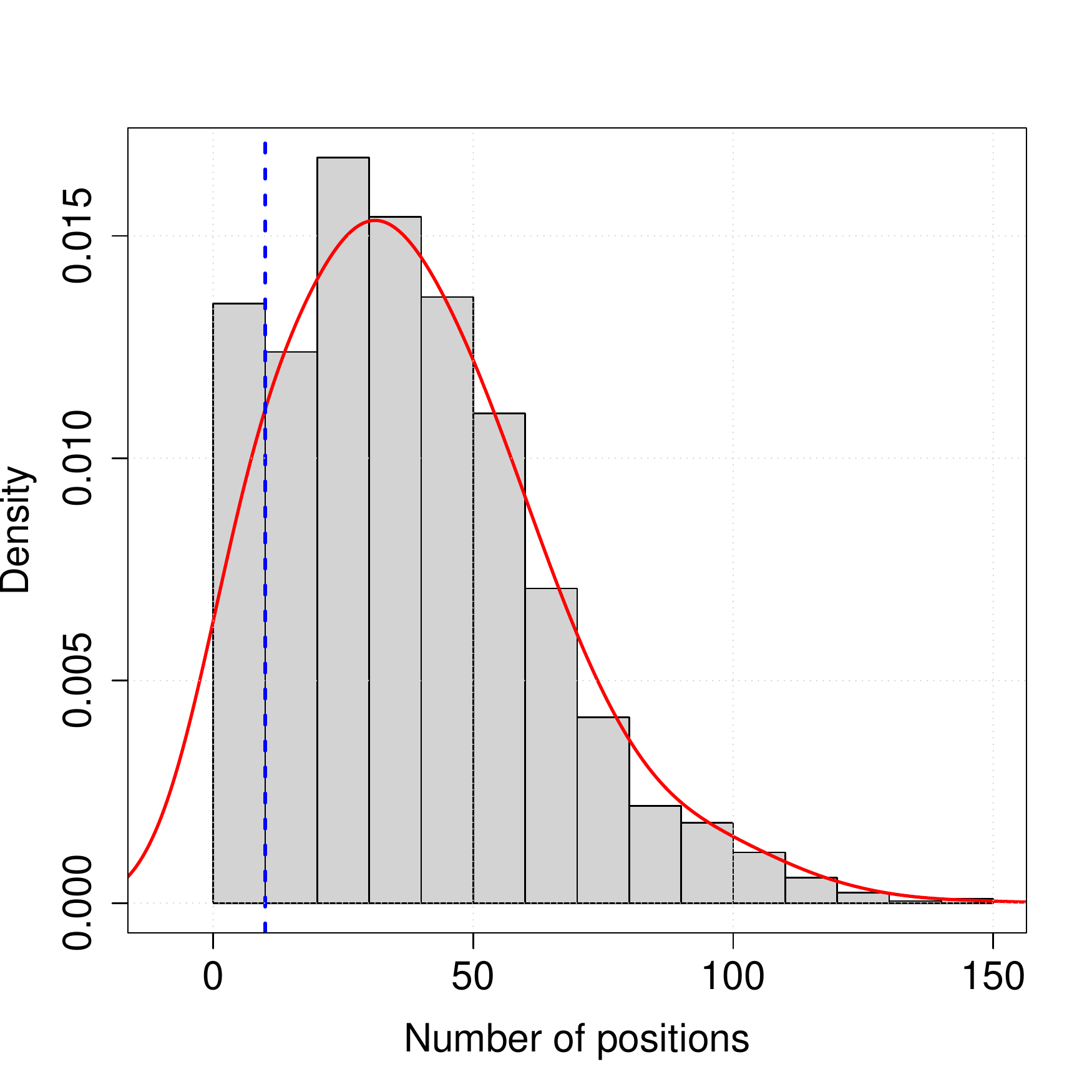}
\includegraphics[width=0.49\textwidth]{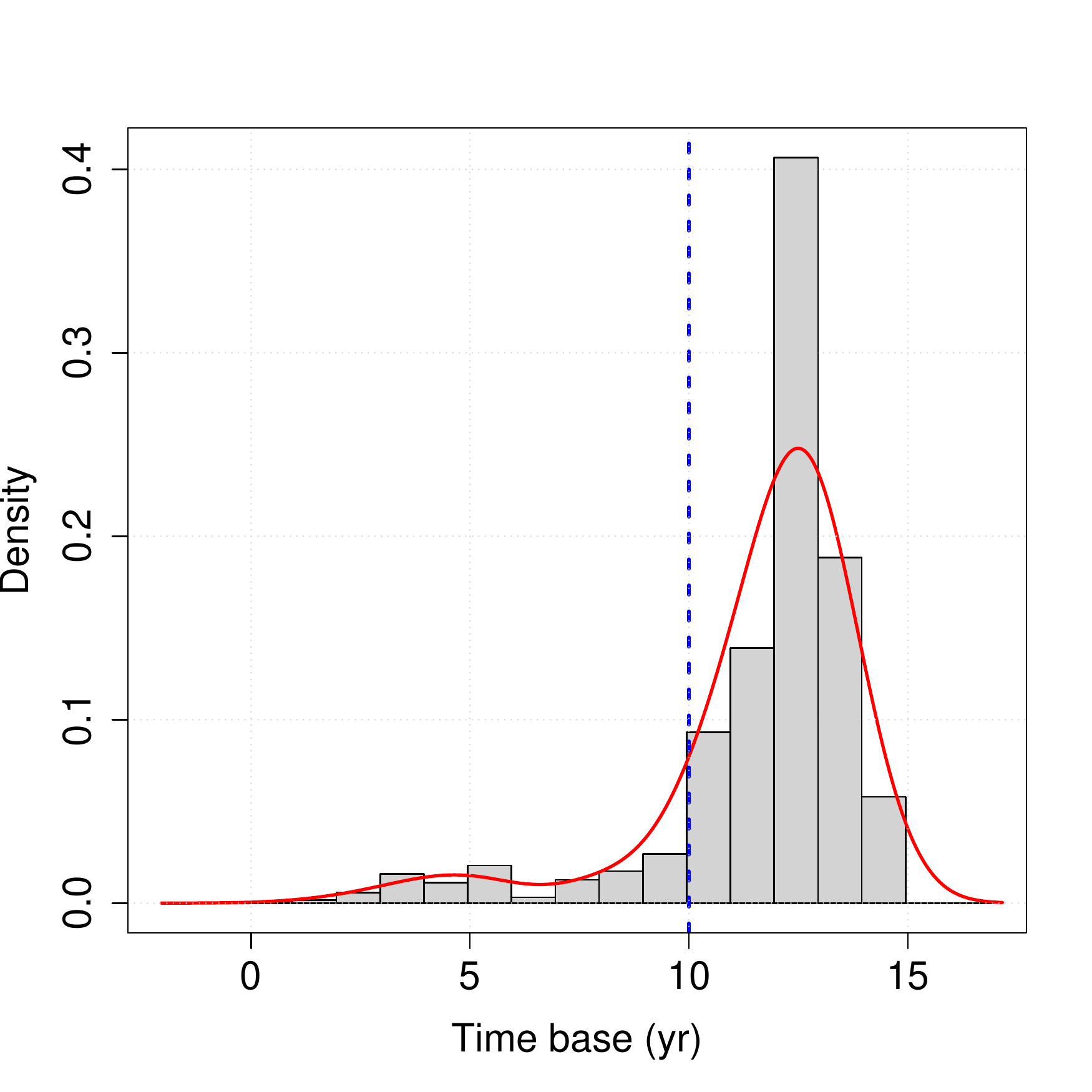}
\caption{
\label{fig3}
Number of positions \textit{(left panel)} and time-base \textit{(right panel)} used to compute the stellar proper motions for Pleiades stars in the DANCe project. The red solid lines indicate the kernel density estimator. The blue dashed lines indicate the selection criteria applied to our sample as described in Sect.~4.1. \bigskip\bigskip\bigskip}
\end{center}
\end{figure*}

The DANCe project combines multi-epoch panchromatic images collected with different instruments at various observatories \citep[see][for more details]{Bouy(2013)} to derive accurate proper motions over a wide field from the ground. The proper motions were computed using, on average, about 40 stellar positions collected at different epochs and a time-base longer than 10~yrs for most stars in our sample (see Fig.~\ref{fig3}). The stars have a median proper motion of $(\mu_{\alpha}\cos\delta,\mu_{\delta})=(+19,-44)$~mas/yr and a median precision better than 1~mas/yr in each component. 

\begin{figure*}[!htp]
\begin{center}
\includegraphics[width=0.49\textwidth]{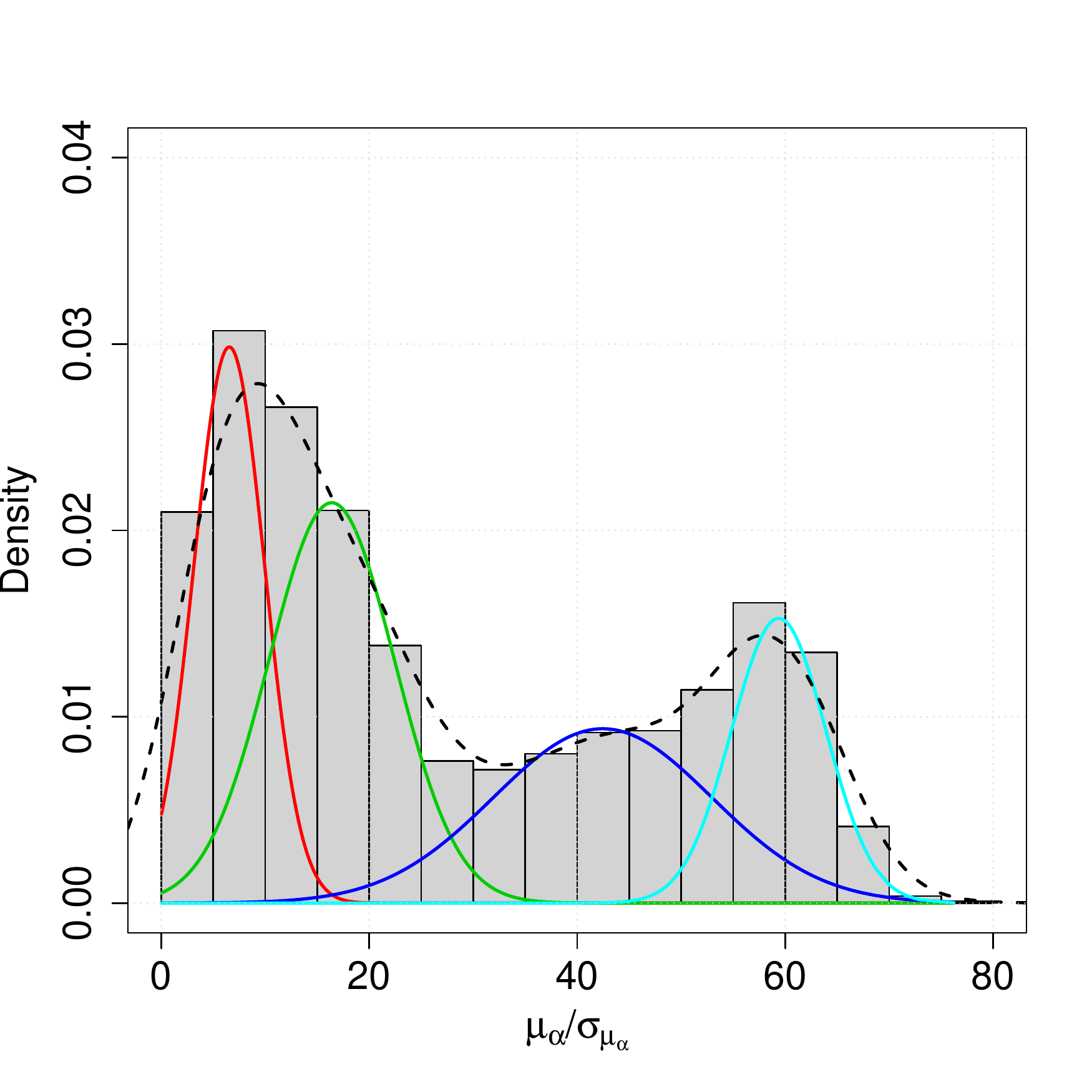}
\includegraphics[width=0.49\textwidth]{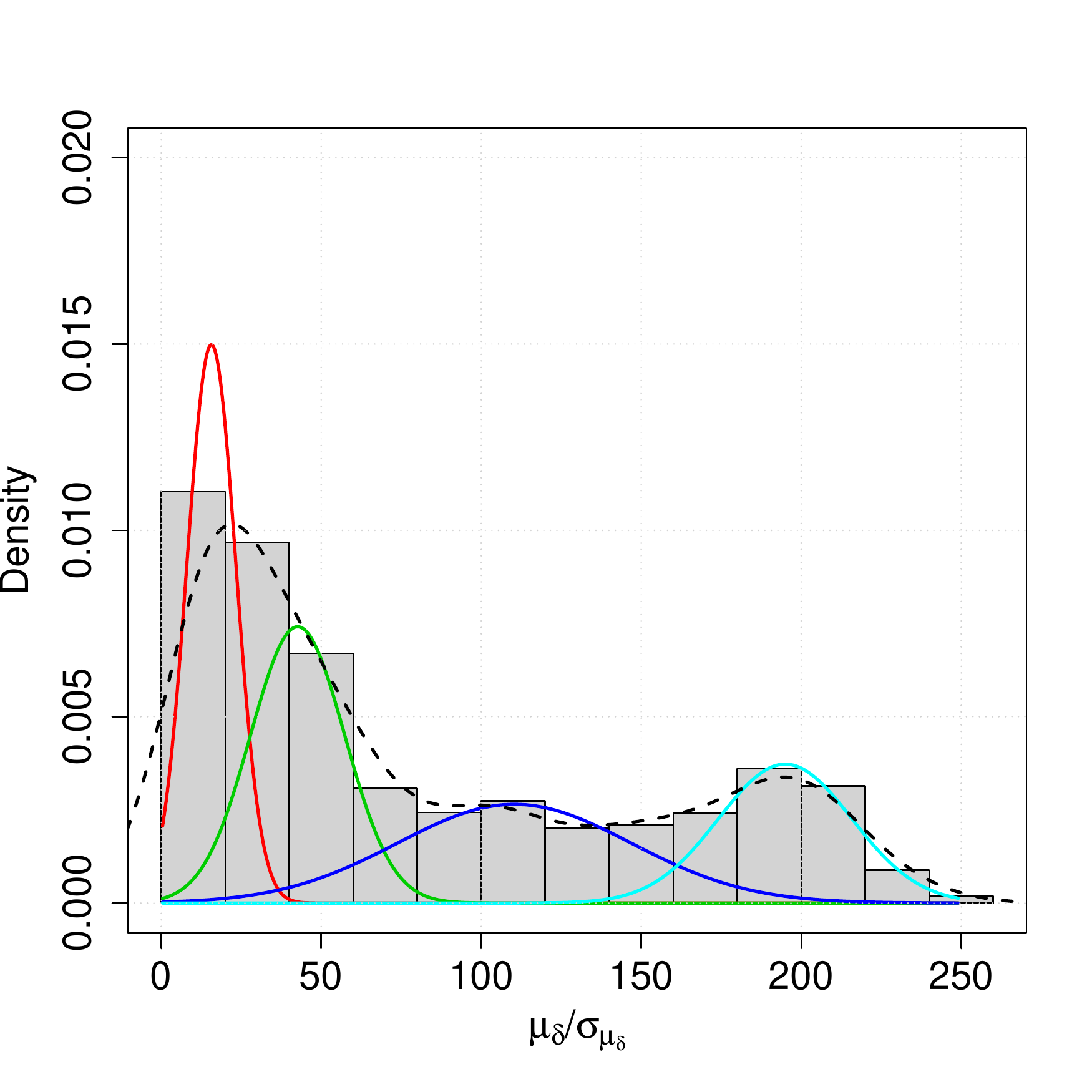}
\caption{
\label{fig4}
Distribution of the relative errors on proper motions (i.e., $\mu/\sigma_{\mu}$) in right ascension \textit{(left panel)} and in declination \textit{(right panel)}. The solid lines indicate the normal components of our mixture model (with $k=4$), and the dashed line indicates the observed distribution of proper motions errors. \bigskip\bigskip\bigskip
}
\end{center}
\end{figure*}

Figure~\ref{fig4} shows the relative errors on proper motions in each component and reveals the existence of different subgroups in our sample regarding the quality of their proper motions measurements. One reason for the observed distribution of proper motion errors is the different observing campaign for each star in the sample, which is based on archival data collected in the DANCe project. 

To identify the stars with best proper motion measurements in our sample (to be used in Sect.~4.2) we model the observed distribution of proper motion errors as a mixture of $k$-components. In this case, each component follows a normal distribution $\mathcal{N}(\mu,\sigma)$ with mean $\mu$ and variance $\sigma^{2}$. The mixture model probability distribution is given by 
\begin{equation}\label{eq1}
p=\sum_{i=1}^{k}\phi_{i}\,\mathcal{N}(\mu_{i},\sigma_{i})\, ,
\end{equation}
where $\phi_{i}$ is the probability that a star belongs to the $k$-component of the model. We use the \texttt{mixtools} package in R programming language to calculate the mixture model parameters $(\phi_{i},\mu_{i},\sigma_{i})$ based on the expectation-maximization algorithm implemented in the \textit{normalmixEM} procedure. 

We perform a Kolmogorov-Smirnov (KS) test between the observed distribution of proper motion errors (see Fig.~\ref{fig4}) and the empirical probability-density function that is given by our mixture model to investigate the number of components (i.e., the value of $k$) that best represent our data. To assess the robustness of this finding, we construct a total of 1000 synthetic samples from Eq.~(\ref{eq1}) using $k=2,3,$ and $4$. Then, we perform a KS test between each synthetic dataset and the observed distribution of proper motion errors. We compute the fraction $f_{0.05}$ of synthetic realizations of our data sample with a $p$-value higher than the adopted significance level of $\alpha=0.05$ \citep[see e.g.,][]{Feigelson(2012)}. Our results from this analysis are summarized in Table~\ref{tab_mixture} and they indicate that the mixture model with four components ($k=4$) is the one that best describes the observed distribution of proper motions errors. These results will be used in Section~4 to construct a control sample to investigate the CP position of the cluster.

\begin{table}[!h]
\centering
\caption{Results of the mixture model applied to the distribution of proper motions errors in right ascension and declination.   
\label{tab_mixture}}
\begin{tabular}{lccc}
\hline
k&$\mu_{i}$&$\sigma_{i}$&$f_{0.05}$\\
&(mas/yr)&(mas/yr)&(\%)\\
\hline
\multicolumn{4}{c}{\textbf{Right Ascension}}\\
\hline
$2$&12.4; 50.3&7.2; 11.2&0.9\\
$3$&10.7; 36.6; 58.8&6.0; 13.2; 4.8&36.0\\
$4$&6.5 ;16.2; 42.9; 59.4&3.4; 5.9; 11.7; 4.3&99.8\\
\hline
\multicolumn{4}{c}{\textbf{Declination}}\\
\hline
$2$&28.4; 145.9&17.0; 53.0&0.0\\
$3$&15.8; 44.0; 154.0&7.7; 17.0; 48.5&10.9\\
$4$&16.0; 43.0; 109.0; 194.0&7.7; 14.7; 35.3; 21.8&99.5\\
\hline

\end{tabular}
\tablefoot{We provide in each case the number $k$ of normal components with the corresponding parameters $(\mu,\sigma)$, and the fraction of simulated datasets with $p>0.05$ as given by the KS-test.
}
\end{table}

\subsection{Radial velocities}

The radial velocity is one important parameter that is needed to derive the kinematic parallax of cluster members in our analysis. In the following, we describe our search in the literature for radial velocity information for the stars in our sample. 

First, we searched the Apache Point Observatory Galactic Evolution Experiment (APOGEE) catalog in the Sloan Digital Sky Survey III (SDSS-III). The APOGEE program completed a systematic and homogeneous spectroscopic survey to build a high-resolution ($R\sim22500$), near-infrared spectra database for more than $10^{5}$ stars in our Galaxy \citep{Majewski(2015)}. We identified a total of 174 stars from our sample in the APOGEE survey. The zeropoint of the APOGEE radial velocities is estimated to be $-0.355\pm0.033$~km/s from comparison with other studies \citep[see][]{Nidever(2015)}. As discussed in Appendix~B of \citet{Galli(2013)}, a small shift in the radial velocities could lead to a more significant offset in the parallaxes derived from the moving cluster method. We verified that a small variation of 0.355~km/s in the radial velocities would produce an offset of 0.45~mas in parallaxes using typical values of proper motions, radial velocities, and the CP position of the Pleiades cluster that will be presented in the forthcoming analysis. Thus, to avoid a systematic bias in the resulting parallaxes we decided to correct the APOGEE radial velocities from their absolute zeropoint before using them in this work. 

Then, we searched the CDS databases to access more radial velocity data for the remaining stars in the sample. The search made use of a query for radial velocity information in script mode using the web-based data mining tools available in the SIMBAD database \citep{SIMBAD}. Our search results for radial velocities is based on: \citet{Wilson(1953)}, \citet{Gontcharov(2006)}, \citet{Kharchenko(2007)}, \citet{White(2007)}, \citet{Mermilliod(2009)} and \citet{Kordopatis(2013)}. Doing so, we retrieved radial velocity information for 241 stars. We note that 21 stars in this list are in common with the APOGEE sample and we calculated the weighted mean of the multiple radial velocity values.

\begin{figure}[!h]
\begin{center}
\includegraphics[width=0.49\textwidth]{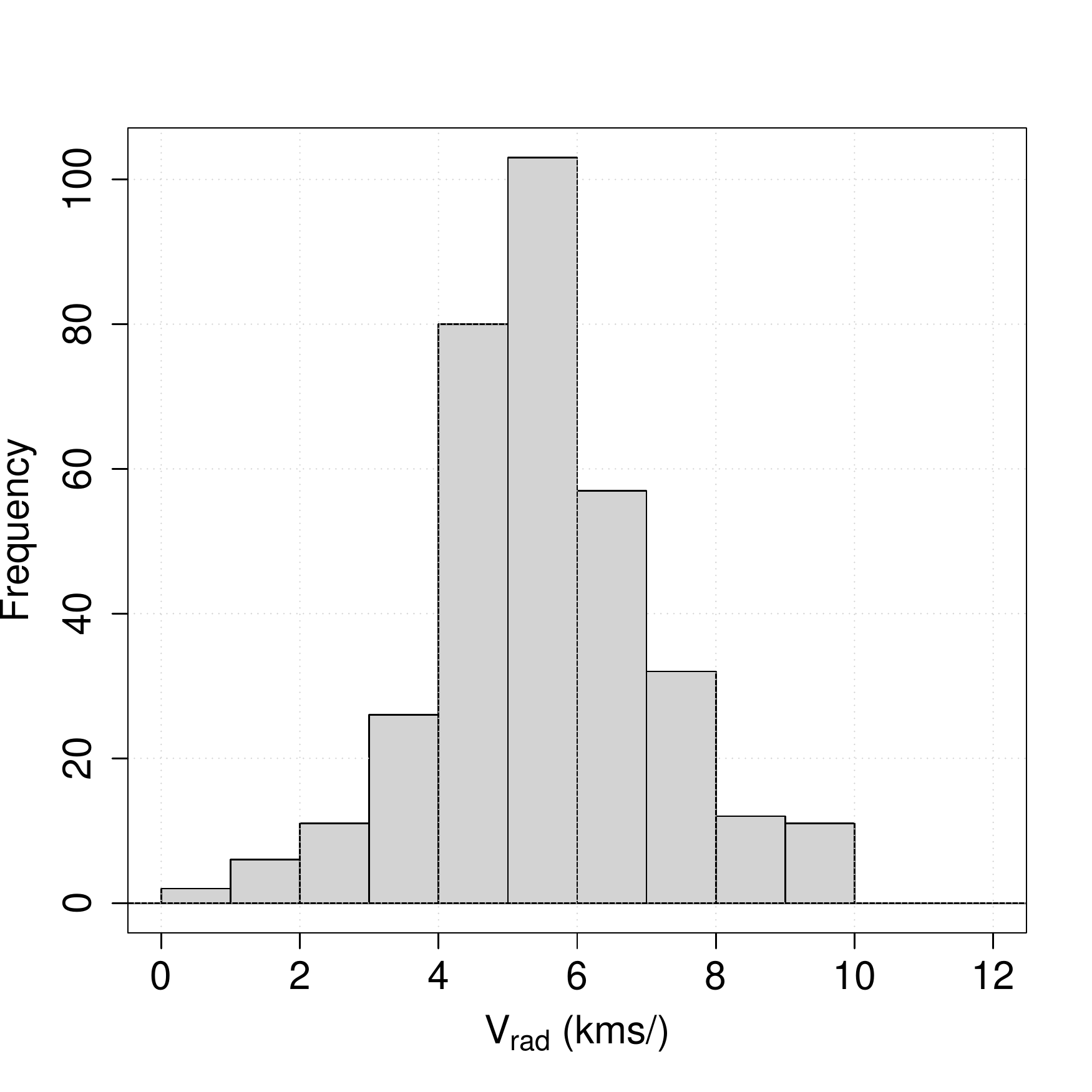}
\caption{
\label{fig5}
Distribution of radial velocities for Pleiades stars collected from the literature. }
\end{center}
\end{figure}

Thus, our search for radial velocity data returned a sample of 394 stars with at least one radial velocity measurement published in the literature. However, this sample also includes stars with poor radial velocity measurements and obvious outliers. After removing obvious outliers from the radial velocity distribution by a $3\sigma$ elimination, we end up with a sample of 340 stars.  Figure~\ref{fig5} shows the distribution of radial velocity data for the stars in our sample. The average radial velocity of this sample is $5.6\pm0.1$~km/s with a median value of $5.4$~km/s. The radial velocities collected in this work will be used in Sect.~5 to derive individual kinematic parallaxes of Pleiades stars from the moving cluster method. 

\subsection{Binaries and multiple systems}

Binaries and multiple systems are common features during star formation and various  surveys in different regions and clusters have been performed to measure the frequency of such systems \citep[see e.g.,][]{Ghez(1993),Duchene(1999),Daemgen(2015)}. Previous studies on the Pleiades cluster indicate that the binary fraction varies from 28\% to 44\%, depending on the mass range, orbital period, and angular separation of the binary systems \citep{Bouvier(1997),Lodieu(2007)}.  

\begin{figure}[!h]
\begin{center}
\includegraphics[width=0.49\textwidth]{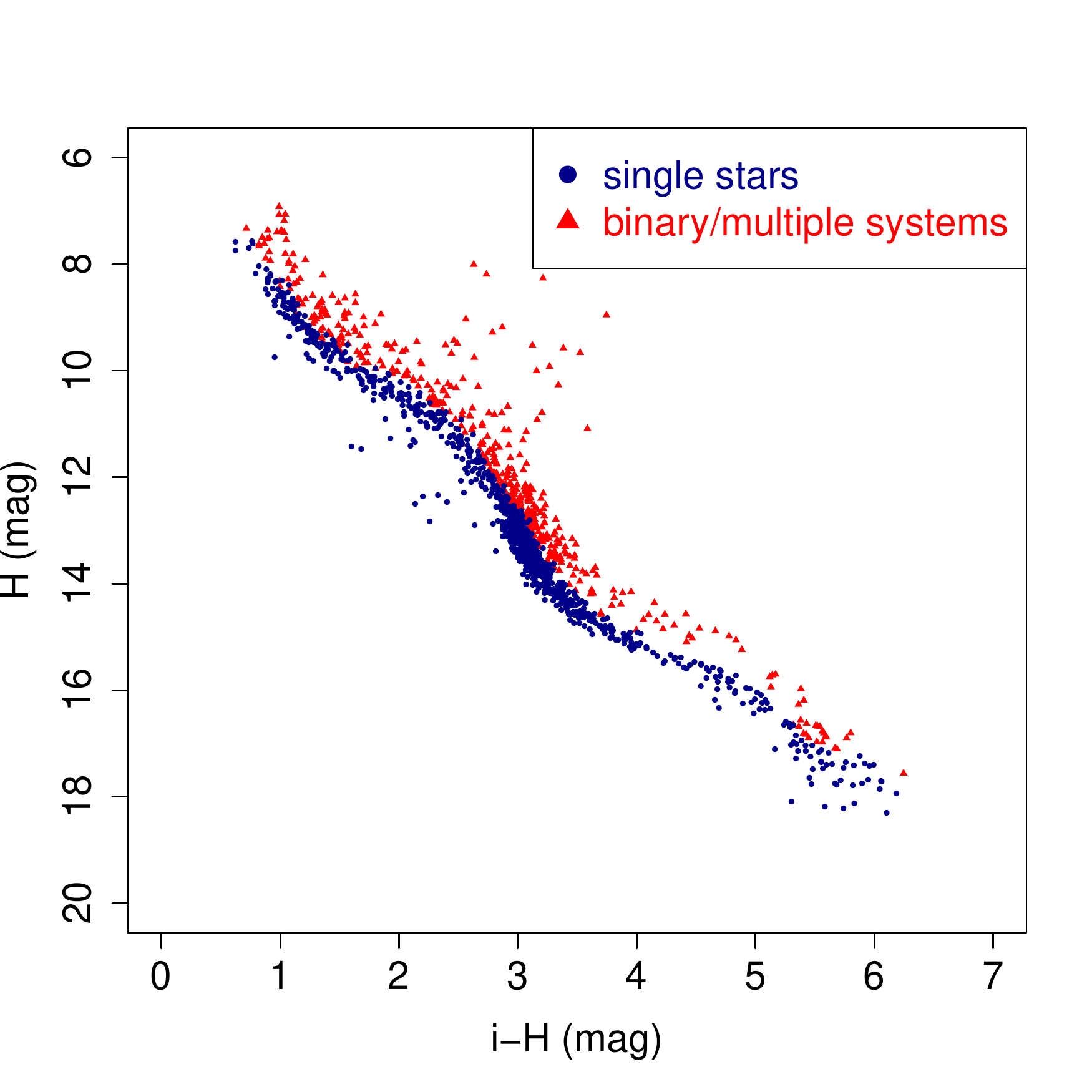}
\caption{
\label{fig6}
CMD to illustrate the locus of potential binaries and multiple systems that might be present in our sample.\bigskip}
\end{center}
\end{figure}

Despite previous efforts, the binarity/multiplicity census of the Pleiades cluster is still far from complete and would require a large observing programme, which is clearly beyond the scope of this paper. However, unresolved binaries and high-order multiple systems ($n\geq3$) appear brighter than single stars and can be identified in a color-magnitude diagram (CMD). To do so, we plot different CMDs using the $i,J,H,K_{s}$ filters that are available for most stars in our sample, and identified those stars that define a binary sequence as potential binaries and multiple system candidates. Figure~\ref{fig6} illustrates this procedure for one CMD used in our analysis. We consider the star to be a binary or multiple system if it is classified as such in more than one CMD. Thus, we found 510 binary or multiple systems that amounts to 24\% of our initial sample. Although this procedure is more likely to remove only the equal-mass binaries, it serves as a first estimate to characterize the binarity/multiplicity in our sample, and it will be useful in our forthcoming analysis to investigate the distance of the cluster. We note that we are not rejecting these stars from the sample. As explained in Sect.~2.2, the parallaxes of binaries and multiple systems derived in this work will be inferred from the average spatial velocity of the cluster (see Sect.~5). 

\section{Convergent point analysis}

In the following, we investigate the CP position of the Pleiades open cluster obtained from different subsets of our initial sample of stars. First, we build two control samples as a quality check of our results and to discuss the velocity dispersion of the cluster. Then, we apply the CPSM to the full sample of stars and investigate our solution based on Monte Carlo simulations. 

\subsection{Preliminary inspection of proper motion data}

The accuracy and precision of proper motions are of ultimate importance in the CP analysis. To ensure the quality of our results presented in the forthcoming sections, we proceed as follows. 

First, we select from our initial sample of 2107 stars only those stars whose proper motion measurements have been computed with a minimum of $n\geq10$ positions  and a time base $\Delta t\geq10$~yr (see Fig.~\ref{fig3}). These selection criteria reduce the sample to 1657~stars. 

Then, we remove the outliers from the proper motion distributions based on a $3\sigma$ elimination in each component. Doing so, we end up with a sample of 1551 stars that will be used in the remaining of this paper to investigate the distance of the cluster.  

\subsection{Control sample 1: the high probability cluster members}

The DANCe catalog is not a homogeneous dataset because the stars in our sample have different astrometric observing campaigns and the multi-band photometry ($u,g,r,i,z,Y,J,H,K_{s}$) is not complete for all stars. The membership probabilities computed from the DANCe project take both astrometric and photometric data into account, and they are more accurate for stars with a complete dataset. 

With this in mind, we apply the following selection criteria (in addition to those described in Sect.~4.1) to identify the most likely cluster members and the stars with more precise proper motion measurements in our sample:
\begin{enumerate}
\item stars with complete data (astrometry + photometry),
\item proper motion errors $(\mu_{\alpha}/\sigma_{\mu_{\alpha}})\geq47$ and $(\mu_{\delta}/\sigma_{\mu_{\delta}})\geq128$,
\item membership probability $p\geq0.9975$ ($3\sigma$).
\end{enumerate} 
Our selection criterium $\#2$ refers to the stars that roughly define the last component of the mixture model applied to the distribution of proper motion errors in Fig.~\ref{fig4}. Our choice of using the membership probability threshold of $99.75\%$ (see criterium $\#3$) is justified to minimize, as much as possible, the contamination rate in our control sample (see Table~4 of \citealt{Sarro(2014)} for more details). Doing so, we end up with a sample of 296 stars. 

Then, we applied the CPSM with $\sigma_{v}=0$ to identify the stars that show strict convergence to the CP of the moving group. We find a sample of 118 stars that defines a core moving group of the cluster (thus, our control sample 1). The corresponding CP is located at 
\begin{equation*}
(\alpha_{cp},\delta_{cp})=(92.9^{\circ},-49.4^{\circ})\pm(1.2^{\circ},1.2^{\circ})\, ,
\end{equation*}
with chi-squared statistics $\chi^{2}_{red}=1.15$ (i.e., $\chi^{2}/\nu=133.0/116$) and correlation coefficient of $\rho=-0.99$. We note that our first CP estimate is already consistent with the solution of $(\alpha_{cp},\delta_{cp})=(92.5^{\circ},-47.9^{\circ})\pm(5.4^{\circ},5.3^{\circ})$ derived by \citet{Makarov(2001)}. However, our result is more precise, which is a result of the more precise proper motion data available nowadays and the improved methodology (described in Sect.~2.2) to calculate the CP position. 

We repeated this analysis after removing the binary (and multiple system) candidates from our sample (as described in  Sect.~3.4), and confirmed that the resulting CP position is in perfect agreement with the solution given above. For clarity of presentation, we provide only the results, including binaries and multiple systems, because they contain more stars in the solution. In Appendix~A we provide an alternative approach to estimate the CP coordinates of the Pleiades that also supports the results presented in this section. 

\subsection{Control sample 2: the nuclear members of the cluster}

\begin{figure*}[!htp]
\begin{center}
\includegraphics[width=0.46\textwidth]{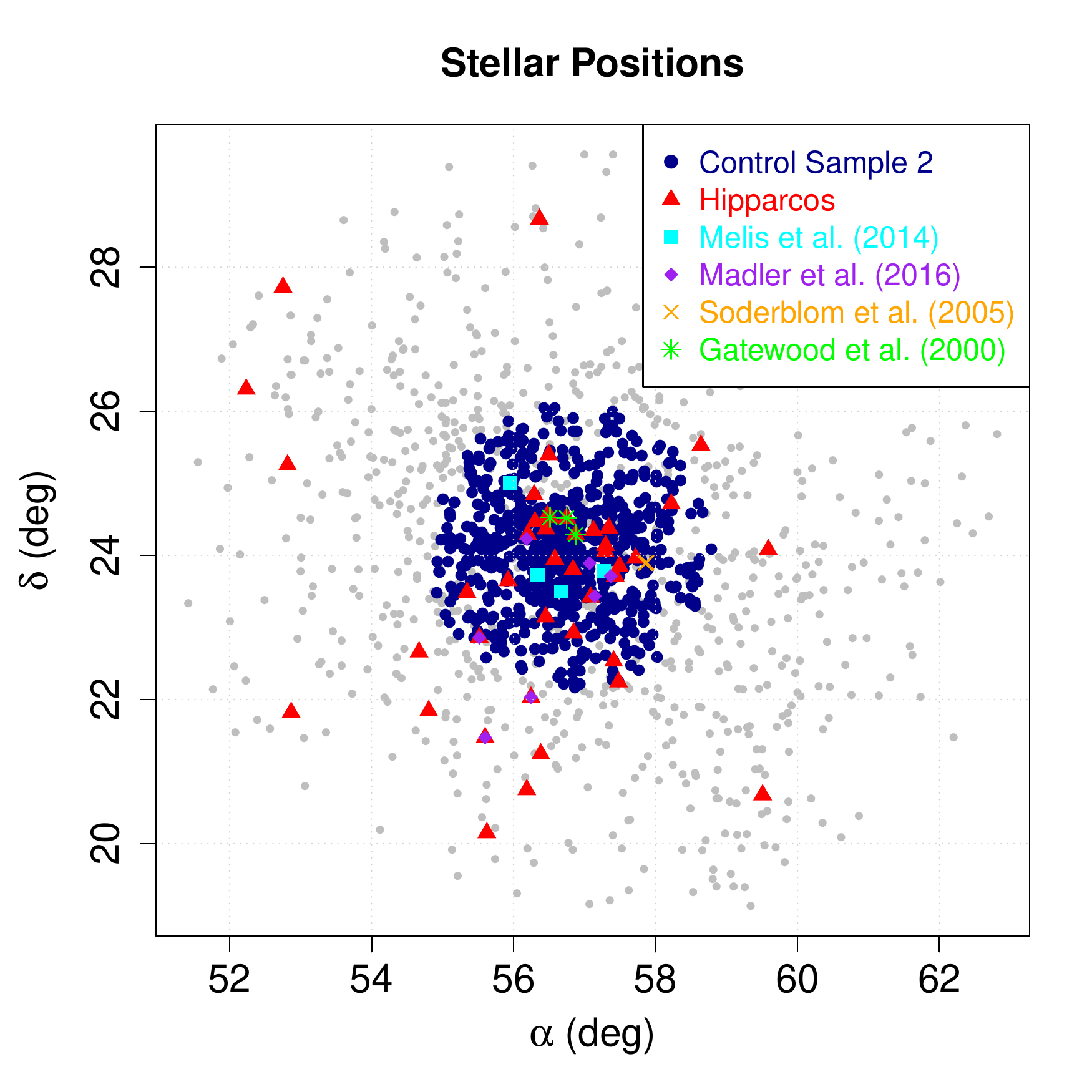}
\includegraphics[width=0.46\textwidth]{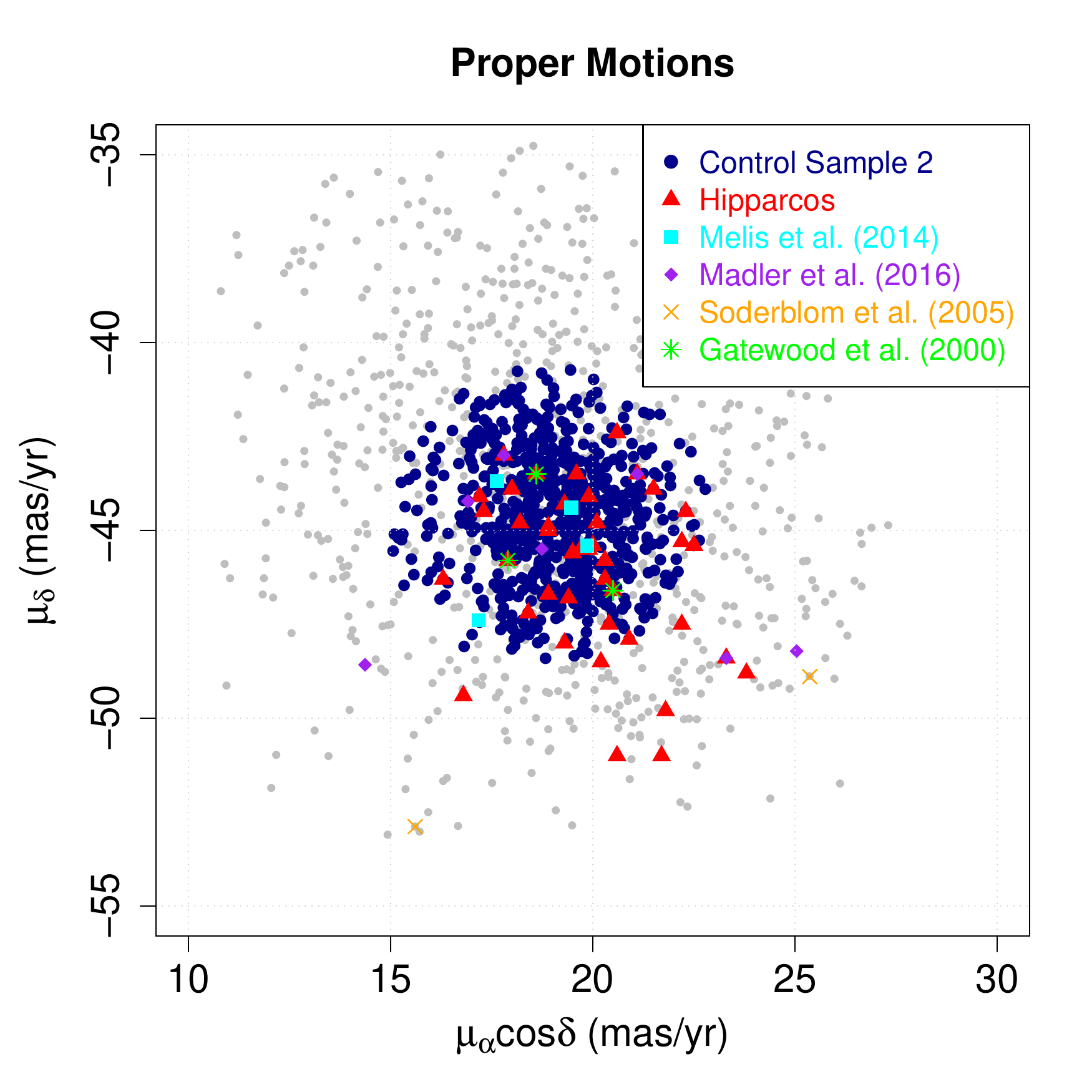}
\caption{
\label{fig7}
Distribution of stellar positions \textit{(left panel)} and proper motions \textit{(right panel)} for the Pleiades stars in the DANCe catalog. The various symbols and colors indicate the nuclear members of the cluster that define our control sample 2, and the stars with individual distances reported in the literature.}
\end{center}
\end{figure*}

The DANCe catalog for the Pleiades open cluster covers an area of $\sim80\deg$ around the cluster center. Ideally, one would expect the contamination rate by field stars to increase with increasing distance to the cluster center. To minimize the existence of field stars (interlopers), we decided to build a second control sample by retaining only the nuclear members of the cluster within $2\deg$ from the cluster center ($\alpha\simeq57^{\circ}$,\,$\delta\simeq+24^{\circ}$) out of the sample of 1551 stars, which was selected as described in Sect.~4.1. Doing so, we end up with a sample of 914 stars that will be used to investigate the typical velocity dispersion in the center of the cluster. In the outer regions, the velocity dispersion required to identify all cluster members might be larger owing to a combination of several effects (e.g. mass segregation, stellar encounters, tidal disruption, and cluster evaporation). Thus, we argue that the cluster members identified by the CPSM in this work represent a minimum moving group of the Pleiades cluster. As illustrated in Fig.~\ref{fig7}, this sample contains most of the stars with known distances in the literature that can be used to compare with our results. 

\begin{figure*}[!htp]
\begin{center}
\includegraphics[width=0.7\textwidth]{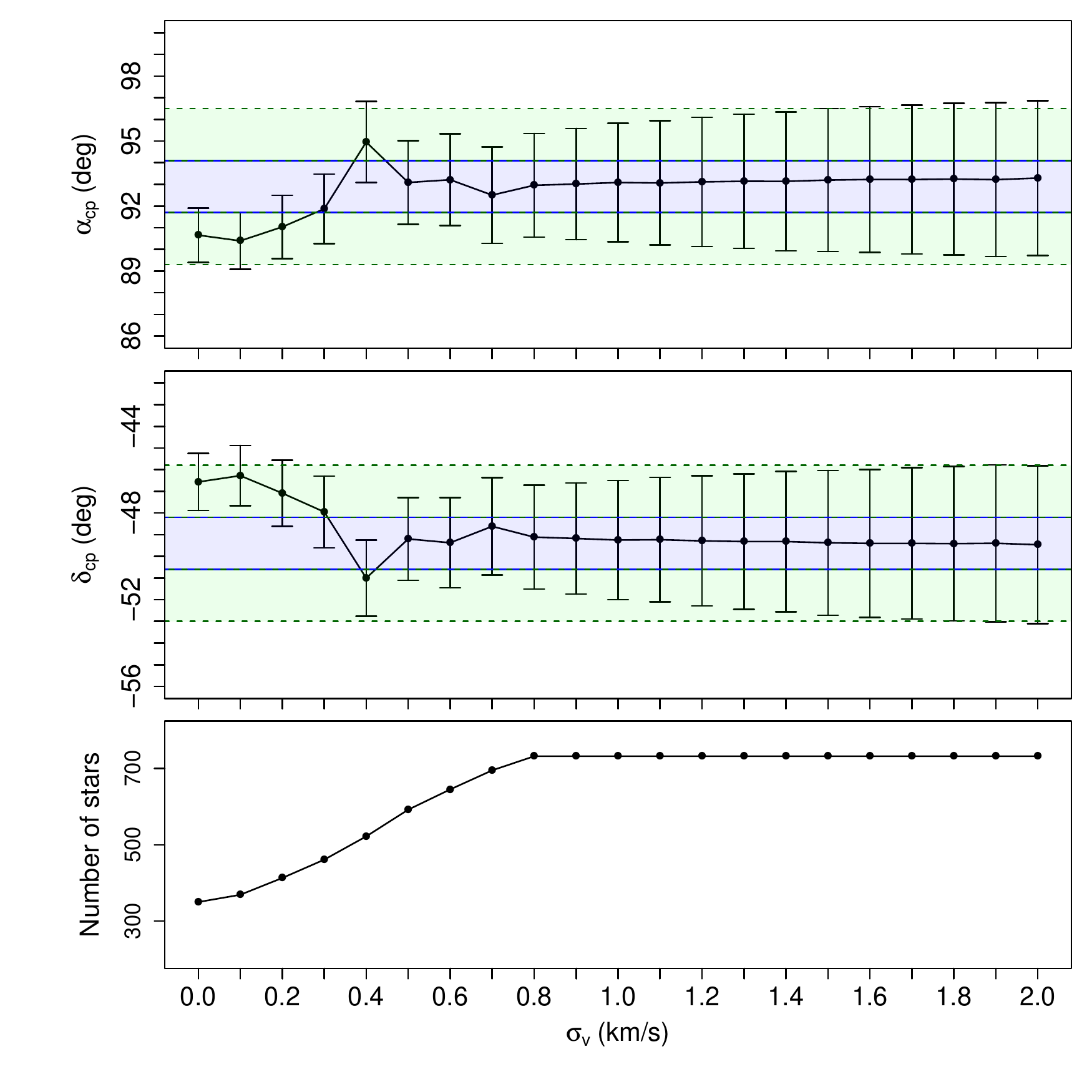}
\caption{
\label{fig8}
Results obtained from the CPSM (CP coordinates and number of moving group members) as a function of the velocity dispersion of the cluster. The solid and dashed lines indicate the CP solution derived from our control sample~1 at $1\sigma$ and $3\sigma$ (see Sect.~4.1).}
\end{center}
\end{figure*}

Then, we run the CPSM using different values for the velocity dispersion of the cluster from 0 to 2~km/s in steps of 0.1~km/s. In each step we calculate the CP position and count the number of stars identified by the CPSM as kinematic moving group members. The results of this analysis are shown in Fig.~\ref{fig8}. We observe that the solution is rather unstable at low values for the velocity dispersion, although the resulting CP positions are still statistically compatible with the results obtained from control sample 1  (see Sect.~4.1). The scatter reduces significantly after $\sigma_{v}=0.5$~km/s with the inclusion of more cluster members in the solution. We note that with $\sigma_{v}=0.8$~km/s, our solution retrieves the maximum number of cluster members (732~stars) that can be identified by the CPSM from this sample of stars, and that the CP coordinates converge to the solution obtained with the control sample 1. Thus, our methodology confirmed $80\%$ of the selected sample as moving group members, as expected for the CPSM \citep[see Fig.~8 of][]{Galli(2012)}. Increasing the velocity dispersion in the CP analysis for values higher than $\sigma_{v}=0.8$~km/s only increases the noise (i.e., errors) in the CP coordinates with no benefit of including more cluster members (see Fig.~\ref{fig8}). Thus, based on this analysis, we infer the velocity dispersion of the cluster to be $\sigma_{v}=0.8\pm0.1$~km/s (see also discussion Sect.~4.5). The corresponding CP solution is located at 
\begin{equation*}
(\alpha_{cp},\delta_{cp})=(93.0^{\circ},-49.1^{\circ})\pm(2.5^{\circ},2.4^{\circ})\, ,
\end{equation*}
with chi-squared statistics $\chi^{2}_{red}=1.03$ (i.e., $\chi^{2}/\nu=753.0/730$) and correlation coefficient of $\rho=-0.99$. Figure~\ref{fig7} illustrates the distribution of the stellar proper motion for the 732 stars selected by the CPSM. These stars define our control sample 2, and the results obtained in this section will be used as a quality check of our final CP solution for the Pleiades cluster (see below).   

\subsection{The CPSM applied to the full sample of Pleiades stars }

Our final analysis consists in running the CPSM on the selected sample of 1551 stars used in this work (see Sect.~4.1). Using a velocity dispersion of $\sigma_{v}=0.8$~km/s, we identify a moving group with 1210 stars and CP located at
\begin{equation*}
(\alpha_{cp},\delta_{cp})=(93.3^{\circ},-49.4^{\circ})\pm(1.4^{\circ},1.4^{\circ})\, ,
\end{equation*}
with chi-squared statistics $\chi^{2}_{red}=0.95$ (i.e., $\chi^{2}/\nu=1142.8/1208$) and correlation coefficient of $\rho=-0.99$. The resulting CP solution is in good agreement with the results presented in Sect.~4.2 and 4.3. Based on this investigation, we consider the 1210 stars identified directly by the CPSM to be confirmed members of the cluster. 

\begin{figure*}[!htp]
\begin{center}
\includegraphics[width=0.7\textwidth]{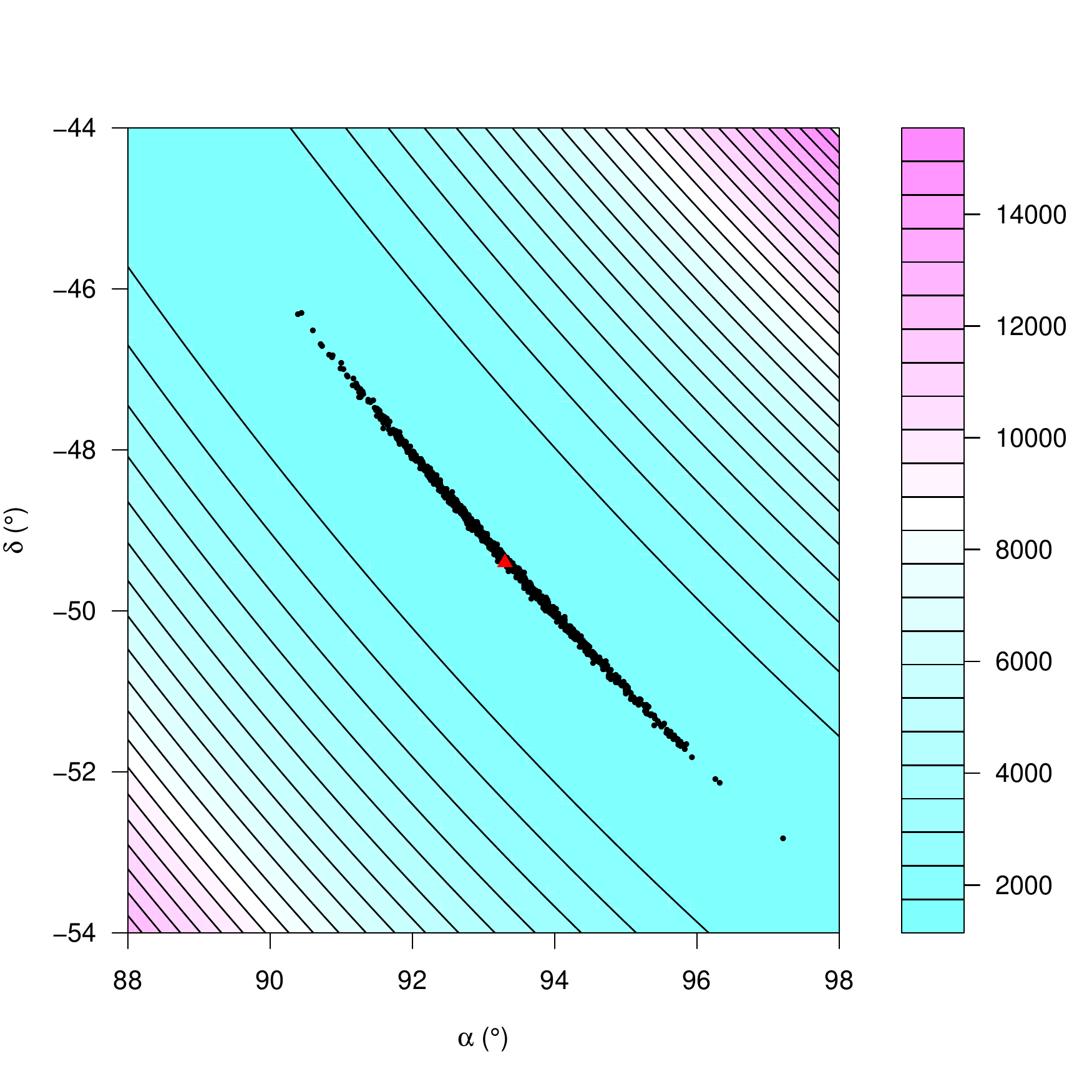}
\caption{
\label{fig9}
CP solutions (black dots) for 1000 Monte Carlo realizations of the Pleiades open cluster overlaid on the $X^{2}$ contour levels (solid lines) indicated by the color code. The red triangle indicates the final CP solution given in Sect.~4.4. }
\end{center}
\end{figure*}

To gain more confidence in the results presented above, we constructed a number of 1000 synthetic samples of the Pleiades cluster based on Monte Carlo simulations. These samples are generated by resampling the stellar proper motions of individual stars from a Gaussian distribution with mean and variance that corresponds to the proper motion measurements and their uncertainties. Then, we run the CPSM for each synthetic realization of the Pleiades cluster and compute the CP position. The results of this analysis are shown in Fig.~\ref{fig9}. The centroid of the Monte Carlo simulations is located at
\begin{equation*}
(\alpha_{cp},\delta_{cp})=(93.3^{\circ},-49.3^{\circ})\pm(1.0^{\circ},1.0^{\circ})\, ,
\end{equation*} 
with $\chi^{2}_{red}=1.19$ and $\rho=-0.99$. This confirms our CP solution showing that it is indeed representative of the Pleiades open cluster.

All the investigations reported in this section, using different samples and techniques to derive the CP position of the Pleiades cluster, make us confident that our solution is well constrained. It will be used in Sect.~5 to calculate individual distances of cluster members.

\subsection{A posteriori assessment of the velocity dispersion}

\begin{figure}[!htp]
\begin{center}
\includegraphics[width=0.45\textwidth]{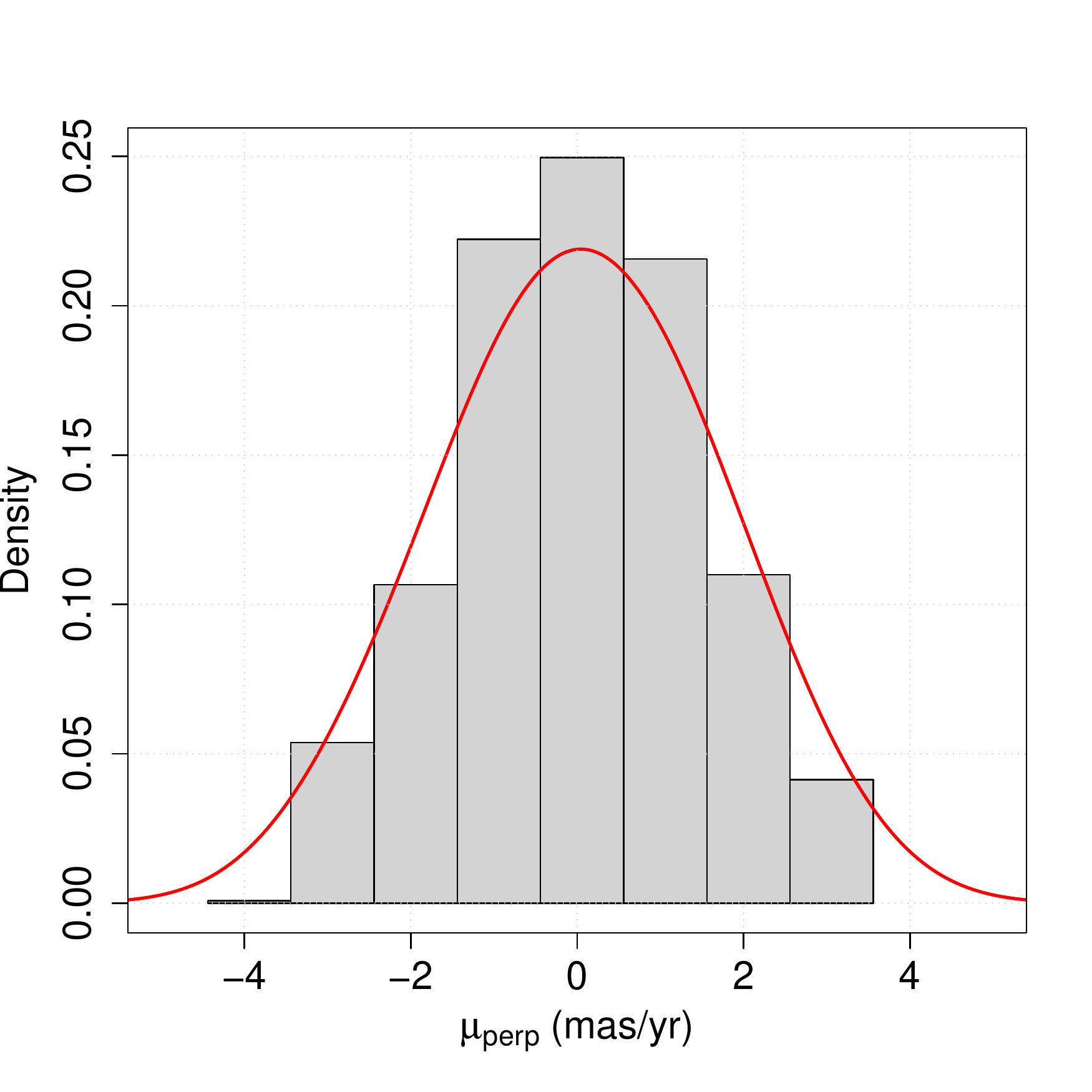}
\caption{
\label{fig10}
Distribution of the $\mu_{\perp}$ values calculated from the CP solution given in Sect.~4.4 for the 1210 cluster members. The red solid line indicates the kernel density estimator.}
\end{center}
\end{figure}

An alternative approach for investigating the velocity dispersion of the cluster consists of analyzing the distribution of the proper motion component that results from the peculiar motions of the stars. To do so, we calculate the proper motion component $\mu_{\perp}$ that is directed perpendicular to the great circle that joins the star and the CP of the moving group following the procedure outlined in Sect.~2.1 of \citet{Galli(2012)}. The distribution of $\mu_{\perp}$ values for the 1210 cluster members identified in Sect.~4.4  is shown in Figure~\ref{fig10}. The average and standard deviation of the distribution are, respectively, 0.03~mas/yr and 1.45~mas/yr. As expected, the average value of $\mu_{\perp}$ is close to zero, which reflects the good convergence of the stellar proper motions to our CP solution. 

The one-dimensional velocity dispersion of the cluster can be roughly estimated from $\sigma_{v}\simeq 4.74\, d\sigma_{\mu}$, where we consider $\sigma_{\mu}$ to be the observed scatter that comes from the $\mu_{\perp}$ statistics. The resulting velocity dispersion depends on the assumed distance $d$ of the cluster that is needed to convert the dispersion of proper motions (in units of mas/yr) to velocity (in units of km/s). Using the \textit{Hipparcos} and VLBI distance estimates of $\sim120$~pc and $\sim136$~pc for the Pleiades we find a velocity dispersion of 0.82~km/s and 0.93~km/s, respectively. These numbers confirm the value of $\sigma_{v}=0.8\pm0.1$~km/s inferred from the analysis presented in Sect.~4.3. 
  
\section{Moving-cluster distance to the Pleiades}

In the following we discuss the distance of the Pleiades based on the CP solution given in Sect.~4.4 and the  sample of 1210 cluster members identified in that analysis. First, we use the sample of stars with known radial velocity to calculate individual parallaxes and the spatial velocity of each star. Then, we calculate approximate parallaxes for other group members from the spatial velocity of the cluster. 

\subsection{Kinematic parallax and spatial velocity for cluster members with known radial velocity}

We calculate the individual kinematic parallax for the stars with known radial velocity from Eq.~(\ref{eq_plxind}), and use them to calculate the three-dimensional Galactic \textit{UVW} velocities from the procedure described by \citet{Johnson(1987)}. In a recent paper, \citet{Bailer-Jones(2015)} presents the common approach of  estimating distances from parallaxes as an inference problem that requires the use of prior assumptions. Consequently, we first apply the methodology described in Sect.~7 of his paper, using an exponentially decreasing volume density prior (with $L=10^{4}$) to avoid biases in the computed UVW velocities. One direct consequence of this approach is that the UVW error bars of individual stars are not symmetric. Then, we perform an iterative clip on the distribution of spatial velocities using Chauvenet's criterion \citep[see e.g.][]{Bevington(2003)} to remove obvious outliers and spot possible errors on the parallaxes derived from radial velocities. This leaves us with a sample of 64 stars with known radial velocity, which is used to accurately constrain the spatial velocity of the cluster. Binaries and multiple systems (see Sect.~3.4) are excluded from this count, because it is not possible to derive their kinematic parallaxes based on a single radial velocity measurement. 

Table~\ref{tab_64stars} lists the parallaxes derived in this work for cluster members with known radial velocity. The mean parallax is $7.44\pm 0.08$~mas (median of 7.38~mas and standard deviation of 0.66~mas). This is consistent with a distance estimate of $134.4^{+2.9}_{-2.8}$~pc given a confidence interval of 95\%. This result is in good agreement with the VLBI distance determination of $136.2\pm1.2$~pc reported by \citet{Melis(2014)}, but it is not consistent with the \textit{\textit{Hipparcos}} distance of $120.2\pm1.9$~pc \citep{vanLeeuwen(2009)}. 

\begin{table*}[!btp]
\centering
\caption{Parallaxes and spatial velocities for the sample of 64~stars with known radial velocities.  
\label{tab_64stars}}
\resizebox{17cm}{!} {
\begin{tabular}{ccccccccccc}
\hline
DANCe&$\alpha$&$\delta$&$\mu_{\alpha}\cos\delta$ & $\mu_{\delta}$&$V_{r}$&Ref.&$\pi$&$U$&$V$&$W$\\
&(h:m:s)&($^{\circ}$ $^\prime$ $^\prime$$^\prime$)&(mas/yr)&(mas/yr)&(km/s)&&(mas)&(km/s)&(km/s)&(km/s)\\
\hline

J034051.27+233554.3	&	03 40 51.27	&	23 35 54.3	&$	15.8	\pm	1.2	$&$	-39.9	\pm	1.1	$&$	4.90	\pm	0.10	$&	1	&$	7.0	\pm	0.9	$&$	-5.3	_{	-0.2	}^{+	0.2	}$&$	-26.5	_{	-3.0	}^{+	4.0	}$&$	-14.5	_{	-1.4	}^{+	1.9	}$\\
J034203.30+243213.3	&	03 42  03.30	&	24 32 13.3	&$	20.3	\pm	0.3	$&$	-44.9	\pm	0.2	$&$	4.90	\pm	0.20	$&	1	&$	7.4	\pm	1.1	$&$	-6.6	_{	-0.4	}^{+	0.5	}$&$	-29.5	_{	-3.6	}^{+	5.0	}$&$	-14.5	_{	-1.6	}^{+	2.1	}$\\
J034204.72+225130.8	&	03 42  04.72	&	22 51 30.8	&$	21.1	\pm	0.9	$&$	-43.5	\pm	0.9	$&$	5.72	\pm	1.42	$&	2	&$	7.3	\pm	2.0	$&$	-7.2	_{	-1.5	}^{+	2.0	}$&$	-34.9	_{	-4.3	}^{+	12.7	}$&$	-15.7	_{	-2.3	}^{+	5.4	}$\\
J034208.84+233517.0	&	03 42  08.84	&	23 35 17.0	&$	16.8	\pm	0.4	$&$	-44.7	\pm	0.3	$&$	4.60	\pm	0.28	$&	1	&$	8.3	\pm	1.2	$&$	-4.8	_{	-0.3	}^{+	0.4	}$&$	-24.6	_{	-2.9	}^{+	4.0	}$&$	-13.9	_{	-1.5	}^{+	2.0	}$\\
J034221.54+243952.7	&	03 42 21.54	&	24 39 52.7	&$	18.6	\pm	0.4	$&$	-47.4	\pm	0.3	$&$	4.24	\pm	0.09	$&	1	&$	8.7	\pm	1.2	$&$	-5.1	_{	-0.2	}^{+	0.3	}$&$	-25.3	_{	-3.0	}^{+	4.2	}$&$	-13.9	_{	-1.5	}^{+	2.0	}$\\
J034227.60+250249.1	&	03 42 27.60	&	25  02 49.1	&$	18.9	\pm	0.4	$&$	-42.3	\pm	0.3	$&$	5.19	\pm	0.47	$&	1,3	&$	6.2	\pm	0.9	$&$	-7.3	_{	-0.4	}^{+	0.5	}$&$	-32.9	_{	-4.1	}^{+	5.7	}$&$	-16.3	_{	-1.7	}^{+	2.4	}$\\
J034228.65+250100.4	&	03 42 28.65	&	25  01  00.4	&$	20.7	\pm	0.3	$&$	-47.5	\pm	0.2	$&$	4.34	\pm	0.09	$&	1	&$	8.3	\pm	1.2	$&$	-6.0	_{	-0.3	}^{+	0.4	}$&$	-27.3	_{	-3.4	}^{+	4.7	}$&$	-13.8	_{	-1.5	}^{+	2.0	}$\\
J034312.12+244445.4	&	03 43 12.12	&	24 44 45.4	&$	20.2	\pm	0.3	$&$	-46.9	\pm	0.2	$&$	5.19	\pm	0.14	$&	1	&$	7.1	\pm	1.0	$&$	-6.8	_{	-0.4	}^{+	0.5	}$&$	-31.4	_{	-3.8	}^{+	5.2	}$&$	-16.0	_{	-1.7	}^{+	2.3	}$\\
J034335.21+252431.0	&	03 43 35.21	&	25 24 31.1	&$	21.5	\pm	0.3	$&$	-47.2	\pm	0.2	$&$	5.03	\pm	0.22	$&	1	&$	7.0	\pm	1.1	$&$	-7.5	_{	-0.6	}^{+	0.7	}$&$	-33.1	_{	-4.2	}^{+	6.1	}$&$	-16.0	_{	-1.8	}^{+	2.6	}$\\
J034336.60+232714.0	&	03 43 36.60	&	23 27 14.1	&$	19.8	\pm	0.3	$&$	-43.8	\pm	0.2	$&$	4.79	\pm	0.05	$&	1	&$	8.3	\pm	1.0	$&$	-5.7	_{	-0.2	}^{+	0.2	}$&$	-25.2	_{	-2.7	}^{+	3.6	}$&$	-12.4	_{	-1.1	}^{+	1.4	}$\\
J034336.92+242338.1	&	03 43 36.92	&	24 23 38.1	&$	18.9	\pm	0.3	$&$	-43.8	\pm	0.2	$&$	5.11	\pm	0.09	$&	1	&$	7.0	\pm	1.0	$&$	-6.4	_{	-0.3	}^{+	0.4	}$&$	-29.5	_{	-3.4	}^{+	4.6	}$&$	-15.0	_{	-1.5	}^{+	2.0	}$\\
J034347.97+250311.1	&	03 43 47.97	&	25  03 11.1	&$	18.1	\pm	0.9	$&$	-44.3	\pm	0.8	$&$	4.79	\pm	0.10	$&	1	&$	7.0	\pm	1.0	$&$	-6.2	_{	-0.3	}^{+	0.4	}$&$	-29.6	_{	-3.6	}^{+	5.1	}$&$	-15.6	_{	-1.7	}^{+	2.3	}$\\
J034355.70+242534.9	&	03 43 55.70	&	24 25 34.9	&$	22.5	\pm	0.4	$&$	-46.7	\pm	0.3	$&$	4.47	\pm	0.18	$&	1	&$	8.8	\pm	1.2	$&$	-6.2	_{	-0.4	}^{+	0.5	}$&$	-26.3	_{	-3.1	}^{+	4.3	}$&$	-12.1	_{	-1.3	}^{+	1.7	}$\\
J034356.69+251544.1	&	03 43 56.69	&	25 15 44.1	&$	19.6	\pm	0.3	$&$	-46.6	\pm	0.2	$&$	4.19	\pm	0.10	$&	1	&$	8.3	\pm	1.2	$&$	-5.7	_{	-0.3	}^{+	0.4	}$&$	-26.6	_{	-3.3	}^{+	4.6	}$&$	-13.7	_{	-1.5	}^{+	2.1	}$\\
J034356.71+245936.3	&	03 43 56.71	&	24 59 36.3	&$	21.9	\pm	0.4	$&$	-46.3	\pm	0.3	$&$	5.33	\pm	0.14	$&	1	&$	6.8	\pm	1.0	$&$	-7.8	_{	-0.5	}^{+	0.6	}$&$	-33.3	_{	-4.1	}^{+	5.6	}$&$	-15.5	_{	-1.6	}^{+	2.3	}$\\
J034420.87+233339.9	&	03 44 20.87	&	23 33 39.9	&$	18.8	\pm	0.3	$&$	-46.0	\pm	0.2	$&$	6.13	\pm	0.39	$&	1	&$	6.7	\pm	0.9	$&$	-6.7	_{	-0.5	}^{+	0.5	}$&$	-32.3	_{	-3.7	}^{+	5.1	}$&$	-17.1	_{	-1.8	}^{+	2.4	}$\\
J034435.42+240004.5	&	03 44 35.42	&	24  00  04.5	&$	20.3	\pm	0.5	$&$	-44.2	\pm	0.5	$&$	5.63	\pm	0.06	$&	1	&$	6.9	\pm	0.9	$&$	-7.2	_{	-0.3	}^{+	0.4	}$&$	-31.2	_{	-3.5	}^{+	4.6	}$&$	-14.9	_{	-1.4	}^{+	1.8	}$\\
J034458.92+220156.8	&	03 44 58.92	&	22  01 56.8	&$	17.8	\pm	1.1	$&$	-43.0	\pm	1.0	$&$	5.80	\pm	0.40	$&	4	&$	7.6	\pm	1.0	$&$	-5.6	_{	-0.4	}^{+	0.4	}$&$	-26.6	_{	-2.9	}^{+	3.9	}$&$	-13.9	_{	-1.4	}^{+	1.8	}$\\
J034458.96+232319.9	&	03 44 58.96	&	23 23 19.9	&$	18.8	\pm	0.3	$&$	-47.1	\pm	0.2	$&$	5.23	\pm	0.10	$&	1,3	&$	8.2	\pm	1.0	$&$	-5.5	_{	-0.2	}^{+	0.2	}$&$	-26.6	_{	-2.8	}^{+	3.7	}$&$	-14.3	_{	-1.3	}^{+	1.7	}$\\
J034509.74+245021.3	&	03 45  09.74	&	24 50 21.3	&$	19.7	\pm	0.8	$&$	-45.5	\pm	0.8	$&$	4.80	\pm	0.80	$&	4	&$	7.6	\pm	1.6	$&$	-6.5	_{	-1.0	}^{+	1.3	}$&$	-30.8	_{	-4.4	}^{+	8.2	}$&$	-15.3	_{	-2.2	}^{+	3.8	}$\\
J034516.14+240716.0	&	03 45 16.14	&	24  07 16.0	&$	17.0	\pm	0.5	$&$	-43.2	\pm	0.4	$&$	4.95	\pm	0.22	$&	1	&$	7.4	\pm	1.0	$&$	-5.6	_{	-0.3	}^{+	0.4	}$&$	-27.1	_{	-3.1	}^{+	4.3	}$&$	-14.6	_{	-1.5	}^{+	2.0	}$\\
J034522.19+232818.1	&	03 45 22.19	&	23 28 18.1	&$	19.3	\pm	0.4	$&$	-44.2	\pm	0.3	$&$	5.02	\pm	0.10	$&	1	&$	8.1	\pm	1.0	$&$	-5.8	_{	-0.2	}^{+	0.3	}$&$	-26.0	_{	-2.8	}^{+	3.6	}$&$	-13.0	_{	-1.2	}^{+	1.5	}$\\
J034529.58+234537.6	&	03 45 29.58	&	23 45 37.6	&$	18.3	\pm	0.3	$&$	-46.6	\pm	0.2	$&$	5.54	\pm	0.09	$&	1	&$	7.4	\pm	0.9	$&$	-6.0	_{	-0.2	}^{+	0.2	}$&$	-29.1	_{	-3.2	}^{+	4.2	}$&$	-15.8	_{	-1.5	}^{+	1.9	}$\\
J034530.23+241845.2	&	03 45 30.23	&	24 18 45.2	&$	17.8	\pm	0.5	$&$	-46.5	\pm	0.4	$&$	4.89	\pm	0.43	$&	1	&$	7.9	\pm	1.2	$&$	-5.5	_{	-0.5	}^{+	0.6	}$&$	-27.6	_{	-3.5	}^{+	5.0	}$&$	-15.1	_{	-1.8	}^{+	2.5	}$\\
J034539.04+251327.6	&	03 45 39.04	&	25 13 27.6	&$	20.2	\pm	0.3	$&$	-43.5	\pm	0.3	$&$	4.51	\pm	0.04	$&	1	&$	7.5	\pm	1.1	$&$	-6.6	_{	-0.3	}^{+	0.4	}$&$	-28.3	_{	-3.4	}^{+	4.8	}$&$	-13.3	_{	-1.4	}^{+	1.9	}$\\
J034544.08+240426.7	&	03 45 44.08	&	24  04 26.7	&$	17.8	\pm	0.3	$&$	-44.4	\pm	0.2	$&$	5.00	\pm	0.13	$&	1	&$	7.6	\pm	1.0	$&$	-5.7	_{	-0.2	}^{+	0.3	}$&$	-27.1	_{	-3.0	}^{+	4.0	}$&$	-14.4	_{	-1.4	}^{+	1.8	}$\\
J034548.95+235110.2	&	03 45 48.95	&	23 51 10.2	&$	19.7	\pm	0.4	$&$	-44.3	\pm	0.3	$&$	5.58	\pm	0.07	$&	1	&$	7.1	\pm	0.9	$&$	-6.8	_{	-0.3	}^{+	0.3	}$&$	-30.0	_{	-3.3	}^{+	4.3	}$&$	-14.7	_{	-1.3	}^{+	1.8	}$\\
J034606.52+235020.2	&	03 46  06.52	&	23 50 20.2	&$	19.6	\pm	0.3	$&$	-45.7	\pm	0.2	$&$	4.97	\pm	0.18	$&	1	&$	8.2	\pm	1.0	$&$	-5.8	_{	-0.3	}^{+	0.4	}$&$	-26.5	_{	-2.9	}^{+	3.9	}$&$	-13.4	_{	-1.3	}^{+	1.7	}$\\
J034607.52+242227.4	&	03 46  07.52	&	24 22 27.4	&$	18.9	\pm	0.5	$&$	-44.9	\pm	0.4	$&$	4.69	\pm	0.06	$&	1	&$	8.1	\pm	1.1	$&$	-5.7	_{	-0.2	}^{+	0.3	}$&$	-26.3	_{	-3.0	}^{+	4.0	}$&$	-13.4	_{	-1.3	}^{+	1.7	}$\\
J034617.95+244109.2	&	03 46 17.95	&	24 41  09.2	&$	18.1	\pm	0.7	$&$	-41.3	\pm	0.7	$&$	5.07	\pm	0.24	$&	1	&$	6.7	\pm	1.0	$&$	-6.6	_{	-0.5	}^{+	0.6	}$&$	-29.6	_{	-3.6	}^{+	4.9	}$&$	-14.6	_{	-1.6	}^{+	2.1	}$\\
J034627.01+242713.9	&	03 46 27.01	&	24 27 13.9	&$	16.2	\pm	1.7	$&$	-42.8	\pm	1.7	$&$	5.68	\pm	0.62	$&	1	&$	6.2	\pm	1.1	$&$	-6.4	_{	-0.7	}^{+	0.8	}$&$	-32.6	_{	-4.4	}^{+	6.7	}$&$	-17.9	_{	-2.3	}^{+	3.4	}$\\
J034628.64+244532.0	&	03 46 28.64	&	24 45 32.0	&$	17.7	\pm	1.0	$&$	-46.2	\pm	0.9	$&$	4.38	\pm	0.06	$&	1	&$	8.4	\pm	1.2	$&$	-5.1	_{	-0.2	}^{+	0.2	}$&$	-25.2	_{	-3.0	}^{+	4.0	}$&$	-13.8	_{	-1.4	}^{+	1.9	}$\\
J034632.87+231819.1	&	03 46 32.87	&	23 18 19.1	&$	21.1	\pm	0.3	$&$	-44.7	\pm	0.3	$&$	5.77	\pm	0.10	$&	1	&$	7.4	\pm	0.9	$&$	-6.9	_{	-0.3	}^{+	0.3	}$&$	-29.4	_{	-3.1	}^{+	4.0	}$&$	-13.6	_{	-1.2	}^{+	1.5	}$\\
J034635.92+235800.9	&	03 46 35.92	&	23 58  00.9	&$	17.1	\pm	0.7	$&$	-41.7	\pm	0.7	$&$	4.68	\pm	0.08	$&	1,3	&$	7.8	\pm	1.0	$&$	-5.3	_{	-0.2	}^{+	0.2	}$&$	-25.0	_{	-2.8	}^{+	3.6	}$&$	-13.0	_{	-1.2	}^{+	1.6	}$\\
J034639.39+240146.7	&	03 46 39.39	&	24  01 46.7	&$	19.5	\pm	0.4	$&$	-44.9	\pm	0.3	$&$	5.26	\pm	0.76	$&	1,3	&$	7.5	\pm	0.9	$&$	-6.3	_{	-0.2	}^{+	0.3	}$&$	-28.4	_{	-3.1	}^{+	4.1	}$&$	-14.1	_{	-1.3	}^{+	1.7	}$\\
J034643.59+235942.4	&	03 46 43.59	&	23 59 42.4	&$	18.9	\pm	0.3	$&$	-44.9	\pm	0.2	$&$	5.12	\pm	0.20	$&	1	&$	7.7	\pm	1.0	$&$	-6.0	_{	-0.3	}^{+	0.4	}$&$	-27.5	_{	-3.1	}^{+	4.1	}$&$	-14.0	_{	-1.4	}^{+	1.8	}$\\
J034649.21+243559.8	&	03 46 49.21	&	24 35 59.8	&$	20.2	\pm	0.4	$&$	-43.0	\pm	0.3	$&$	5.19	\pm	0.11	$&	1	&$	7.0	\pm	0.9	$&$	-7.1	_{	-0.4	}^{+	0.5	}$&$	-30.1	_{	-3.5	}^{+	4.7	}$&$	-13.9	_{	-1.4	}^{+	1.8	}$\\
J034653.27+225251.0	&	03 46 53.27	&	22 52 51.0	&$	19.5	\pm	0.6	$&$	-48.3	\pm	0.5	$&$	6.01	\pm	0.59	$&	1,3	&$	7.8	\pm	0.9	$&$	-6.1	_{	-0.2	}^{+	0.2	}$&$	-28.8	_{	-2.9	}^{+	3.7	}$&$	-15.2	_{	-1.3	}^{+	1.6	}$\\
J034654.03+251444.8	&	03 46 54.03	&	25 14 44.8	&$	18.9	\pm	0.3	$&$	-45.6	\pm	0.2	$&$	5.45	\pm	0.09	$&	1	&$	6.5	\pm	0.9	$&$	-7.2	_{	-0.4	}^{+	0.5	}$&$	-33.3	_{	-4.0	}^{+	5.5	}$&$	-17.1	_{	-1.8	}^{+	2.5	}$\\
J034709.45+234431.6	&	03 47  09.45	&	23 44 31.7	&$	19.6	\pm	0.3	$&$	-44.4	\pm	0.2	$&$	4.95	\pm	0.26	$&	1	&$	8.2	\pm	1.1	$&$	-5.9	_{	-0.4	}^{+	0.4	}$&$	-26.1	_{	-2.9	}^{+	3.9	}$&$	-12.8	_{	-1.3	}^{+	1.7	}$\\
J034713.54+234251.1	&	03 47 13.54	&	23 42 51.1	&$	17.4	\pm	0.4	$&$	-45.2	\pm	0.4	$&$	5.94	\pm	0.17	$&	1,3	&$	6.8	\pm	0.8	$&$	-6.3	_{	-0.2	}^{+	0.2	}$&$	-30.5	_{	-3.3	}^{+	4.3	}$&$	-16.6	_{	-1.5	}^{+	2.0	}$\\
J034720.97+234812.0	&	03 47 20.97	&	23 48 12.0	&$	16.3	\pm	1.0	$&$	-46.3	\pm	1.0	$&$	4.90	\pm	0.80	$&	4	&$	8.3	\pm	1.7	$&$	-4.9	_{	-0.8	}^{+	0.8	}$&$	-26.8	_{	-3.8	}^{+	6.7	}$&$	-15.3	_{	-2.2	}^{+	3.6	}$\\
J034722.99+245055.9	&	03 47 22.99	&	24 50 55.9	&$	18.4	\pm	0.8	$&$	-47.6	\pm	0.7	$&$	4.47	\pm	0.33	$&	1	&$	8.5	\pm	1.3	$&$	-5.3	_{	-0.5	}^{+	0.5	}$&$	-26.1	_{	-3.3	}^{+	4.7	}$&$	-14.1	_{	-1.7	}^{+	2.3	}$\\
J034725.35+240256.7	&	03 47 25.35	&	24  02 56.7	&$	18.8	\pm	0.5	$&$	-43.3	\pm	0.5	$&$	5.86	\pm	0.16	$&	1	&$	6.5	\pm	0.8	$&$	-7.1	_{	-0.3	}^{+	0.4	}$&$	-31.6	_{	-3.5	}^{+	4.6	}$&$	-15.6	_{	-1.5	}^{+	2.0	}$\\
J034730.60+242213.7	&	03 47 30.60	&	24 22 13.7	&$	19.7	\pm	0.4	$&$	-46.4	\pm	0.3	$&$	4.95	\pm	0.10	$&	1	&$	8.0	\pm	1.0	$&$	-6.0	_{	-0.3	}^{+	0.3	}$&$	-27.4	_{	-3.1	}^{+	4.1	}$&$	-13.8	_{	-1.3	}^{+	1.8	}$\\
J034733.47+234132.7	&	03 47 33.47	&	23 41 32.7	&$	19.9	\pm	0.4	$&$	-41.6	\pm	0.3	$&$	5.71	\pm	0.56	$&	1	&$	6.8	\pm	1.0	$&$	-7.2	_{	-0.8	}^{+	0.9	}$&$	-30.7	_{	-3.8	}^{+	5.5	}$&$	-13.9	_{	-1.7	}^{+	2.3	}$\\
J034733.69+244102.8	&	03 47 33.69	&	24 41  02.8	&$	17.8	\pm	0.6	$&$	-47.1	\pm	0.6	$&$	5.42	\pm	0.07	$&	1	&$	7.1	\pm	0.9	$&$	-6.2	_{	-0.2	}^{+	0.3	}$&$	-30.5	_{	-3.5	}^{+	4.7	}$&$	-16.8	_{	-1.7	}^{+	2.2	}$\\
J034738.02+232804.8	&	03 47 38.02	&	23 28  04.8	&$	21.3	\pm	0.3	$&$	-41.9	\pm	0.2	$&$	5.16	\pm	0.08	$&	1	&$	7.8	\pm	0.9	$&$	-6.7	_{	-0.3	}^{+	0.3	}$&$	-26.7	_{	-2.8	}^{+	3.6	}$&$	-11.5	_{	-1.0	}^{+	1.3	}$\\
J034738.05+244911.1	&	03 47 38.05	&	24 49 11.1	&$	18.5	\pm	0.3	$&$	-41.9	\pm	0.2	$&$	4.98	\pm	0.22	$&	1	&$	6.9	\pm	1.0	$&$	-6.6	_{	-0.4	}^{+	0.5	}$&$	-29.2	_{	-3.5	}^{+	4.8	}$&$	-14.2	_{	-1.5	}^{+	2.0	}$\\
J034740.46+242152.1	&	03 47 40.46	&	24 21 52.1	&$	15.9	\pm	0.7	$&$	-45.2	\pm	0.6	$&$	4.96	\pm	0.32	$&	1,3	&$	7.6	\pm	1.0	$&$	-5.2	_{	-0.2	}^{+	0.2	}$&$	-26.8	_{	-3.1	}^{+	4.1	}$&$	-15.4	_{	-1.6	}^{+	2.1	}$\\
J034741.19+234424.7	&	03 47 41.19	&	23 44 24.7	&$	20.1	\pm	0.4	$&$	-44.2	\pm	0.4	$&$	5.81	\pm	0.36	$&	1	&$	7.0	\pm	0.9	$&$	-7.0	_{	-0.5	}^{+	0.6	}$&$	-30.7	_{	-3.5	}^{+	4.8	}$&$	-14.6	_{	-1.5	}^{+	2.0	}$\\
J034750.95+243018.7	&	03 47 50.95	&	24 30 18.7	&$	17.6	\pm	0.3	$&$	-46.4	\pm	0.2	$&$	4.88	\pm	0.32	$&	1	&$	7.9	\pm	1.1	$&$	-5.5	_{	-0.4	}^{+	0.5	}$&$	-27.2	_{	-3.3	}^{+	4.6	}$&$	-14.9	_{	-1.7	}^{+	2.3	}$\\
J034755.28+231905.7	&	03 47 55.28	&	23 19  05.8	&$	16.6	\pm	0.3	$&$	-39.5	\pm	0.2	$&$	5.83	\pm	0.21	$&	1	&$	6.4	\pm	0.8	$&$	-6.4	_{	-0.3	}^{+	0.3	}$&$	-29.0	_{	-3.1	}^{+	4.0	}$&$	-14.7	_{	-1.4	}^{+	1.7	}$\\
J034757.61+263745.2	&	03 47 57.61	&	26 37 45.2	&$	20.2	\pm	2.0	$&$	-41.5	\pm	1.9	$&$	4.22	\pm	0.16	$&	3	&$	6.7	\pm	1.2	$&$	-7.5	_{	-0.6	}^{+	0.9	}$&$	-31.4	_{	-4.3	}^{+	6.5	}$&$	-13.9	_{	-1.7	}^{+	2.6	}$\\
J034805.83+230202.8	&	03 48  05.83	&	23  02  02.8	&$	18.7	\pm	0.3	$&$	-43.4	\pm	0.2	$&$	5.91	\pm	0.22	$&	1	&$	7.2	\pm	0.8	$&$	-6.4	_{	-0.3	}^{+	0.3	}$&$	-28.7	_{	-3.0	}^{+	3.9	}$&$	-14.3	_{	-1.3	}^{+	1.7	}$\\
J034807.97+234423.4	&	03 48  07.97	&	23 44 23.4	&$	17.6	\pm	0.4	$&$	-42.7	\pm	0.3	$&$	5.54	\pm	0.29	$&	1	&$	7.0	\pm	0.9	$&$	-6.2	_{	-0.4	}^{+	0.4	}$&$	-28.7	_{	-3.2	}^{+	4.3	}$&$	-14.8	_{	-1.5	}^{+	2.0	}$\\
J034813.78+233759.1	&	03 48 13.78	&	23 37 59.1	&$	15.5	\pm	0.5	$&$	-42.2	\pm	0.5	$&$	5.24	\pm	0.20	$&	1	&$	7.3	\pm	0.9	$&$	-5.3	_{	-0.2	}^{+	0.3	}$&$	-26.4	_{	-2.9	}^{+	3.8	}$&$	-14.7	_{	-1.4	}^{+	1.8	}$\\
J034817.30+243015.7	&	03 48 17.30	&	24 30 15.7	&$	18.4	\pm	0.3	$&$	-49.2	\pm	0.2	$&$	4.74	\pm	0.17	$&	1	&$	8.7	\pm	1.2	$&$	-5.3	_{	-0.3	}^{+	0.3	}$&$	-26.1	_{	-3.0	}^{+	4.0	}$&$	-14.4	_{	-1.5	}^{+	2.0	}$\\
J034819.84+233611.7	&	03 48 19.84	&	23 36 11.7	&$	17.0	\pm	0.7	$&$	-43.1	\pm	0.7	$&$	5.01	\pm	0.35	$&	1	&$	7.9	\pm	1.1	$&$	-5.3	_{	-0.4	}^{+	0.4	}$&$	-25.5	_{	-3.0	}^{+	4.0	}$&$	-13.6	_{	-1.5	}^{+	1.9	}$\\
J034822.81+244853.5	&	03 48 22.81	&	24 48 53.5	&$	19.5	\pm	0.3	$&$	-46.4	\pm	0.2	$&$	5.81	\pm	0.22	$&	1,3	&$	6.6	\pm	0.9	$&$	-7.3	_{	-0.4	}^{+	0.5	}$&$	-33.5	_{	-3.9	}^{+	5.3	}$&$	-16.9	_{	-1.8	}^{+	2.4	}$\\
J034834.52+232604.9	&	03 48 34.52	&	23 26  04.9	&$	16.9	\pm	1.5	$&$	-44.2	\pm	1.4	$&$	5.73	\pm	0.23	$&	1,3	&$	7.2	\pm	0.9	$&$	-5.8	_{	-0.2	}^{+	0.2	}$&$	-28.3	_{	-3.0	}^{+	3.9	}$&$	-15.4	_{	-1.4	}^{+	1.8	}$\\
J034902.35+231508.4	&	03 49  02.35	&	23 15  08.4	&$	15.6	\pm	1.5	$&$	-43.5	\pm	1.4	$&$	5.87	\pm	0.19	$&	3	&$	7.0	\pm	0.8	$&$	-5.6	_{	-0.2	}^{+	0.2	}$&$	-28.3	_{	-3.0	}^{+	3.9	}$&$	-16.0	_{	-1.5	}^{+	1.9	}$\\
J034928.76+234243.6	&	03 49 28.76	&	23 42 43.6	&$	18.7	\pm	0.5	$&$	-45.5	\pm	0.5	$&$	5.73	\pm	0.22	$&	3	&$	7.3	\pm	0.9	$&$	-6.3	_{	-0.3	}^{+	0.4	}$&$	-29.2	_{	-3.2	}^{+	4.1	}$&$	-14.9	_{	-1.4	}^{+	1.8	}$\\
J035020.92+242800.0	&	03 50 20.92	&	24 28  00.0	&$	19.1	\pm	1.2	$&$	-46.5	\pm	1.2	$&$	5.60	\pm	0.85	$&	5	&$	7.2	\pm	1.4	$&$	-6.7	_{	-1.0	}^{+	1.2	}$&$	-32.2	_{	-4.5	}^{+	7.7	}$&$	-16.3	_{	-2.3	}^{+	3.7	}$\\

\hline

\end{tabular}
}

\tablefoot{We provide for each star the DANCe identifier, position, proper motion, radial velocity, source of radial velocity in the literature (see references below), parallax derived in this work and the Galactic spatial velocity components.}

\tablebib{
Radial velocity sources:
(1)~\citet{Majewski(2015)};
(2)~\citet{Kordopatis(2013)};
(3)~\citet{Mermilliod(2009)};
(4)~\citet{Gontcharov(2006)};
(5)~\citet{White(2007)}.\bigskip
}
\end{table*}

\begin{figure}[!htp]
\begin{center}
\includegraphics[width=0.24\textwidth]{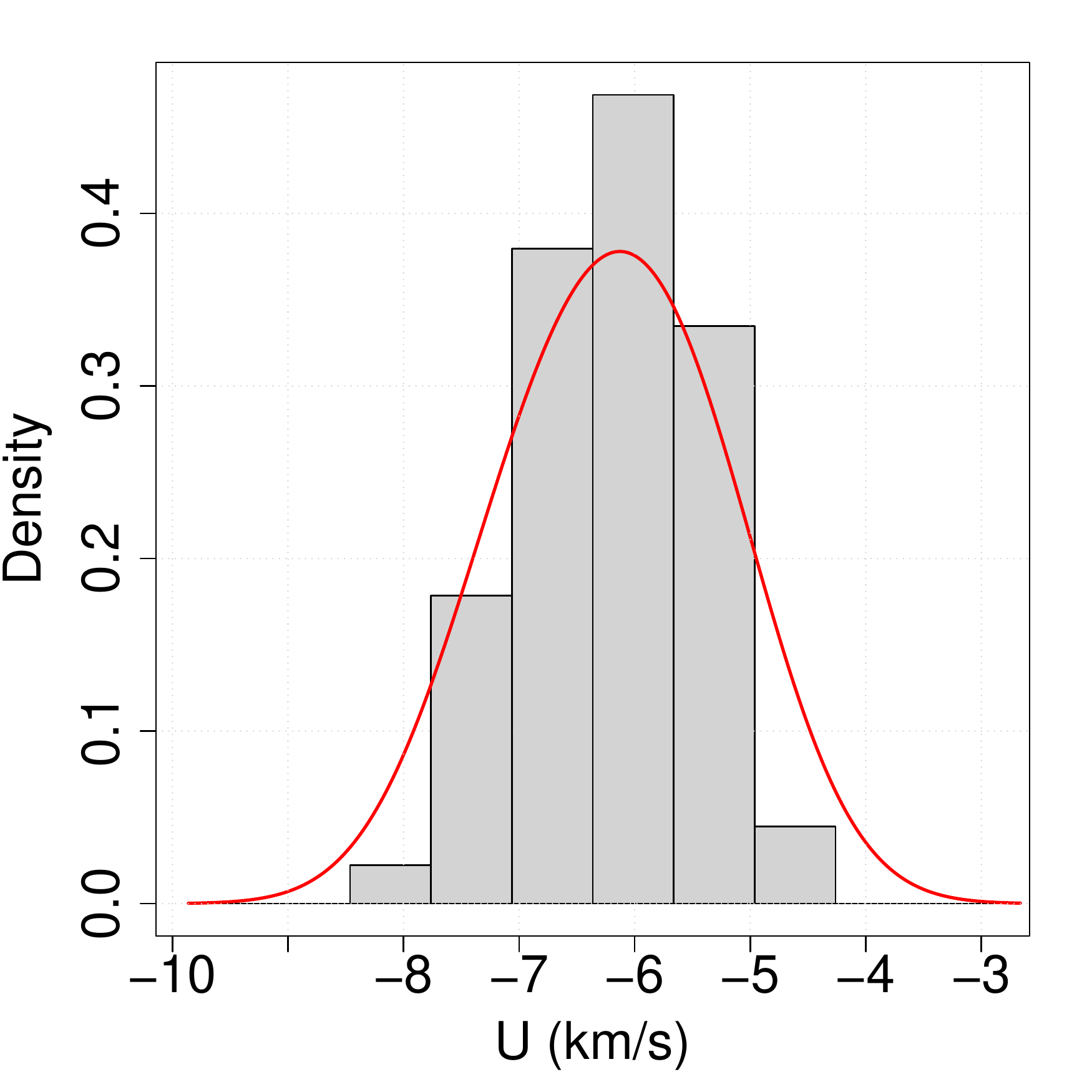}
\includegraphics[width=0.24\textwidth]{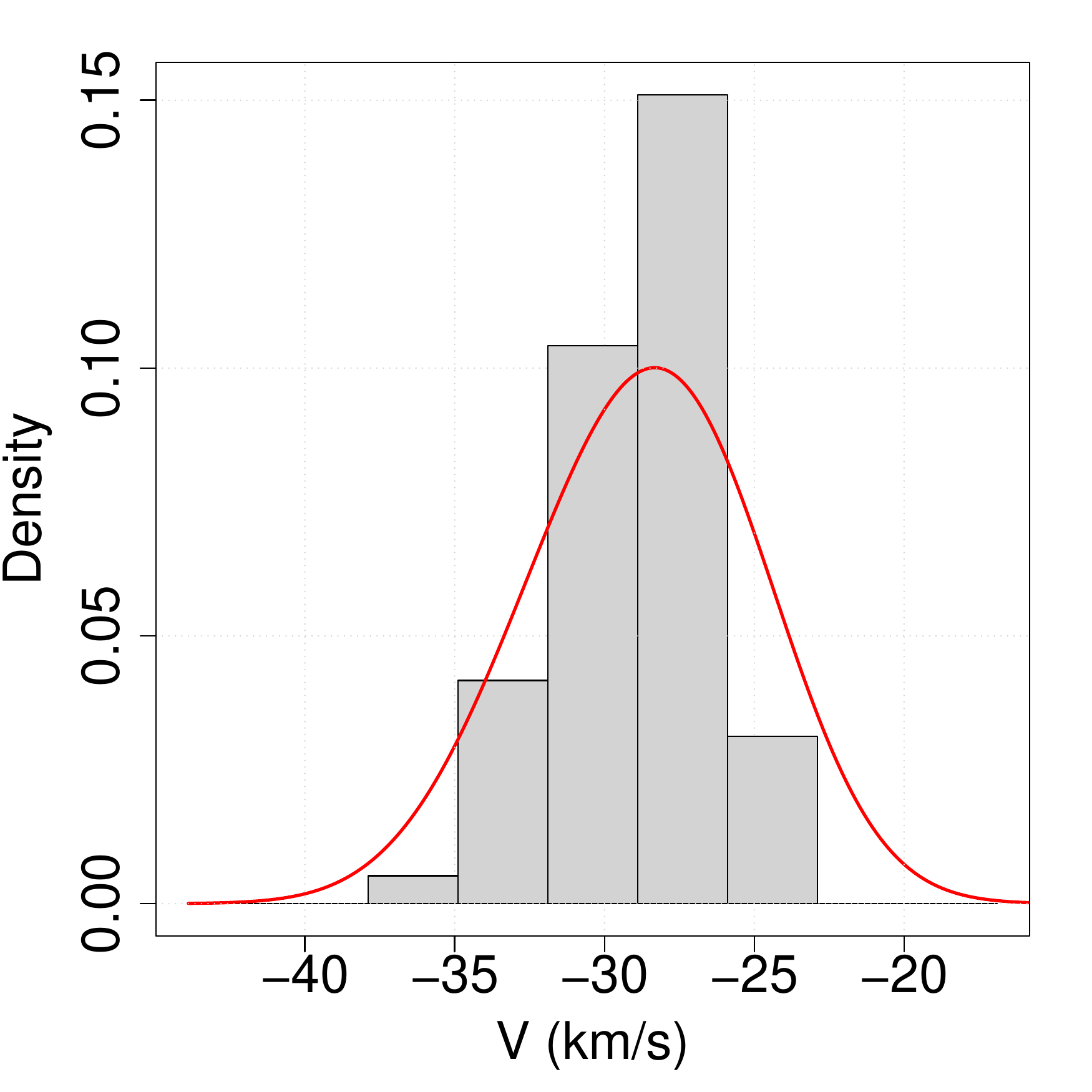}
\includegraphics[width=0.24\textwidth]{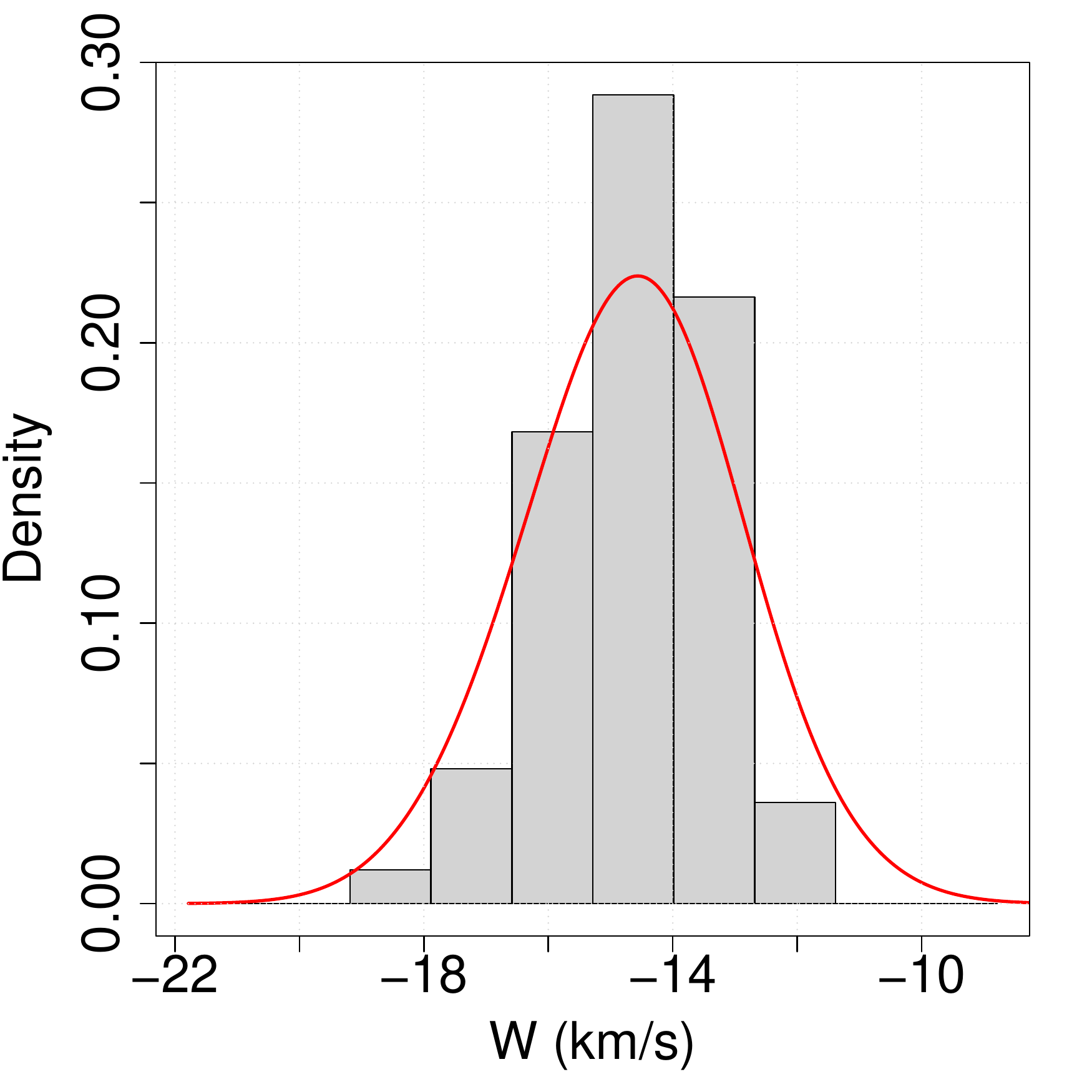}
\includegraphics[width=0.24\textwidth]{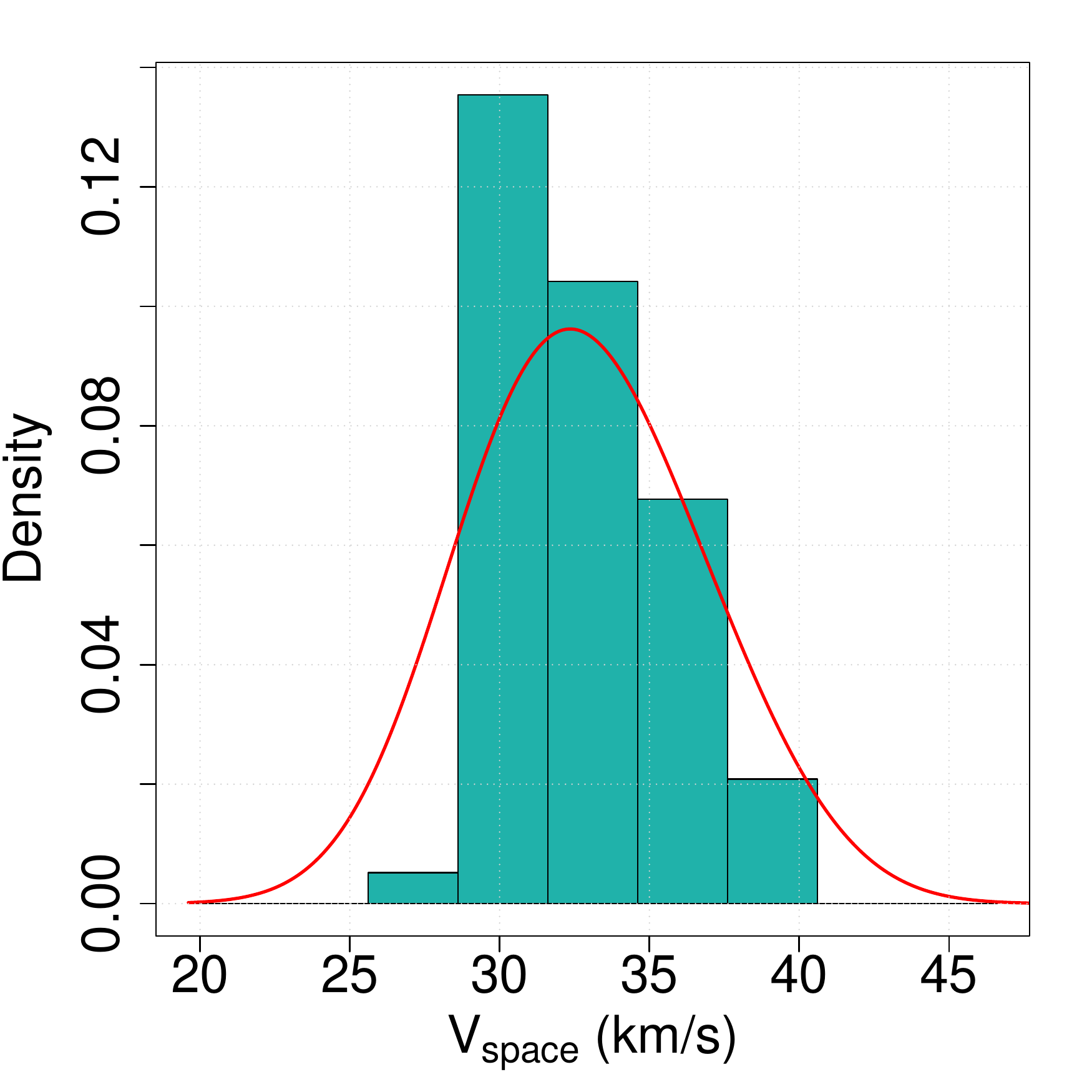}

\caption{
\label{fig11}
Histogram of the Galactic space velocity components for the 64 cluster members with known radial velocities. The red solid line indicates the kernel density estimator.  }
\end{center}
\end{figure}

Individual values for the space velocity of each star are also listed in Table~\ref{tab_64stars}. Figure~\ref{fig11} illustrates the distribution of the Galactic space velocity in each component and Table~\ref{tab_vel} lists their average values.  In a previous paper, \citet{Robichon(1999)} report the space velocity of $(U,V,W)=(-6.4,-24.4,-13.0)\pm(0.5,0.7,0.4)$~km/s (not corrected for the Solar motion) based on \textit{\textit{Hipparcos}} parallaxes. The difference of $(\Delta U,\Delta V,\Delta W)= (0.2,-4.3,-1.7)\pm(0.5,0.8,0.4)$~km/s between these two estimates is significant and comes from the different parallaxes used for each star in both studies.

\begin{table}[!h]
\centering
\caption{Spatial velocity of the Pleiades cluster derived in this work from the sample of 64 stars with known radial velocities.   
\label{tab_vel}}
\begin{tabular}{lcccc}
\hline
&Mean $\pm$ SEM&Median&Mode& SD\\
&(km/s)&(km/s)&(km/s)&(km/s)\\
\hline
$U$&$-6.2\pm 0.1$&-6.2&-6.1&0.7\\
$V$&$-28.7\pm 0.3$&-28.5&-28.2&2.5\\
$W$&$-14.7\pm 0.2$&-14.6&-14.6&1.3\\
\hline
$V_{space}$&$32.9\pm 0.3$&32.8&31.3&2.8\\
\hline
\end{tabular}
\tablefoot{We provide for each velocity component the mean, standard error of the mean (SEM), median, mode and standard deviation (SD) values. 
}
\end{table}

\subsection{Kinematic parallax for other cluster members}

\begin{figure}[!htp]
\begin{center}
\includegraphics[width=0.49\textwidth]{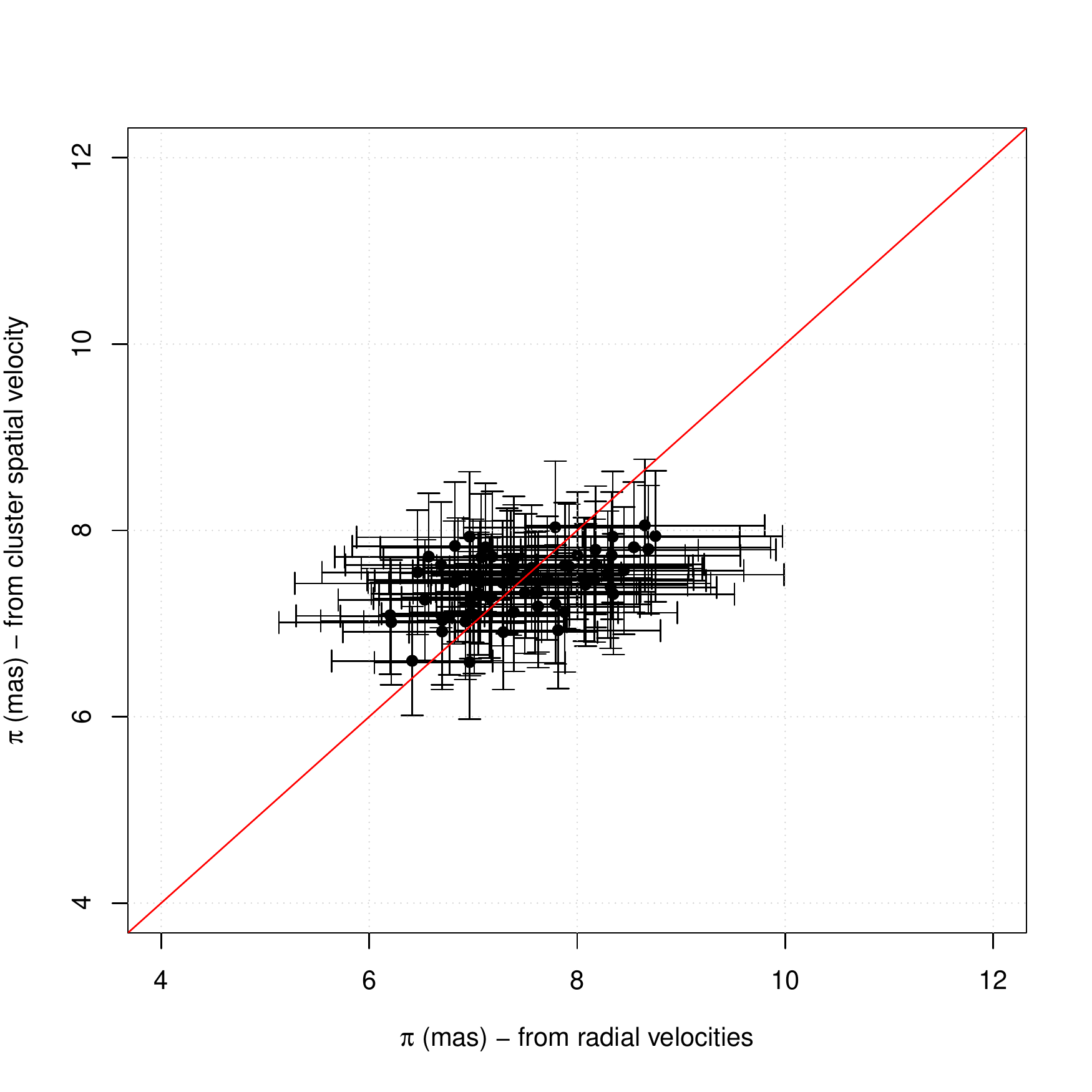}
\caption{
\label{fig12}
Comparison of the parallaxes derived from individual radial velocities and the mode of the cluster spatial velocity for the sample of 64 stars. The red solid line indicates perfect correlation.
\bigskip}
\end{center}
\end{figure}

Once the spatial velocity of the cluster is defined we can derive approximate parallaxes to the remaining group members under the assumption that all stars in a moving group share the same space motion. As a consistency check, we first compare the parallaxes derived from Eqs.~(1) and (2) for the sample of 64 stars with known radial velocities. Figure~\ref{fig12} illustrates this comparison. We find a mean difference and root mean square (rms) of 0.01~mas and 0.58~mas, respectively, between the two methods to compute kinematic parallaxes. These numbers are smaller than the typical error (about 1.0~mas) of the parallaxes listed in Table~\ref{tab_64stars} and confirm that both techniques yield consistent results. We performed a consistency check of the parallaxes derived from Eq.~(2) using the mean, median, and mode of the cluster spatial velocity. We verified that the agreement between both methods to compute kinematic parallaxes is better when we use the mode instead of the mean (or median) of the cluster spatial velocity owing to the asymmetry observed in the distribution for the total spatial velocity $V_{space}$ (see Fig.~\ref{fig11}).  

\begin{table*}[!btp]
\centering
\caption{Approximate parallaxes for the sample of 1146 stars.
\label{tab_1154stars}}
\resizebox{13cm}{!} {
\begin{tabular}{cccccc}
\hline
DANCe&$\alpha$&$\delta$&$\mu_{\alpha}\cos\delta$ & $\mu_{\delta}$&$\pi$\\
&(h:m:s)&($^{\circ}$ $^\prime$ $^\prime$$^\prime$)&(mas/yr)&(mas/yr)&(mas)\\

\hline		

J032727.94+240458.8	&	03 27 27.94	&	24  04 58.8	&$	20.5	\pm	2.1	$&$	-49.0	\pm	2.1	$&$	8.1	\pm	1.3	$\\	
J032733.90+264340.6	&	03 27 33.90	&	26 43 40.6	&$	19.7	\pm	6.0	$&$	-42.1	\pm	5.5	$&$	7.1	\pm	1.5	$\\	
J032801.57+230442.3	&	03 28  01.57	&	23  04 42.3	&$	18.5	\pm	0.4	$&$	-38.4	\pm	0.3	$&$	6.5	\pm	1.1	$\\	
J032816.07+222732.1	&	03 28 16.07	&	22 27 32.1	&$	24.4	\pm	2.7	$&$	-46.5	\pm	2.9	$&$	8.1	\pm	1.3	$\\	
J032835.63+240043.7	&	03 28 35.63	&	24  00 43.7	&$	23.4	\pm	2.4	$&$	-44.4	\pm	2.5	$&$	7.7	\pm	1.3	$\\	
J032856.70+261831.0	&	03 28 56.70	&	26 18 31.0	&$	21.8	\pm	1.1	$&$	-49.8	\pm	1.1	$&$	8.3	\pm	1.3	$\\	
J032907.68+252143.5	&	03 29  07.68	&	25 21 43.5	&$	22.6	\pm	2.0	$&$	-40.5	\pm	2.0	$&$	7.1	\pm	1.2	$\\	
J032911.26+270952.5	&	03 29 11.26	&	27  09 52.6	&$	20.5	\pm	2.6	$&$	-39.1	\pm	2.7	$&$	6.7	\pm	1.2	$\\	
J032933.69+214237.8	&	03 29 33.69	&	21 42 37.8	&$	20.5	\pm	3.7	$&$	-38.2	\pm	4.0	$&$	6.7	\pm	1.2	$\\	
J032958.76+232218.2	&	03 29 58.76	&	23 22 18.2	&$	19.6	\pm	2.2	$&$	-41.7	\pm	2.1	$&$	7.1	\pm	1.2	$\\	
J033016.90+213525.6	&	03 30 16.90	&	21 35 25.6	&$	21.7	\pm	4.8	$&$	-45.3	\pm	4.8	$&$	7.7	\pm	1.4	$\\	
J033023.16+250223.1	&	03 30 23.16	&	25  02 23.1	&$	24.2	\pm	1.7	$&$	-45.0	\pm	1.7	$&$	7.8	\pm	1.3	$\\	
J033109.77+261113.2	&	03 31  09.77	&	26 11 13.2	&$	24.5	\pm	2.2	$&$	-47.3	\pm	2.2	$&$	8.1	\pm	1.3	$\\	
J033111.29+241356.2	&	03 31 11.29	&	24 13 56.2	&$	23.1	\pm	1.5	$&$	-44.6	\pm	1.4	$&$	7.7	\pm	1.2	$\\	
J033115.95+251519.7	&	03 31 15.95	&	25 15 19.7	&$	22.2	\pm	0.8	$&$	-45.3	\pm	0.8	$&$	7.7	\pm	1.2	$\\	
J033116.63+264303.1	&	03 31 16.63	&	26 43  03.1	&$	21.2	\pm	2.2	$&$	-40.3	\pm	2.2	$&$	6.9	\pm	1.2	$\\	
J033117.94+260143.4	&	03 31 17.94	&	26  01 43.4	&$	19.6	\pm	2.3	$&$	-43.7	\pm	2.3	$&$	7.3	\pm	1.2	$\\	
J033120.70+255733.6	&	03 31 20.70	&	25 57 33.6	&$	22.4	\pm	2.0	$&$	-43.9	\pm	2.0	$&$	7.5	\pm	1.2	$\\	
J033126.72+271936.0	&	03 31 26.72	&	27 19 36.0	&$	21.2	\pm	3.5	$&$	-40.7	\pm	3.5	$&$	7.0	\pm	1.3	$\\	
J033127.36+263009.5	&	03 31 27.36	&	26 30  09.5	&$	18.7	\pm	2.3	$&$	-38.6	\pm	2.3	$&$	6.5	\pm	1.1	$\\	

\hline

\end{tabular}
}

\tablefoot{We provide for each star the DANCe identifier, position, proper motion and the parallax obtained in this work. This is a short version of the table. The complete version will be available in electronic format. 
}

\end{table*}

\begin{figure}[!htp]
\begin{center}
\includegraphics[width=0.49\textwidth]{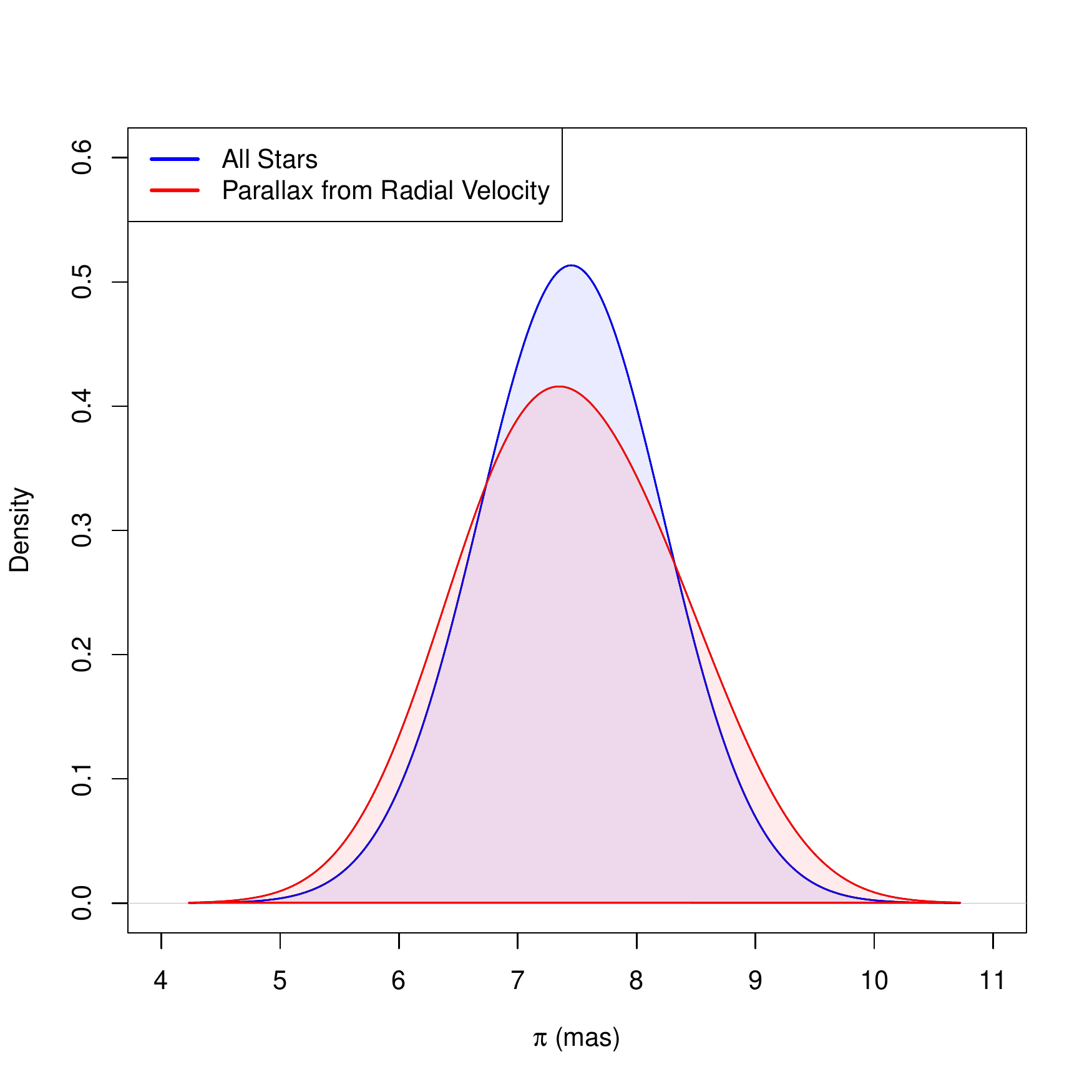}
\caption{
\label{fig13}
Probability density function for the parallaxes obtained from the sample of members with known radial velocities (64 stars) and all cluster members (1210 stars).\bigskip} 
\end{center}
\end{figure}


Then, we calculate from Eq.~(\ref{eq_plxapp}) the kinematic parallax for the remaining 1146 cluster members with unknown (or poor) radial velocities, binaries and multiple systems in our sample of stars selected by the CPSM. Table~\ref{tab_1154stars} lists the resulting kinematic parallaxes for these stars. As explained in Sect.~2.1, the associated uncertainties take into account the errors on proper motions, the CP position, and the average error of the individual spatial velocities derived for the 64~stars with known radial velocities (see Sect.~5.1). The rms of 0.58~mas that results from the analysis shown in Figure~\ref{fig12} is added quadratically to the parallax uncertainties to consider for possible systematic errors that could arise from using the  spatial velocity of the cluster to derive individual kinematic parallaxes. 

The mean parallax of the complete sample of cluster members (1210 stars) is $7.44\pm 0.01$~mas (median of 7.46~mas and standard deviation of 0.42~mas). This result is consistent with our first estimate using radial velocities  (see Sect.~5.1), but it is more precise, which comes naturally from the more significant number of stars used to derive the standard error of the mean. Figure~\ref{fig13} illustrates the density distribution of parallaxes obtained for these two samples. We note that the observed dispersion of parallaxes is also smaller for the complete sample. This mostly reflects the spread of the stellar proper motions since radial velocities are available for only a small fraction of the sample. However, we retain the results given in Sect~5.1 as our final solution in this work, because the individual parallaxes computed from radial velocities, when available, are better constrained.

\section{Discussion}

\subsection{Parallaxes for different samples of the Pleiades cluster}

\begin{figure*}[!htbp]
\begin{center}
\includegraphics[width=0.49\textwidth]{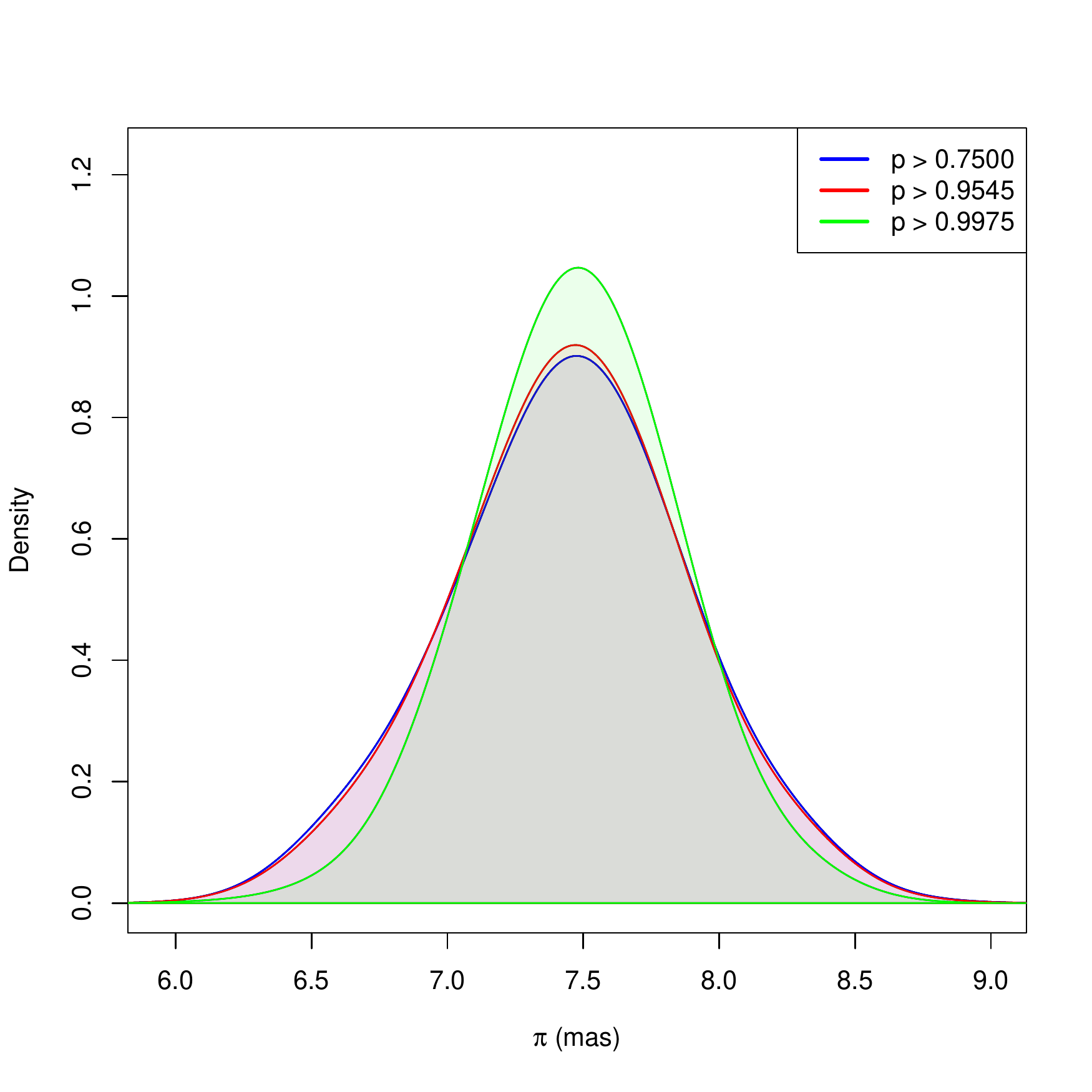}
\includegraphics[width=0.49\textwidth]{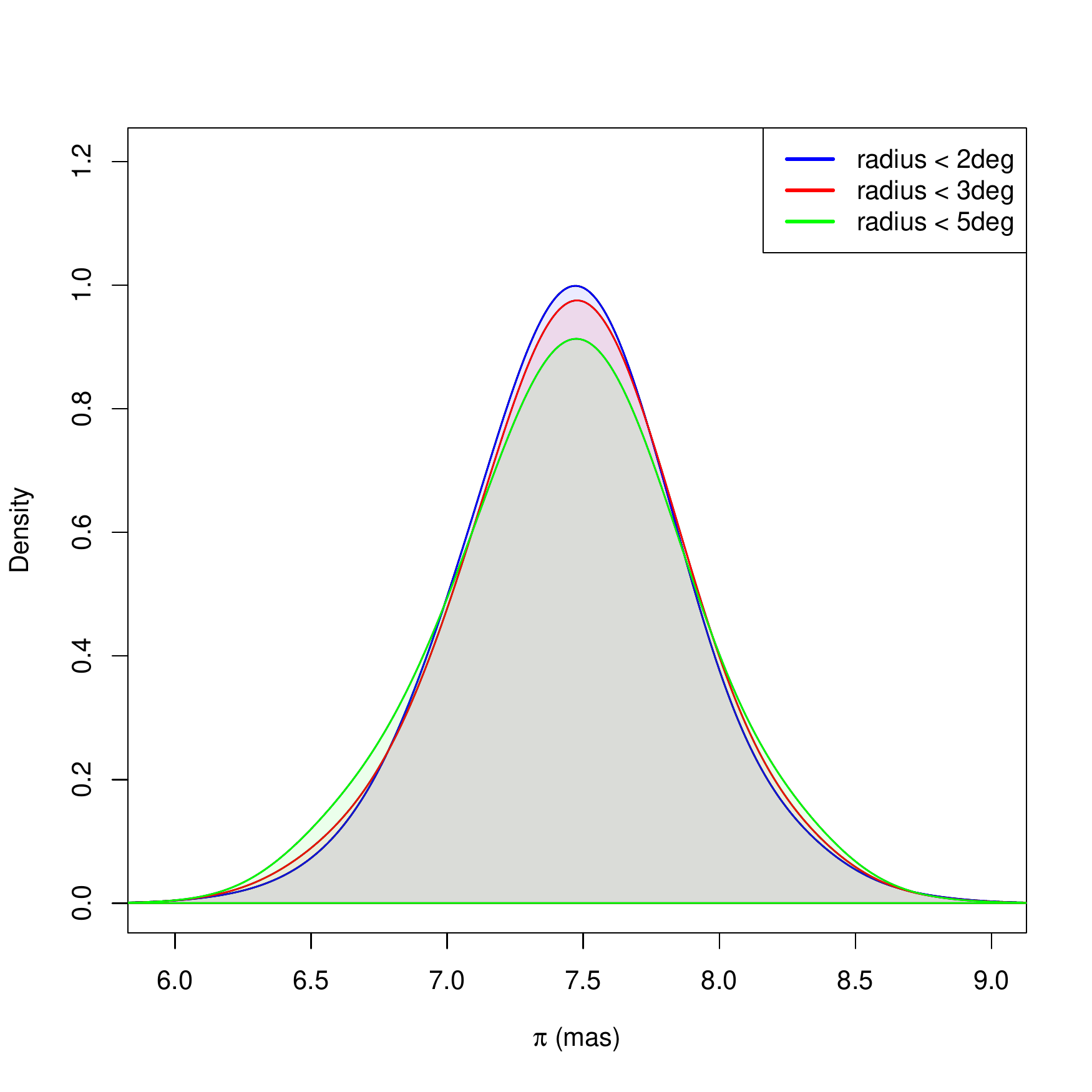}

\caption{Probability density function for the parallaxes of different samples of cluster members selected from their membership probability \textit{(left panel)} and distance to the cluster center \textit{(right panel)}.
\label{fig14}\bigskip\bigskip
}
\end{center}
\end{figure*}

In Sect.~4 we built two control samples based on the membership probability of cluster members and their distance to the cluster center to guide our CP analysis. Here, we perform a posteriori assessment of our results to investigate the distance of the Pleiades given by different subsets of our sample of cluster members selected by the CPSM (see Sect.~4.4). We construct these samples by selecting the stars within a given radius from the cluster center and membership probability threshold. Table~\ref{tab_plx_samples} lists the basic properties and Figure~\ref{fig14} illustrates the parallax distribution of each sample.  

It is apparent from this analysis that the parallax results obtained with different samples of cluster members are consistent between themselves. In particular, we note that the average parallaxes listed in Table~\ref{tab_plx_samples} are all compatible at the $3\sigma$ level. This gives us more confidence in the distance determination of the Pleiades open cluster derived in this paper (see Sect.~5.1) and confirms that it is not dependent on the adopted selection criteria.

\begin{figure}[!htp]
\begin{center}
\includegraphics[width=0.49\textwidth]{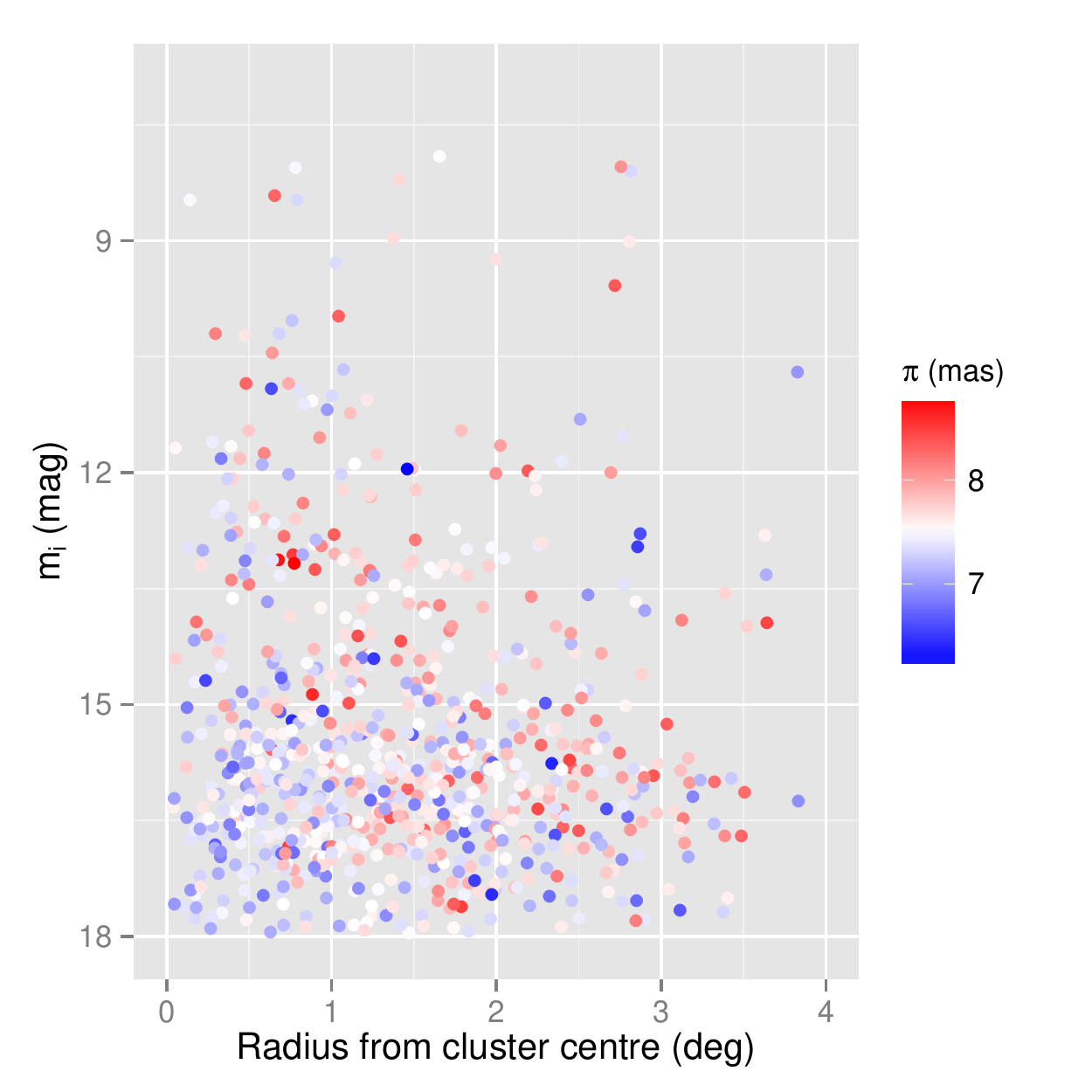}

\caption{Apparent magnitudes of cluster members as a function of the distance to the cluster center and parallaxes for individual stars.
\label{fig15}
}
\end{center}
\end{figure}

We also observe from Table~\ref{tab_plx_samples} that the standard deviation of parallaxes increases with the distance to the cluster center. Figure~\ref{fig15} shows the parallax dispersion as expected from the cluster depth along the line of sight as a function of the apparent magnitudes (from the \textit{i}-filter) and radius from cluster centre. This diagram provides information on the spatial distribution and segregation of stars within the cluster. We note a clear decrease of the stellar magnitudes with increasing radius and parallax dispersion. This confirms the effect of mass segregation in the cluster (using the stellar magnitudes as a proxy for spectral types) as anticipated in previous studies \citep[see e.g.][]{Raboud(1998),Palmer(2014)}. 

\begin{table}[!htp]
\centering
\caption{Kinematic parallax of the Pleiades derived from different subgroups in our sample of cluster members selected by the CPSM. 
\label{tab_plx_samples}}
\begin{tabular}{lcccc}

\hline
Sample&Stars&Mean $\pm$ SEM&Median&SD\\
&&(mas)&(mas)&(mas)\\
\hline
$p\geq0.7500$&1209&$7.44\pm 0.01$&7.46&0.42\\
$p\geq0.9545$&1146&$7.44\pm 0.01$&7.46&0.41\\
$p\geq0.9975$&789&$7.47\pm 0.01$&7.47&0.34\\
\hline
radius $\leq 2^{\circ}$&809&$7.46\pm 0.01$&7.46&0.37\\
radius $\leq 3^{\circ}$&1024&$7.46\pm 0.01$&7.46&0.39\\
radius $\leq 5^{\circ}$&1189&$7.45\pm 0.01$&7.46&0.41\\
\hline

\end{tabular}
\tablefoot{We provide for each sample the number of stars, mean, standard error of the mean (SEM), median and standard deviation (SD) values.  
}
\end{table}

\subsection{Comparison with previous parallaxes of individual stars}

Here, we compare the parallaxes derived in this work with previous results published in the literature for individual stars. Fortunately, our sample of 1210 cluster members contains 64 stars from various sources in the literature that applied different techniques to measure the stellar parallaxes. Their parallaxes are listed in Table~\ref{tab_plx_lit}. This comparison includes results from \citet{Gatewood(2000),Melis(2014),Madler(2016)}, \textit{Hipparcos} catalog \citep{vanLeeuwen(2007)} and the first results delivered by the \textit{Gaia} satellite. In this context, it is important to mention that the binary system HII~2147~NW+HII~2147~SE from the VLBI study of \citet{Melis(2014)} is not resolved in our sample. To make a better comparison with our results we decided to work with the weighted average of the proper motions and parallaxes of the two components.  

\begin{table*}[!htp]
\centering
\caption{Parallax results for cluster members in common with other studies. 
\label{tab_plx_lit}}
\resizebox{18cm}{!} {
\begin{tabular}{clcccccc}

\hline
&&\textbf{Kinematic parallax}&\textbf{\textit{\textit{Hipparcos}}}&\textbf{TGAS}&\textbf{VLBI project}&\textbf{Twin stars}&\textbf{Trigonometric parallax}\\
&&(This Work)&\citep{vanLeeuwen(2007)}&\citep{GaiaDR1(2016)}&\citep{Melis(2014)}&\citep{Madler(2016)}&\citep{Gatewood(2000)}\\
\hline
DANCe&Other name&$\pi$&$\pi$&$\pi$&$\pi$&$\pi$&$\pi$\\
&&(mas)&(mas)&(mas)&(mas)&(mas)&(mas)\\
\hline
J032856.70+261831.0	&	HIP~16217	&$	8.3	\pm	1.3	$&$	8.51	\pm	0.92	$&$	7.92	\pm	0.31	$&				&				&				\\
J032933.69+214237.8	&	TYC2~ 1246-808-1	&$	6.7	\pm	1.2	$&				&$	10.72	\pm	0.29	$&				&				&				\\
J033016.90+213525.6	&	TYC2~ 1264-617-1	&$	7.7	\pm	1.4	$&				&$	8.28	\pm	0.46	$&				&				&				\\
J033115.95+251519.7	&	HIP~16407	&$	7.7	\pm	1.2	$&$	6.75	\pm	0.85	$&$	7.57	\pm	0.34	$&				&				&				\\
J033128.34+214918.9	&	HIP~16423	&$	8.3	\pm	1.3	$&$	8.20	\pm	1.32	$&$	7.81	\pm	0.26	$&				&				&				\\
J033458.66+233148.5	&	TYC2~ 1798-596-1	&$	7.2	\pm	1.2	$&				&$	7.43	\pm	0.36	$&				&				&				\\
J033724.05+222103.6	&	TYC2~ 1247-211-1	&$	7.6	\pm	1.2	$&				&$	6.38	\pm	0.35	$&				&				&				\\
J033840.72+223934.6	&	HIP~17000	&$	7.6	\pm	1.2	$&$	8.12	\pm	0.51	$&				&				&				&				\\
J033913.20+215035.7	&	HIP~17043	&$	7.3	\pm	1.2	$&$	7.33	\pm	0.61	$&$	7.18	\pm	0.32	$&				&				&				\\
J033951.16+251141.5	&	TYC2~ 1803-1236-1	&$	7.2	\pm	1.2	$&				&$	7.27	\pm	0.33	$&				&				&				\\
J034122.97+232913.0	&	HIP~17225	&$	7.6	\pm	1.2	$&$	8.10	\pm	1.06	$&$	7.85	\pm	0.24	$&				&				&				\\
J034204.72+225130.8	&	HIP~17289	&$	7.3	\pm	2.0	$&$	7.65	\pm	1.50	$&$	7.36	\pm	0.25	$&				&$	7.53	\pm	0.19	$&				\\
J034224.00+212824.6	&	HIP~17316	&$	8.3	\pm	1.3	$&$	7.27	\pm	1.59	$&$	7.84	\pm	0.23	$&				&$	7.94	\pm	0.15	$&				\\
J034229.87+200859.9	&	HIP~17325	&$	7.9	\pm	1.3	$&$	7.30	\pm	1.00	$&$	7.75	\pm	0.47	$&				&				&				\\
J034255.11+242935.1	&	TYC2~ 1803-478-1	&$	7.3	\pm	1.2	$&				&$	7.61	\pm	0.31	$&				&				&				\\
J034341.53+233856.9	&	HIP~17401	&$	7.4	\pm	1.2	$&$	7.58	\pm	0.90	$&$	8.19	\pm	0.52	$&				&				&				\\
J034348.34+250015.7	&	TYC2~ 1803-8-1	&$	7.6	\pm	1.2	$&				&$	7.39	\pm	0.30	$&$	7.418	\pm	0.025	$&				&				\\
J034425.72+242341.0	&	TYC2~ 1803-188-1	&$	8.2	\pm	1.3	$&				&$	8.14	\pm	0.32	$&				&				&				\\
J034444.86+204452.8	&	HIP~17481	&$	7.4	\pm	1.2	$&$	9.44	\pm	1.03	$&$	7.61	\pm	0.45	$&				&				&				\\
J034448.21+241722.1	&	HIP~17489	&$	7.5	\pm	1.2	$&$	8.65	\pm	0.36	$&				&				&				&				\\
J034458.92+220156.8	&	HIP~17511	&$	7.6	\pm	1.0	$&$	10.67	\pm	1.37	$&$	6.98	\pm	0.29	$&				&$	7.40	\pm	0.24	$&				\\
J034509.74+245021.3	&	HIP~17527	&$	7.6	\pm	1.6	$&$	7.97	\pm	0.37	$&				&				&				&				\\
J034512.49+242802.2	&	HIP~17531	&$	7.5	\pm	1.2	$&$	7.97	\pm	0.33	$&				&				&				&				\\
J034521.19+234338.8	&	HII625	&$	7.4	\pm	1.2	$&				&				&$	7.223	\pm	0.057	$&				&				\\
J034531.99+211448.1	&	HIP~17552	&$	8.0	\pm	1.3	$&$	11.04	\pm	0.93	$&$	7.79	\pm	0.35	$&				&				&				\\
J034539.90+224140.1	&	TYC2~ 1799-1102-1	&$	7.4	\pm	1.2	$&				&$	7.80	\pm	0.31	$&				&				&				\\
J034548.82+230849.7	&	HIP~17572	&$	8.1	\pm	1.3	$&$	8.24	\pm	0.75	$&$	7.83	\pm	0.77	$&				&				&				\\
J034549.61+242203.9	&	HIP~17573	&$	7.5	\pm	1.2	$&$	8.51	\pm	0.28	$&				&				&				&				\\
J034554.48+243316.2	&	HIP~17579	&$	7.6	\pm	1.2	$&$	8.77	\pm	0.54	$&$	7.95	\pm	0.87	$&				&				&				\\
J034555.73+274900.8	&	TYC2~ 1807-955-1	&$	6.9	\pm	1.2	$&				&$	3.63	\pm	0.28	$&				&				&				\\
J034559.14+252354.9	&	HIP~17583	&$	7.7	\pm	1.2	$&$	8.00	\pm	0.89	$&$	7.57	\pm	0.34	$&				&				&				\\
J034602.90+243140.4	&	HIP~17588	&$	7.3	\pm	1.2	$&$	8.58	\pm	0.56	$&$	7.54	\pm	0.64	$&				&				&$	7.40	\pm	0.61	$\\
J034619.57+235654.1	&	HIP~17608	&$	7.7	\pm	1.2	$&$	8.58	\pm	0.37	$&				&				&				&				\\
J034627.28+241518.0	&	TYC2~ 1800-1908-1	&$	7.7	\pm	1.2	$&				&$	7.74	\pm	0.32	$&				&				&				\\
J034634.20+233726.5	&	TYC2~ 1800-1621-1	&$	7.2	\pm	1.2	$&				&$	7.81	\pm	0.26	$&				&				&				\\
J034640.25+232951.7	&	HII1136	&$	7.7	\pm	1.2	$&				&				&$	7.382	\pm	0.031	$&				&				\\
J034659.40+243112.4	&	HIP~17664	&$	7.8	\pm	1.3	$&$	7.66	\pm	0.66	$&$	7.58	\pm	0.29	$&				&				&$	7.25	\pm	0.58	$\\
J034704.21+235942.8	&	TYC2~ 1800-1579-1	&$	7.4	\pm	1.2	$&				&$	7.13	\pm	0.32	$&				&				&				\\
J034720.97+234812.0	&	HIP~17692	&$	8.3	\pm	1.7	$&$	8.90	\pm	0.77	$&$	7.17	\pm	0.38	$&				&				&				\\
J034721.04+240658.6	&	TYC2~ 1800-2201-1	&$	7.4	\pm	1.2	$&				&$	7.88	\pm	0.46	$&				&				&				\\
J034722.90+225519.6	&	HIP~17694	&$	8.0	\pm	1.3	$&$	8.62	\pm	0.84	$&$	6.62	\pm	0.66	$&				&				&				\\
J034729.45+241718.0	&	HIP~17704	&$	7.5	\pm	1.2	$&$	9.42	\pm	0.75	$&$	7.57	\pm	0.40	$&				&				&$	8.97	\pm	0.68	$\\
J034820.82+232516.5	&	HIP~17776	&$	7.8	\pm	1.2	$&$	8.45	\pm	0.39	$&				&				&				&				\\
J034830.10+242043.9	&	HIP~17791	&$	7.3	\pm	1.2	$&$	7.87	\pm	1.32	$&				&				&				&				\\
J034834.52+232604.9	&	HII1924	&$	7.2	\pm	0.9	$&				&				&				&$	6.98	\pm	0.78	$&				\\
J034906.12+234652.3	&	HII2147NW+SE	&$	7.2	\pm	1.2	$&				&				&$	7.322	\pm	0.021	$&				&				\\
J034909.74+240312.3	&	HIP~17847	&$	7.5	\pm	1.2	$&$	8.53	\pm	0.39	$&				&				&				&				\\
J034911.22+240812.2	&	HIP~17851	&$	7.9	\pm	1.3	$&$	8.54	\pm	0.31	$&				&				&				&				\\
J034911.26+223634.1	&	TYC2~ 1800-118-1	&$	7.3	\pm	1.2	$&				&$	7.68	\pm	0.35	$&				&				&				\\
J034912.19+235312.5	&	TYC2~ 1800-1406-1	&$	7.3	\pm	1.2	$&				&$	7.84	\pm	0.29	$&				&				&				\\
J034921.75+242251.4	&	HIP~17862	&$	7.4	\pm	1.2	$&$	8.18	\pm	0.59	$&$	7.22	\pm	0.46	$&				&				&				\\
J034925.98+241451.7	&	TYC2~ 1800-901-1	&$	7.5	\pm	1.2	$&				&$	7.33	\pm	0.31	$&				&				&				\\
J034928.76+234243.6	&	HII2311	&$	7.3	\pm	0.9	$&				&				&				&$	7.52	\pm	0.84	$&				\\
J034938.18+223200.5	&	HIP~17892	&$	7.5	\pm	1.2	$&$	8.30	\pm	0.66	$&$	7.65	\pm	0.38	$&				&				&				\\
J034940.92+232029.7	&	TYC2~ 1800-628-1	&$	7.4	\pm	1.2	$&				&$	7.21	\pm	0.31	$&				&				&				\\
J034943.53+234242.7	&	HIP~17900	&$	7.3	\pm	1.2	$&$	8.72	\pm	0.60	$&$	7.88	\pm	0.54	$&				&				&				\\
J034955.07+221438.9	&	HIP~17921	&$	7.8	\pm	1.2	$&$	8.86	\pm	0.42	$&$	7.52	\pm	0.53	$&				&				&				\\
J034956.60+242056.4	&	TYC2~ 1800-985-1	&$	7.4	\pm	1.2	$&				&$	7.41	\pm	0.35	$&				&				&				\\
J034958.05+235055.3	&	HIP~17923	&$	7.8	\pm	1.2	$&$	6.81	\pm	0.72	$&$	7.24	\pm	0.44	$&				&				&				\\
J035051.45+231944.5	&	TYC2~ 1800-586-1	&$	7.7	\pm	1.2	$&				&$	7.07	\pm	0.28	$&				&				&				\\
J035052.43+235741.3	&	HIP~17999	&$	7.6	\pm	1.2	$&$	9.93	\pm	0.75	$&$	3.40	\pm	0.73	$&				&				&				\\
J035253.48+244256.6	&	HIP~18154	&$	7.3	\pm	1.2	$&$	10.13	\pm	1.66	$&				&				&				&				\\
J035801.69+204036.5	&	HIP~18544	&$	8.2	\pm	1.3	$&$	8.20	\pm	1.44	$&$	7.70	\pm	0.46	$&				&				&				\\
J035820.90+240452.0	&	HIP~18559	&$	8.0	\pm	1.3	$&$	6.71	\pm	1.27	$&$	7.31	\pm	0.43	$&				&				&				\\
\hline

\end{tabular}
}
\tablefoot{We provide for each star the DANCe identifier, an alternative identifier and the parallax given in each study. 
}

\end{table*}

\begin{figure*}[!htp]
\begin{center}
\includegraphics[width=0.67\textwidth]{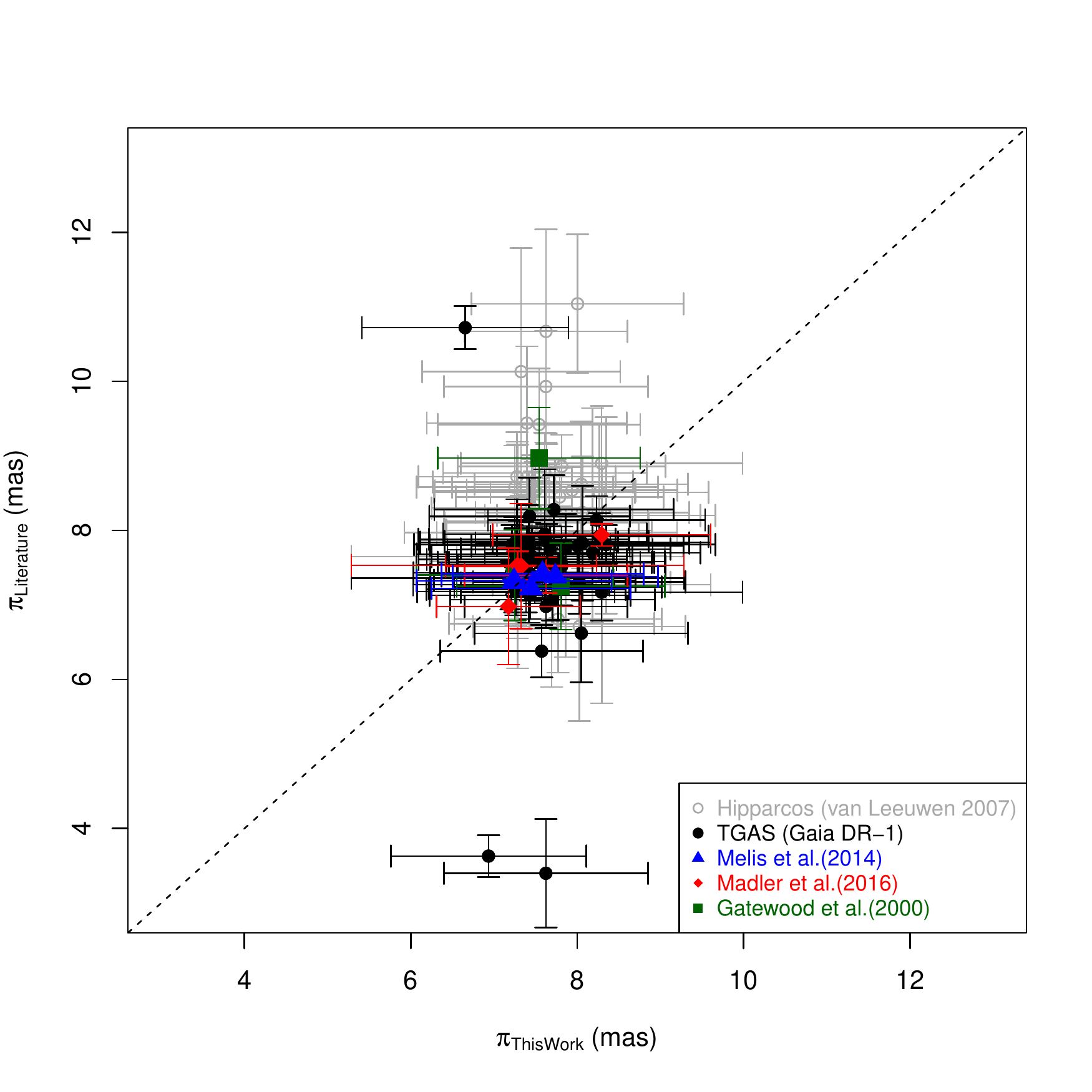}
\caption{Comparison between the kinematic parallaxes derived in this work and parallaxes from the literature \citep{vanLeeuwen(2007),Melis(2014),Madler(2016),Gatewood(2000),GaiaDR1(2016)}. The black dashed line indicates the expected distribution for equal results. 
\label{fig16}
}
\end{center}
\end{figure*}

Figure~\ref{fig16} compares our kinematic parallaxes with published results in the literature, and Table~\ref{tab_rms} presents the mean difference and rms that result from the comparison with each source. This comparison confirms the good accuracy of our results when compared to the individual parallaxes of twin stars obtained by \citet{Madler(2016)} and the very precise multi-epoch VLBI trigonometric parallaxes from \citet{Melis(2014)} given that these works are independent. The comparison with the trigonometric parallaxes measured by \citet{Gatewood(2000)} returned that the parallax for one star is not consistent with our results, although we observe a good agreement with the other two stars in common. In general our results do not reproduce the parallaxes from the \textit{\textit{Hipparcos}} catalog. But, interestingly, a few stars in our sample with parallaxes between 7.0~mas and 8.0~mas show a reasonable agreement with the \textit{\textit{Hipparcos}} parallaxes. The question arises whether the discrepancy of \textit{\textit{Hipparcos}} parallaxes indeed applies to all cluster members as commonly reported in the literature.  

\begin{table}[!h]
\centering
\caption{Comparison of the parallaxes derived in this work with other studies in the literature.  
\label{tab_rms}}
\begin{tabular}{lccc}

\hline
Sample&Stars&Difference&rms\\
&&(mas)&(mas)\\
\hline
\citet{Melis(2014)}&4&0.16&0.26\\
\citet{Madler(2016)}&5&0.07&0.28\\
\citet{Gatewood(2000)}&3&-0.34&1.09\\
\textit{\textit{Hipparcos}} \citep{vanLeeuwen(2007)}&39&-0.69&1.25\\
TGAS (from \textit{Gaia} DR1)&48&0.16&1.09\\
\hline

\end{tabular}
\tablefoot{We provide for each comparison sample the number of stars in common with this work, the mean difference and the rms of the comparison. 
}
\end{table}


During the submission process of the current manuscript, the first \textit{Gaia} data release (\textit{Gaia} DR1) became available providing astrometry and photometry for 1 billion sources brighter than magnitude 20.7 in the satellite's white-light photometric band \citep{GaiaDR1(2016)}. \textit{Gaia} DR1 also includes trigonometric parallaxes and mean proper motions for about 2 million stars from the \textit{Tycho-Gaia} astrometric solution (TGAS). We use this first dataset of trigonometric parallaxes, which has just been delivered by the \textit{Gaia} collaboration, to make a comparison with the results obtained from the moving cluster method in this work. As expected, this is the source of trigonometric parallaxes with the largest number of stars in common with our sample of Pleiades stars (see Table~\ref{tab_rms}). This gives us the possibility to perform a more detailed comparison of the results  delivered by these studies.

As illustrated in Figure~\ref{fig16}, the TGAS trigonometric parallaxes are in good agreement with the kinematic parallaxes derived from the moving-cluster method and the other independent studies discussed before. Indeed, this confirms the long-term suspicion that the \textit{\textit{Hipparcos}} parallaxes for Pleiades stars are biased. However, there are still a few stars in our sample that exhibit discrepant proper motions and/or parallaxes with respect to the TGAS results (at the 3$\sigma$ level). We summarize their properties in Table~\ref{tab_discrepant}.

\begin{table*}[!htp]
\centering
\caption{Comparison of the proper motions and parallaxes with TGAS results from \textit{Gaia} DR1 for discrepant stars. 
\label{tab_discrepant}}
\resizebox{17cm}{!} {
\begin{tabular}{cccccccc}

\hline
&\multicolumn{3}{c}{\textbf{DANCe Proper Motions + Kinematic Parallaxes}}&&\multicolumn{3}{c}{\textbf{TGAS (\textit{Gaia} DR1) Results}}\\
\hline
DANCe&$\mu_{\alpha}\cos\delta$&$\mu_{\delta}$&$\pi$&TYC2/HIP&$\mu_{\alpha}\cos\delta$&$\mu_{\delta}$&$\pi$\\
&(mas/yr)&(mas/yr)&(mas)&&(mas/yr)&(mas/yr)&(mas)\\
\hline
J035052.43+235741.3	&$	20.0	\pm	0.9	$&$	-45.4	\pm	0.9	$&$	7.6	\pm	1.2	$&	HIP~17999	&$	20.092	\pm	0.047	$&$	-45.355	\pm	0.023	$&$	3.40	\pm	0.73	$\\
J032933.69+214237.8	&$	20.5	\pm	3.7	$&$	-38.2	\pm	4.0	$&$	6.7	\pm	1.2	$&	TYC2~1246-808-1	&$	6.794	\pm	1.001	$&$	-37.726	\pm	0.490	$&$	10.72	\pm	0.29	$\\
J034555.73+274900.8	&$	15.1	\pm	1.9	$&$	-43.1	\pm	1.9	$&$	6.9	\pm	1.2	$&	TYC2~1807-955-1	&$	4.019	\pm	1.620	$&$	-43.033	\pm	0.761	$&$	3.63	\pm	0.28	$\\
\hline
J034634.20+233726.5	&$	16.8	\pm	1.0	$&$	-43.9	\pm	1.0	$&$	7.2	\pm	1.2	$&	TYC2~1800-1621-1	&$	21.338	\pm	0.562	$&$	-42.786	\pm	0.411	$&$	7.81	\pm	0.26	$\\
J034911.26+223634.1	&$	18.6	\pm	1.1	$&$	-43.2	\pm	1.1	$&$	7.3	\pm	1.2	$&	TYC2~1800-118-1	&$	19.484	\pm	0.361	$&$	-47.537	\pm	0.389	$&$	7.68	\pm	0.35	$\\
J034955.07+221438.9	&$	22.5	\pm	0.7	$&$	-45.4	\pm	0.7	$&$	7.8	\pm	1.2	$&	HIP~17921	&$	20.169	\pm	0.030	$&$	-44.225	\pm	0.015	$&$	7.52	\pm	0.53	$\\
J035820.90+240452.0	&$	16.8	\pm	0.9	$&$	-49.4	\pm	0.9	$&$	8.0	\pm	1.3	$&	HIP~18559	&$	19.021	\pm	0.080	$&$	-45.223	\pm	0.031	$&$	7.31	\pm	0.43	$\\
J033016.90+213525.6	&$	21.7	\pm	4.8	$&$	-45.3	\pm	4.8	$&$	7.7	\pm	1.4	$&	TYC2 1246-617-1	&$	31.661	\pm	1.349	$&$	-44.733	\pm	0.610	$&$	8.28	\pm	0.46	$\\
\hline
\end{tabular}
}
\tablefoot{We provide for each star the DANCe identifier, proper motions from the DANCe survey, kinematic parallaxes derived from the moving cluster method, Tycho-2/\textit{\textit{Hipparcos}} identifiers, proper motions and parallaxes from the TGAS catalog.  \bigskip
}
\end{table*}

\begin{figure*}[!htbp]
\begin{center}
\includegraphics[width=0.49\textwidth]{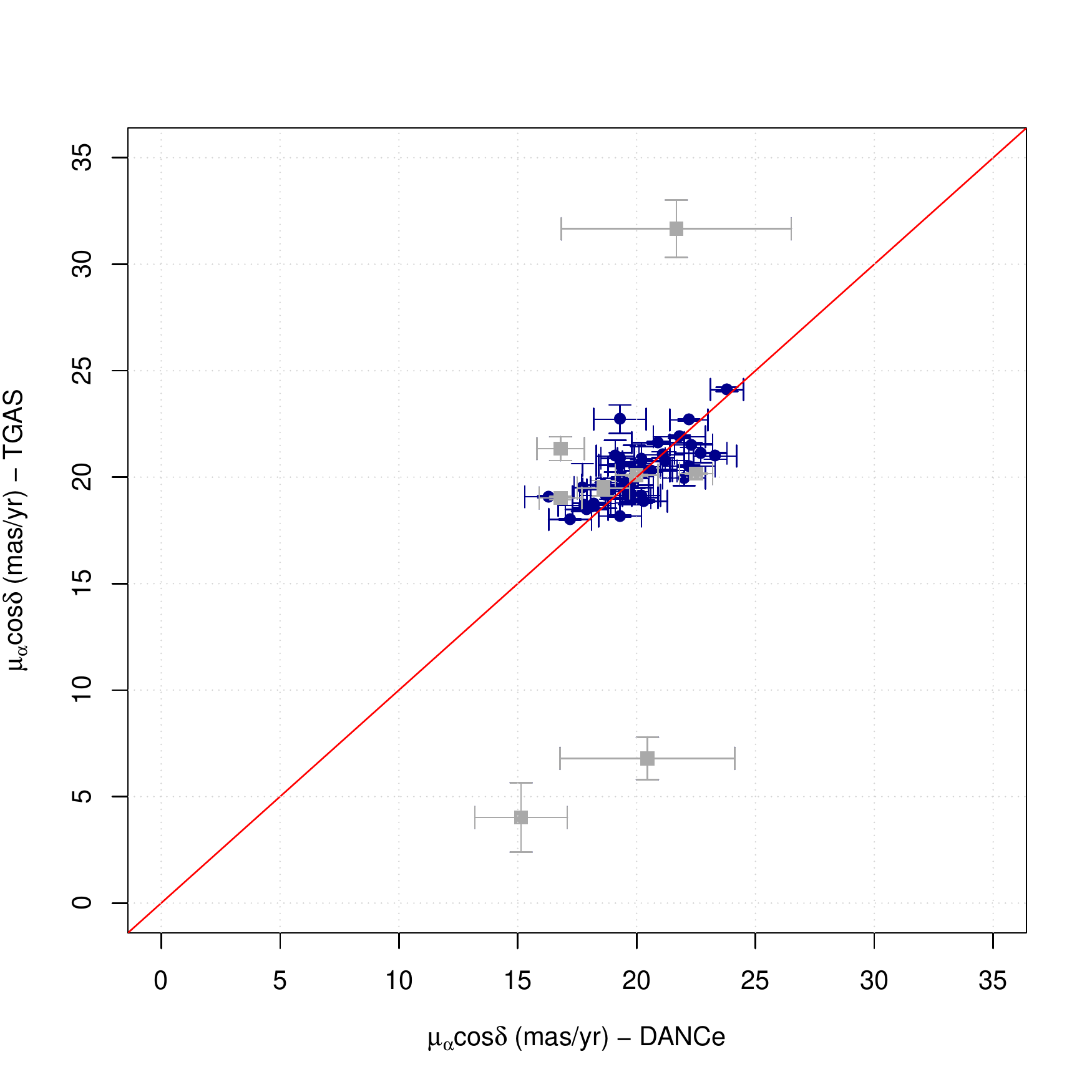}
\includegraphics[width=0.49\textwidth]{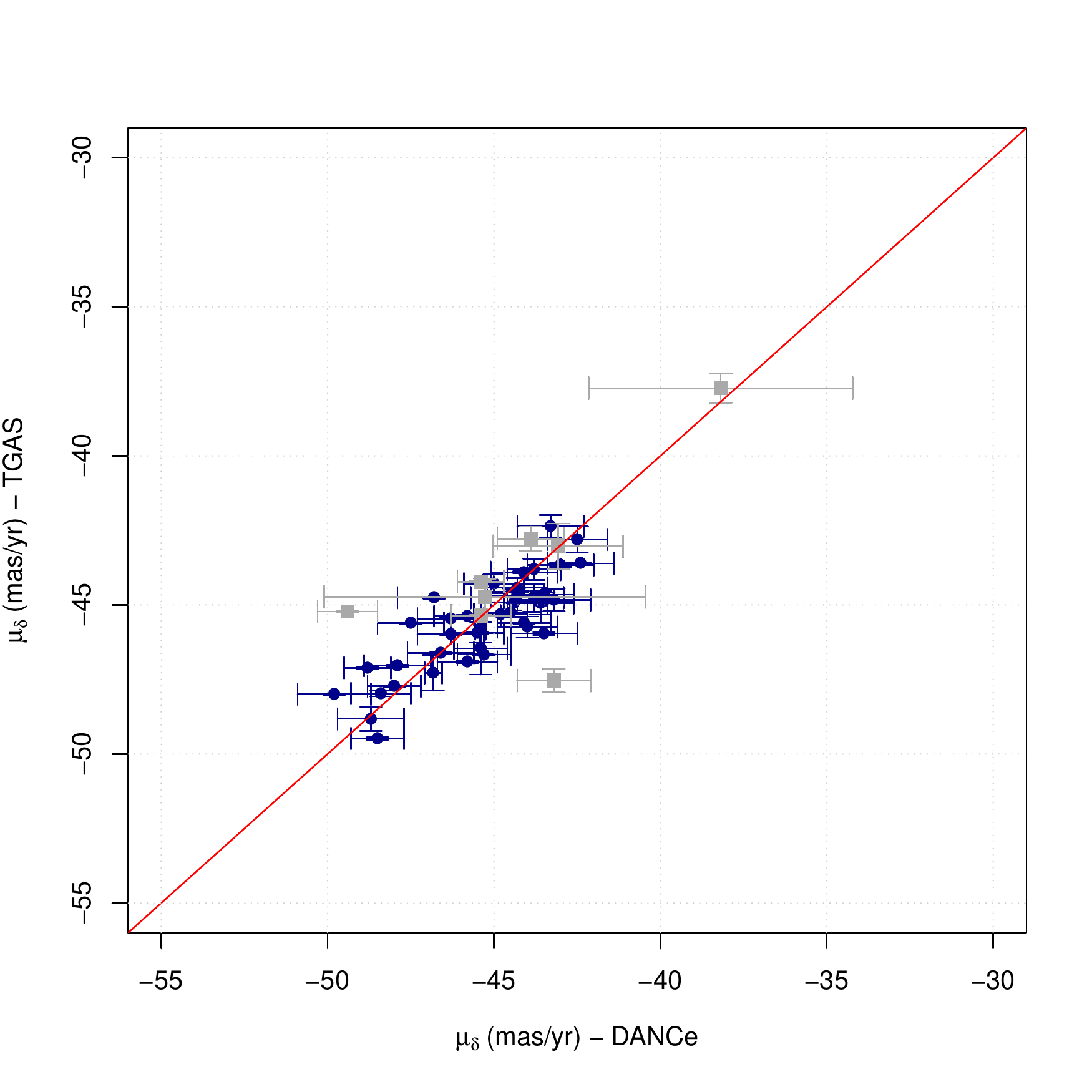}
\caption{Comparison between the proper motions measurements in right ascension \textit{(left panel)} and declination \textit{(right panel)} from the DANCe project and the TGAS catalog. The gray squares indicate the discrepant stars  listed in Table~\ref{tab_discrepant}. 
\label{fig17}
}
\end{center}
\end{figure*}

We note from Figure~\ref{fig16} that the kinematic parallaxes derived in this work for three stars (HIP~17999, TYC2~1246-808-1, and TYC2~1807-955-1) are not consistent with TGAS.  It is clear from Table~\ref{tab_discrepant} that the proper motions obtained for HIP~17999 in both surveys are consistent between each other and also in good agreement  with the observed proper motions for other cluster members (see, for example, Fig.~\ref{fig7}). Interestingly, the TGAS trigonometric parallax of $\pi=3.40\pm0.73$~mas is also not consistent with the results reported by the \textit{\textit{Hipparcos}} satellite: $\pi=9.83\pm1.00$~mas \citep{ESA(1997)} and $\pi=9.93\pm0.75$~mas \citep{vanLeeuwen(2007)}. So, if we assume that HIP~17999 is indeed a cluster member, as suggested by its proper motion, then the TGAS trigonometric parallax for this target needs further clarification.  On the other hand, TYC2~1246-808-1 and TYC2~1807-955-1 exhibit different proper motion values in the right ascension component that justifies the different parallax results in both studies. We searched for proper motion measurements for these two stars in the UCAC4 \citep{UCAC4} and PPMXL \citep{PPMXL} catalogs, and verified that they are more consistent with the TGAS results. As discussed in previous papers of the DANCe project \citep[see e.g.][]{Bouy(2013),Bouy(2015)} the proper motion measurements close to the saturation limit of the instruments (that is considered to be the limit of sensitivity of the Tycho-2 catalog) suffer from biases and incompleteness, such that they should be considered with caution. Replacing the DANCe proper motions for these two stars by the TGAS results will force them to be rejected in our membership analysis. We verified that removing TYC2~1246-808-1 and TYC2~1807-955-1 from our sample has negligible impact on the results presented in the previous sections. 

Figure~\ref{fig17} compares the DANCe and TGAS proper motions for the stars in common. In general there is a good agreement between the proper motions in both surveys, but another four stars (TYC2~1800-1621-1, TYC2 1800-118-1, HIP~17921, and HIP~18559) exhibit proper motion measurements that are not consistent at the $3\sigma$ level (see also Table~\ref{tab_discrepant}). However, the parallaxes derived in this work are still compatible with the TGAS results. The shift (or absolute difference) in the observed proper motions is smaller compared, for example, to TYC2~1246-808-1 and TYC2~1807-955-1 to produce a more significant bias in the parallaxes. In addition, we note that the proper motions for TYC2~1246-617-1 are still compatible within $3\sigma$, because of the large error bars of the DANCe measurement, but the difference in the right ascension component is 10.0~mas/yr and it clearly stands out in the comparison shown in Figure~\ref{fig17}. The systematic shift (or mean difference) between the two proper motion datasets (after removing the discrepant stars) is -0.2~mas/yr in right ascension and 0.3~mas/yr in declination. The rms of the comparison is 1.2~mas/yr in right ascension and 1.1~mas/yr in declination. This represents a good agreement between the DANCe and TGAS proper motions.

After removing the discrepant stars listed in Table~\ref{tab_discrepant} from our sample of stars in common with TGAS, the mean difference and rms given in Table~\ref{tab_rms} become 0.12~mas and 0.49~mas, respectively. The mean parallax derived from the TGAS results is $7.53\pm0.19$~mas with median of 7.58~mas and standard deviation of 0.38~mas. However, as reported in the paper describing the summary of  the \textit{Gaia} DR1 properties, there is a systematic uncertainty of about 0.3~mas, which is correlated over small spatial scales and must be added to the parallax uncertainties. Thus, we report the mean parallax of $7.53\pm 0.30$~mas as a more conservative solution obtained from the TGAS results for the stars in common with our sample. This is consistent with a distance estimate of $133\pm5$~pc for the cluster. However, this result is somewhat different from the value of $134\pm6$~pc reported in Sect.~5.5 of the \textit{Gaia}~DR1 paper \citep[see][]{GaiaDR1(2016)}, because of the different sample of stars used in each case. So, we conclude that the distance to the Pleiades cluster derived from the TGAS trigonometric parallaxes is consistent with the distance determination obtained in this work and previous studies. 

To summarize, the investigations reported above using parallaxes derived from independent studies and different samples of Pleiades stars show that there is a good agreement with the results obtained in this work, and this make us confident that our distance estimate (see Sect.~5.1) is indeed representative of the cluster. 
 

\subsection{Revisiting the distance of the Pleiades open cluster}

\begin{figure*}[!htp]
\begin{center}
\includegraphics[width=1.0\textwidth]{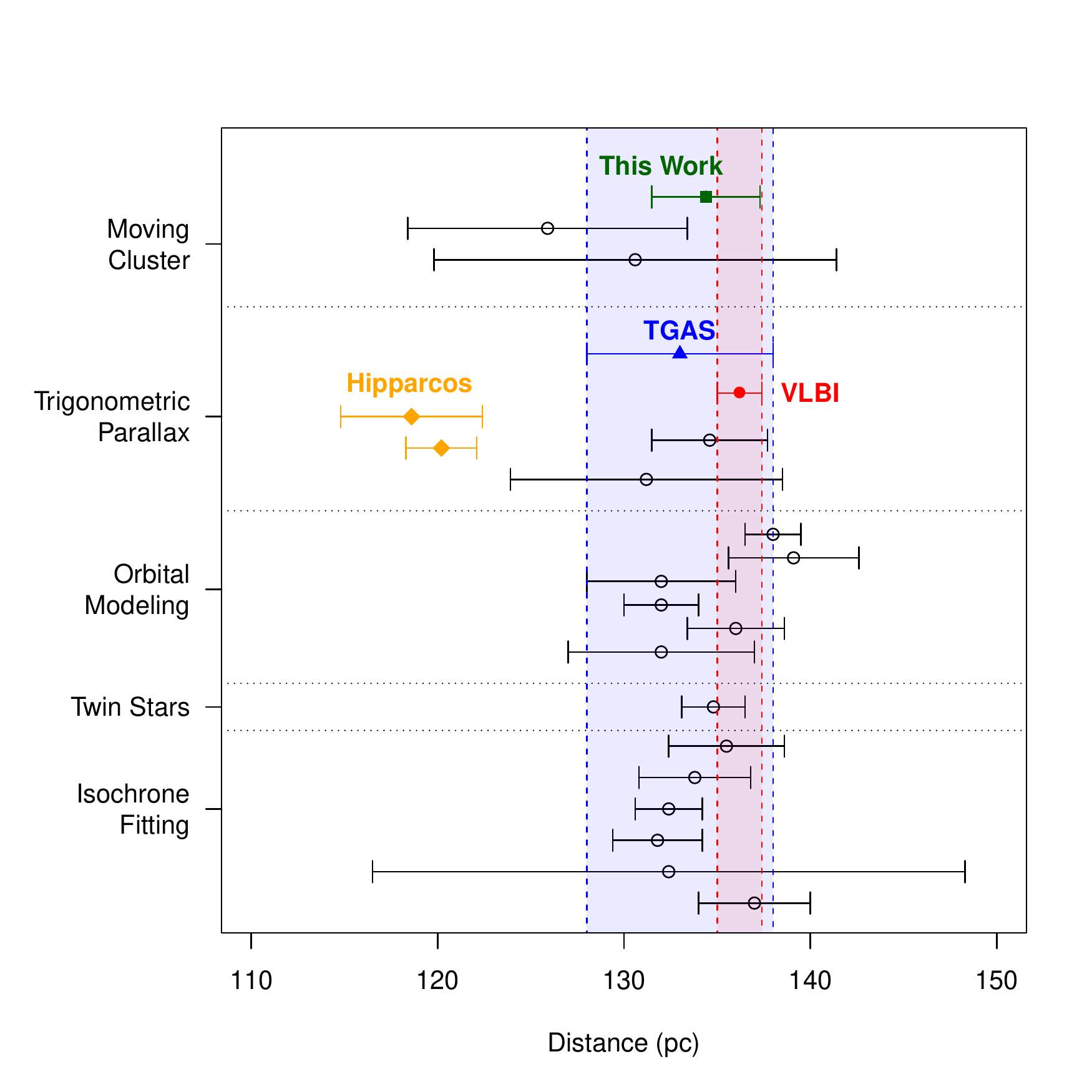}
\caption{Summary of distance measurements in the literature for the Pleiades open cluster obtained from different methods. The results presented here come from: (i) \textit{moving cluster} - This Work (see Sect.~5.1), \citet{Roeser(2013)}, \citet{Narayanan(1999)}, (ii) \textit{trigonometric parallax}~- \citet{vanLeeuwen(1999)}, \citet{vanLeeuwen(2009)}, \citet{Melis(2014)}, \citet{Soderblom(2005)}, \citet{Gatewood(2000)}, TGAS \citep{GaiaDR1(2016)} - using result reported in this work (see Sect.~6.2), (iii) \textit{orbital modeling} - \citet{Groenewegen(2007)}, \citet{Southworth(2005)}, \citet{Zwahlen(2004)}, \citet{Munari(2004)}, \citet{Pan(2004)}, \citet{David(2016)}, (iv) \textit{twin stars}- \citet{Madler(2016)}, (v) \textit{isochrone fitting} - \citet{An(2007)}, \citet{Percival(2005)}, \citet{Stello(2001)}, \citet{Pinsonneault(1998)}, \citet{Giannuzzi(1995)} and  \citet{Nicolet(1981)},
\label{fig18}
}
\end{center}
\end{figure*}

Figure~\ref{fig18} summarizes the distances obtained for the Pleiades cluster in the literature by each method.  This is an updated version of Figure~1 from \citet{Melis(2014)} to include the more recent results. 

We note that all methods employed so far yield a distance determination that is larger than the result obtained by \textit{Hipparcos}. However, in many cases the error bars do not allow us to unambiguously reject the \textit{Hipparcos} result. In this context, it is important to distinguish between the methods that yield distances to individual cluster members and the ones that can only be used to estimate the average distance of the cluster.  Among the methods included in this discussion, the main sequence fitting (i.e., isochrone fitting) is the only method that cannot be used to deliver individual distances to cluster members. Although this methodology yields mostly smaller errors in the distance determination, the resulting distance modulus is representative of the cluster photometry, but not individual stars. The error on the average distance obtained in this work is sometimes greater compared to the other distance determinations of the cluster. This reflects the dispersion of individual parallaxes (see Sect.~5.1) in our solution, which is a direct result of the spread in proper motions and radial velocities. 

As illustrated in Fig.~\ref{fig18}, the average distance for the Pleiades cluster obtained in this work clearly supports the recent VLBI distance determination for the cluster, based on the very precise and accurate astrometry by \citet{Melis(2014)}. Our results are also very consistent with other non-\textit{Hipparcos} distances reported  by different studies over the last two decades. The weighted mean distance of the non-\textit{Hipparcos} results listed in Fig.~\ref{fig18} (including this work) yields a mean distance of $135.0\pm 0.6$~pc (median of 133.8~pc and standard deviation of 2.7~pc). This confirms the historical discrepancy between the \textit{\textit{Hipparcos}} distance determination and the results given by other methods (see also Sect.~1). 

Despite our efforts to investigate the distance of the cluster, we emphasize that  individual distances for Pleiades stars are available for only a few cluster members in the literature (as discussed in Sect.~1). The results obtained in this work with the moving cluster method significantly increase the number of Pleiades stars with known individual distances, thus  yielding a sample that is the largest one to date (even if we only consider the 64~stars with parallaxes derived from radial velocities).    

\section{Conclusions}

We have implemented a new version of the CPSM based on the MCMC method that exploits parallelism and exhibits good performance when dealing with rich clusters, such as the Pleiades, with several hundreds of stars. This new implementation of the CPSM was calibrated on the TW~Hydrae association and applied to the Pleiades open cluster. 

We performed extensive investigations to search for the CP position of the Pleiades cluster using (i) different samples of stars, (ii) an alternative approach based on the intersection of great circles defined by the stellar proper motions in the sky, and (iii) Monte Carlo simulations. Our analysis allowed us to accurately constrain the CP position, infer the velocity dispersion, and confirm 1210 stars as kinematic members of the cluster. 

The moving cluster method applied to our sample returned individual parallaxes for 64~stars based on their radial velocities, and approximate parallaxes for the remaining 1146 group members from the spatial velocity of the cluster. We emphasize that this is the largest sample of Pleiades stars with individual distances to date. Our results are in good agreement with previous parallaxes for individual stars published in the literature (excluding \textit{Hipparcos}), and demonstrate the good accuracy of the distance determination obtained in this work. We report in this paper a distance estimate of  $134.4^{+2.9}_{-2.8}$~pc that is consistent with previous distance determinations of the cluster given by different methods. In particular, our results support the recent distance determination of $136.2\pm 1.2$~pc, obtained from the trigonometric parallax method that is based on VLBI multi-epoch observations and that represents the most precise and direct alternative to measure distances nowadays. Our distance determination is also in good agreement with the distance of $133\pm 5$~pc obtained from the TGAS trigonometric parallaxes for the stars in common with our sample.

Our results are not consistent with the \textit{\textit{Hipparcos}} distance determination of $120.2\pm 1.9$~pc for the cluster, and they confirm the discrepancy of these parallaxes as reported extensively in the literature. The weighted mean distance of the non-\textit{\textit{Hipparcos}} methods, including the results of this work, is $135.0\pm 0.6$~pc, which exceeds the \textit{\textit{Hipparcos}} distance by roughly 12\%. 

This study represents one more step towards better constraining the distance of the Pleiades cluster before the upcoming and very precise trigonometric parallaxes from the \textit{Gaia} space mission. Our results are already consistent with the first trigonometric parallaxes included in the TGAS catalog for the brightest stars in the cluster and they will be useful in the next future for a comparative analysis of other cluster members that were not included in the \textit{Gaia} DR1.

\begin{acknowledgements}

We thank the referee for constructive comments that helped us to improve the manuscript. This work was funded by the S\~ao Paulo State Science Foundation (FAPESP). This research has made use of the computing facilities of the Laboratory of Astroinformatics at IAG/USP (S\~ao Paulo, Brazil). It has also made use of the SIMBAD database operated at the CDS (Strasbourg, France). 
\end{acknowledgements}

\bibliographystyle{aa}
\bibliography{references.bib}

\onecolumn
\appendix
\section{Great circle intersections}

An alternative approach for investigating the CP position of a moving group is based on the idea of representing the stellar proper motions of individual cluster members by great circles on the celestial sphere and calculating their intersections. One would expect the density of great circle intersections to be higher close to the CP position and its mirror-point \citep[see e.g.][]{Abad(2003),Galli(2012)}. 

As described in Sect.~3 of \citet{Galli(2012)}, the motion of each cluster member over a great circle on the celestial sphere can be described by the polar vector $\mathbf{p}$ that is given by
\begin{equation}\label{eq.2}
\mathbf{p}=\mathbf{r}\times\dot{\mathbf{r}}=
\mu_{\alpha}\cos\delta
\left(
      \begin{array}{c}
      -\cos\alpha\sin\delta \\
       -\sin\alpha\sin\delta\\
       \cos\delta\\
       \end{array}
\right)
-
\mu_{\delta}
\left(
      \begin{array}{c}
      -\sin\alpha \\
       \cos\alpha\\
       0\\
       \end{array}
\right)\, ,
\end{equation}
where $\mathbf{r}=(x,y,z)=(\cos\alpha\cos\delta,\sin\alpha\cos\delta,\sin\delta)$ is the position of a star with coordinates $(\alpha,\delta)$ and proper motion $(\mu_\alpha, \mu_\delta)$. The intersection of great circles is defined by $\mathbf{p_{i}}\times\mathbf{p_{j}}$, where $i$ and $j$ denote the polar vectors of two cluster members. Thus, the number of great circle intersection points is $N_{int}=n(n-1)$ for a sample of $n$ stars. 

\begin{figure*}[!h]
\begin{center}
\includegraphics[width=0.82\textwidth]{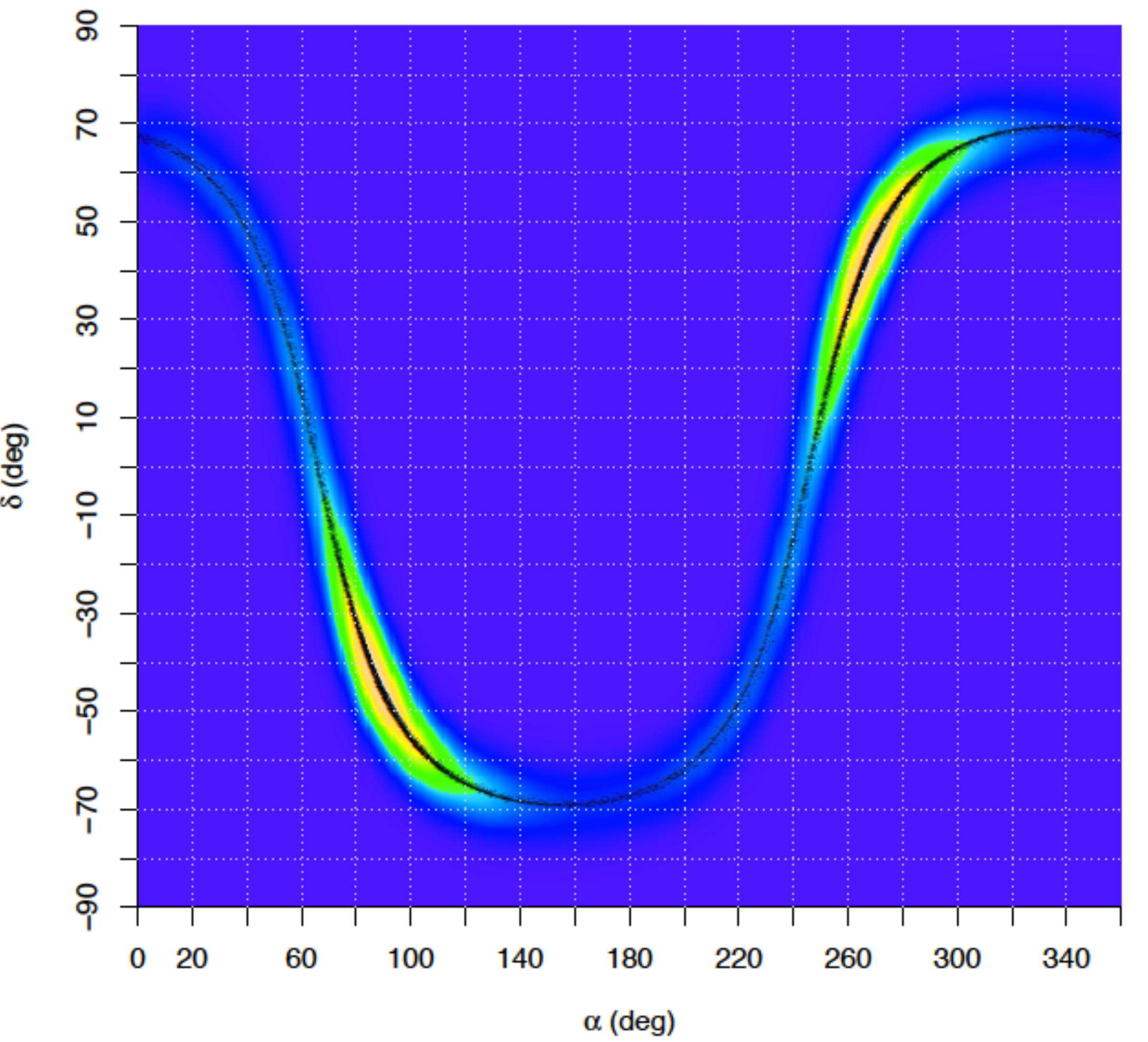}
\caption{
\label{figA1}
Diagram of great circle intersections for the stars in control sample 1. The colors indicate the regions with highest density of great circle intersections that are close to the CP position (and its mirror point). Each point (black dots) represents the intersection of two great circles in the sample.}
\end{center}
\end{figure*}

We calculate the great circle intersections for the stars in our control sample 1 (see Sect.~4.1) and display the results in Fig.~\ref{figA1}. We note that the great circle intersections roughly define a line connecting the CP with its mirror point. The two solutions correspond to the regions of highest density of great circle intersections in the diagram. They appear as an extended region (instead of a point), because of the proper motion errors and the velocity dispersion of the cluster, which prevent perfect parallelism of the stellar motions from occurring. It is clear from this analysis that a preferred direction of motion exists and the search for the CP position can be limited to a small area of the sky to optimize the methodology outlined in Section~2.1. Moreover, the diagram itself confirms the existence of a moving group structure in this sky region, and it can also be used to provide an independent estimate of the CP coordinates. 

\begin{figure*}[!h]
\begin{center}
\includegraphics[width=0.49\textwidth]{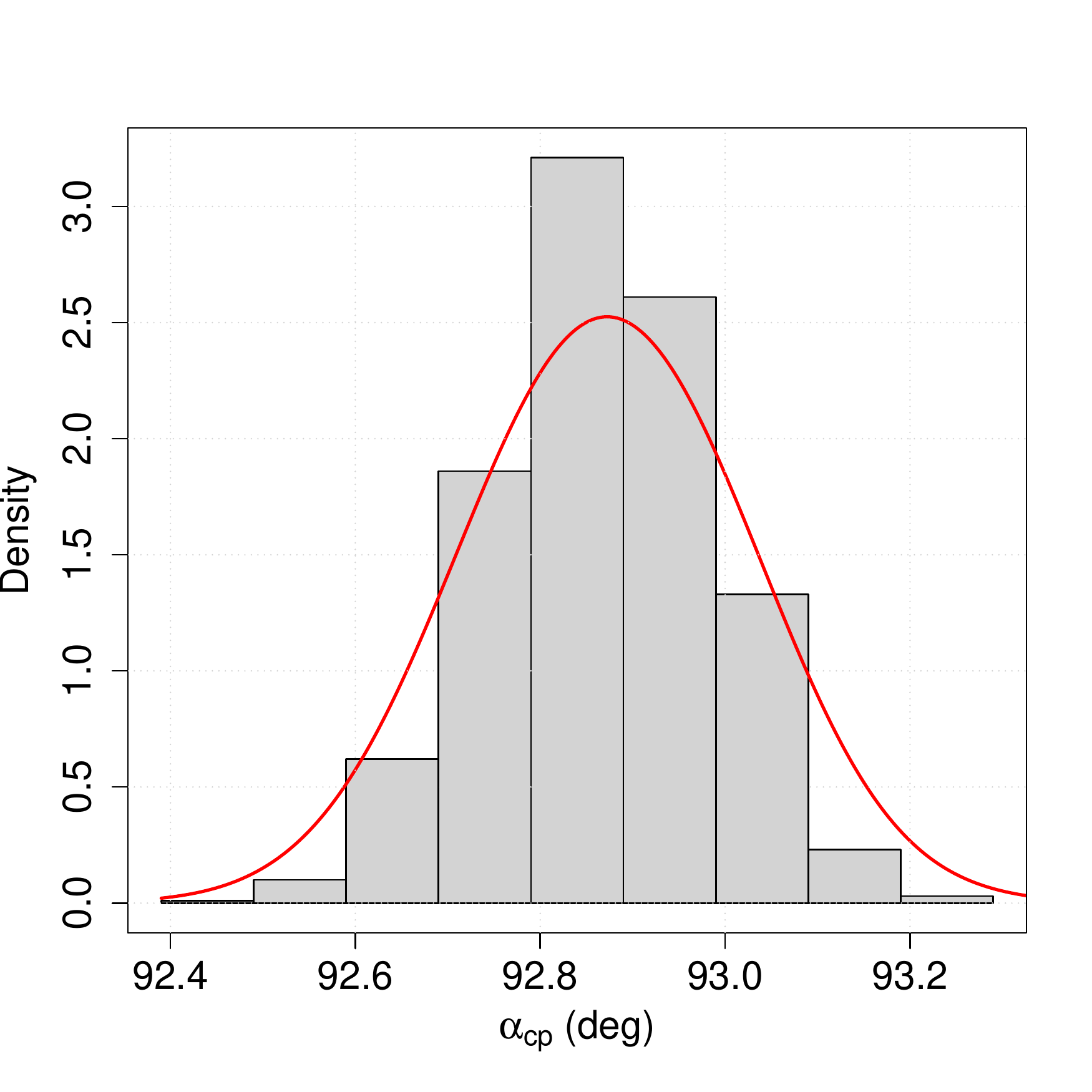}
\includegraphics[width=0.49\textwidth]{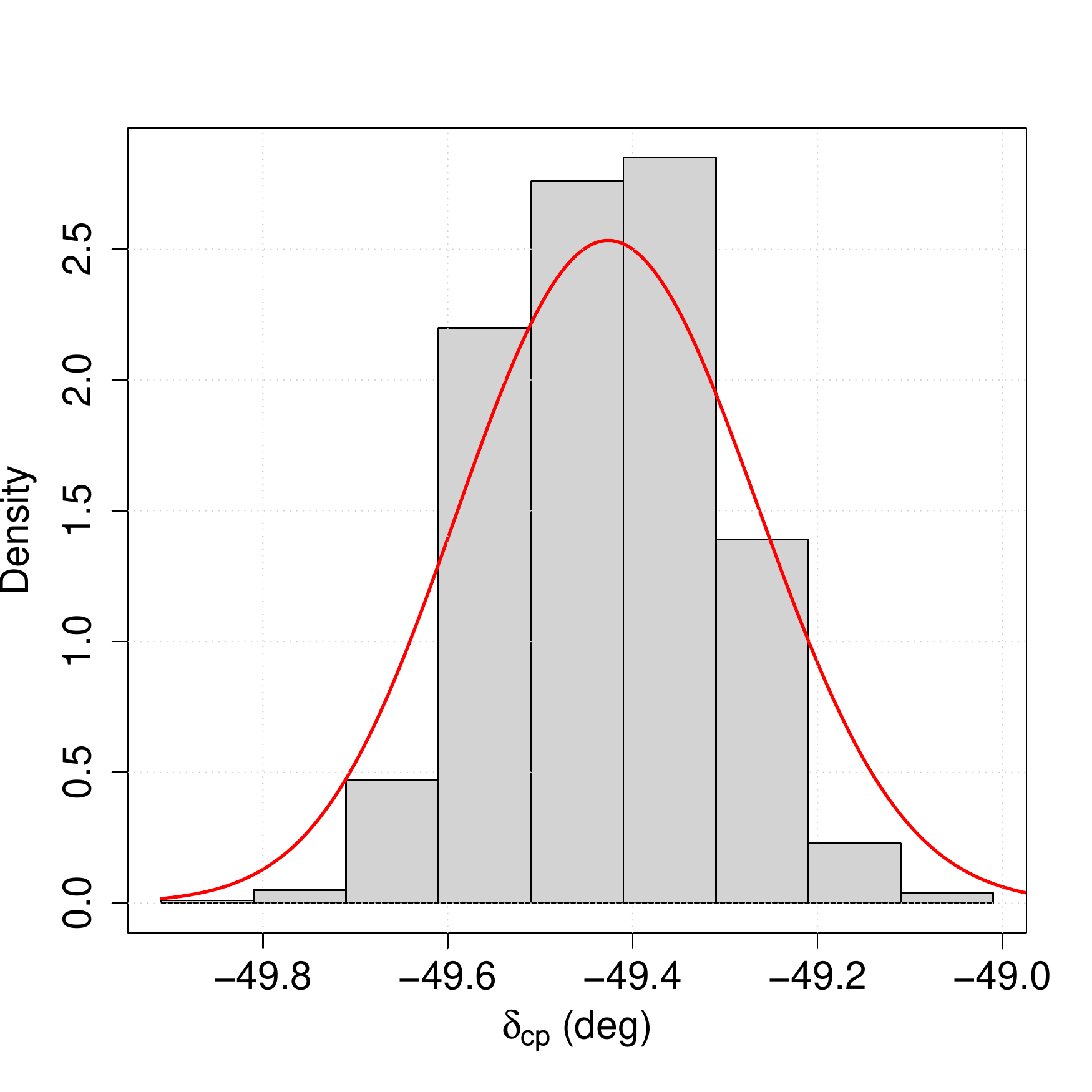}
\caption{
\label{figA2}
Histogram of the CP coordinates derived from the k-NN analysis of the great circle intersections using different number of neighbors (i.e., values of $k$). The red solid line indicates the kernel density estimator.}
\end{center}
\end{figure*}

In this context, one simple approach consists of running a k-NN regression. In this method, the k-NN algorithm \citep{Fix(1951),Venables(2002)} is used to predict the CP position, which is taken to be the weighted mean of the $k$ nearest neighbors (i.e., great circle intersections). We perform a weighted regression to weight the contribution of the neighbors by the inverse of their distance. The resulting CP position obviously depends on the number of nearest neighbors $k$ that is taken into account. To overcome this problem, we varied the value of $k$ in the range of $100\leq k\leq N_{int}$, and run the k-NN regression to compute the CP position as a function of $k$. The lower limit for $k$ is chosen to avoid a noisy CP solution that is derived with only a few neighbors. Figure~\ref{figA2} shows the distribution of the CP coordinates obtained from the k-NN regression. The average CP position of $(\alpha_{cp},\delta_{cp})=(92.2^{\circ},-48.6^{\circ})\pm(0.8^{\circ},0.9^{\circ})$ is in good agreement with the solution derived from the CPSM.

The methodology described in this section, which is based on a k-NN regression analysis of the great circle intersections provides an interesting tool to detect moving group structures in stellar catalogs and to estimate their CP coordinates. It is used to support and confirm our first CP solution obtained for a control sample of the Pleiades cluster. However, we emphasize that the CP position derived directly from the CPSM is likely to be more accurate, because it takes into account the proper motion errors and the velocity dispersion of the cluster.

\end{document}